\documentclass[%
	american,%
	openright,
	DIV=11,pagesize,%
	headinclude=true,headlines=0,footinclude=true,footlines=-5,twoside=semi%
]{scrartcl}%,envcountsame

\def\indraftmode{false}

\input{quickXsort_preamble}

\newboolean{citeproofs}
\setboolean{citeproofs}{false}

\def\mytitle{%
	QuickXsort~-- A Fast Sorting Scheme in Theory and Practice%
}
\title{\mytitle%
	\thanks{%
		Parts of this article have been presented (in preliminary form)
		at the \emph{International Computer Science Symposium in Russia (CSR) 2014}~\protect\cite{EdelkampWeiss2014}
		and at the \emph{International Conference on Probabilistic,
		Combinatorial and Asymptotic Methods for the
		Analysis of Algorithms (AofA) 2018}~\protect\cite{Wild2018a}.%
	}%
}
\author{%
	Stefan Edelkamp\and%
	Armin Weiß\and%
	Sebastian Wild%
}

\hypersetup{
	pdftitle={\mytitle},
	pdfauthor={Stefan Edelkamp,	Armin Weiß, Sebastian Wild},
	final,
	unicode=true, 
	bookmarks=true,
	bookmarksnumbered=true,
	bookmarksdepth=2,
	bookmarksopen=true,
	breaklinks=true,
	hidelinks,
}

\date{\today}

\newcommand{\holdA}{\eta}
\newcommand{\cmtA}{\sigma}
\newcommand{\cmtB}{\tau}
\newcommand{\betL}{\lambda}
\newcommand{\betR}{\rho}

\begin{document}

\maketitle

\begin{abstract}
\noindent\textbf{\textsf{Abstract.}}\;
QuickXsort is a highly efficient in-place sequential sorting scheme that
mixes Hoare's Quicksort algorithm with X, where X can be chosen
from a wider range of other known sorting algorithms, like Heapsort,
Insertionsort and Mergesort. Its major advantage is that QuickXsort can be in-place even if X is not.
In this work we provide general transfer theorems expressing the number of
comparisons of QuickXsort in terms of the number of comparisons of X. More
specifically, if pivots are chosen as medians of (not too fast) growing size samples, the
average number of comparisons of QuickXsort and X differ only by $o(n)$-terms. For
median-of-$k$ pivot selection for some constant $k$, the difference is a linear term whose
coefficient we compute precisely. For instance, median-of-three QuickMergesort uses at
most $n \lg n - 0.8358n + \Oh(\log n)$ comparisons.

Furthermore, we examine the possibility of sorting base cases with some other algorithm 
using even less comparisons. 
By doing so the average-case number of comparisons
can be reduced down to $n \lg n - 1.4106n + o(n)$ 
for a remaining gap
of only $0.0321n$ comparisons to the known lower bound 
(while using only $\Oh(\log n)$ additional space and $\Oh(n\log n)$ time overall).

Implementations of these sorting strategies
show that the algorithms challenge well-established
library implementations like Musser's Introsort.
\end{abstract}

\bigskip
\setlength\columnsep{2em}
\begin{multicols}2
\RedeclareSectionCommands[
	tocpagenumberformat={\smaller[2]},
]{section,subsection,subsubsection}
\smaller
\tableofcontents
\end{multicols}
\markleft{\mytitle} % overwrite toc mark

\section{Introduction}
\label{sec:intro}

Sorting a sequence of $n$ elements remains one of the most frequent tasks carried out by computers. 
In the comparisons model, the well-known lower bound for sorting $n$ distinct elements says
that using fewer than $\lg (n!) = n \lg n - \lg e \cdot n \pm \Oh(\log n) 
	\approx n \lg n - 1.4427 n + \Oh(\log n) $%
\footnote{
	We write $\lg$ for $\log_2$, but 
    use $\log$ to denote an otherwise unspecified
    logarithm in the $\Oh$ notation.
}
comparisons is not possible, both in the worst case
and in the average case.
The average case refers to a uniform distribution of all input permutations
(random-permutation model).

In many practical applications of sorting, element comparisons have a similar running-time
cost as other operations (\eg, element moves or control-flow logic).
Then, a method has to balance costs to be overall efficient.
This explains why Quicksort is generally considered the fastest general purpose sorting method,
despite the fact that its number of comparisons is slightly higher than for other methods.

There are many other situations, however, where comparisons do have significant costs,
in particular, when complex objects are sorted \wrt a order relation defined by a custom procedure.
We are therefore interested in algorithms whose comparison count is 
\emph{optimal up to lower order terms},
\ie, sorting methods that use $n\lg n + o(n \log n)$ or better $n\lg n + \Oh (n)$ comparisons;
moreover, we are interested in bringing the coefficient of the linear term
as close to the optimal $-1.4427$ as possible
(since the linear term is not negligible for realistic input sizes).
Our focus lies on \emph{practical} methods whose running time is competitive 
to standard sorting methods even when comparisons are cheap.
As a consequence, expected (rather than worst case) performance is our main concern.

We propose \QuickXsort as a general template for practical, comparison-efficient 
internal%
\footnote{
	Throughout the
	text, we avoid the (in our context somewhat ambiguous) terms \emph{in-place} or \emph{in-situ}.
	We instead call an algorithm \emph{internal}
	if it needs at most $\Oh(\log n)$ words of space (in addition to
	the array to be sorted). 
	In particular, \algorithmname{Quicksort} is an internal
	algorithm whereas standard \algorithmname{Mergesort} is not (hence called \emph{external}) 
	since it uses a linear amount of buffer space for merges. 
} 
sorting methods.
\QuickXsort we uses the recursive scheme of ordinary Quicksort, but
instead of doing two recursive calls after partitioning, first one of the segments is sorted by some other sorting method ``X''.
Only the second segment is recursively sorted by \QuickXsort.
The key insight is that X can use the second segment as a temporary buffer area;
so X can be an \emph{external} method, but the resulting \QuickXsort is still an \emph{internal} method.
\QuickXsort only requires $\Oh(1)$ words of extra space,
even when X itself requires a linear-size buffer.

We discuss a few concrete candidates for X to illustrate the versatility of \QuickXsort.
We provide a precise analysis of \QuickXsort in the form of ``transfer theorems'':
we express the costs of \QuickXsort in terms of the costs of X, where generally the use of
\QuickXsort adds a certain overhead to the lower order terms of the comparison counts.
Unlike previous analyses for special cases, our results give tight bounds.

A particularly promising (and arguably the most natural) candidate for X is \Mergesort.
\Mergesort is both fast in practice and comparison-optimal up to lower order terms;
but the linear-extra space requirement can make its usage impossible.
With \algorithmname{QuickMergesort} we describe an internal sorting algorithm
that is competitive in terms of number of comparisons and running time.

\paragraph{Outline}

The remainder of this section surveys previous work and summarizes the contributions
of this article.
We then describe \QuickXsort in detail in \wref{sec:quickxsort}.
In \wref{sec:prelims}, we introduce mathematical notation and recall known results 
that are used in our analysis of \QuickXsort. 
In \wref{sec:recurrence}, we postulate the general recurrence for \QuickXsort and 
describe the distribution of subproblem sizes. \wref{sec:analysis-growing-k} contains 
transfer theorems for growing size samples and \wref{sec:analysis-fixed-k} for
constant size samples. In \wref{sec:analysis-for-concrete-X}, we apply these transfer
theorems to \QuickMergesort and \algorithmname{QuickHeapsort} and discuss the results. 
In \wref{sec:variance} contains a transfer theorem for the variance of \QuickXsort. 
Finally, in \wref{sec:experiments} we present our
experimental results and conclude in \wref{sec:conclusion} with some open questions.

\subsection{Related work}

We pinpoint selected relevant works from the vast literature on sorting;
our overview cannot be comprehensive, though.

\paragraph{Comparison-efficient sorting}
There is a wide range of sorting algorithms achieving the bound of $n \lg n + \Oh(n)$ comparisons. The most prominent is \Mergesort, which additionally comes with a small coefficient in the linear term. Unfortunately, \Mergesort requires linear extra space. Concerning the space \algorithmname{UltimateHeapsort} \cite{Katajainen1998} does better, however, with the cost of a quite large linear term. 
Other algorithms, provide even smaller linear terms than \Mergesort.
\prettyref{tab:compare} lists some milestones in the
race for reducing the coefficient in the linear term.
Despite the fundamental nature of the problem, little improvement has been made
(\wrt the worst-case comparisons count) over the Ford
and Johnson's \algorithmname{MergeInsertion} algorithm \cite{FordJ59}~-- which was published 1959!
\algorithmname{MergeInsertion} requires 
$n \lg n - 1.329n + \Oh(\log n)$
comparisons in the worst case~\cite{Knuth1998}.

\begin{table}[tbph]
	\renewcommand{\thempfootnote}{\alph{mpfootnote}}
	\begin{threeparttable}
	\caption{%
		Milestones of comparison-efficient sorting methods. 
		The methods use (at most) $n \lg n + b n + o(n)$ comparisons
		for the given $b$ in worst ($b_{\mathrm{wc}}$) and/or 
		average case ($b_{\mathrm{ac}}$).
		Space is given in machine words (unless indicated otherwise).
	}
	\label{tab:compare}
	\small
	\def\notknown{\multicolumn1c{?}}
	\begin{tabular}{r ll lll}
	\toprule
		                                         \textbf{Algorithm} & {$b_{\mathrm{ac}}$} & {$b_{\mathrm{ac}}$} \textbf{empirical} & {$b_{\mathrm{wc}}$}  & \textbf{Space} & \textbf{Time}   \\
	\midrule
		                                                Lower bound & $-1.44$                  &                                             & $-1.44$                   & $\Oh(1)$       & $\Oh(n \log n)$ \\[1ex]
		                        \algorithmname{Mergesort}~\cite{Knuth1998} & $-1.24      $            &                                             & $-0.91$                   & $\Oh(n)$       & $\Oh(n \log n)$ \\
		                    \algorithmname{Insertionsort}~\cite{Knuth1998} & $-1.38$\tnotex{tn:new}   &                                             & $-0.91$                   & $\Oh(1)$       & $\Oh(n^2)$\tnotex{tn:moreplace}      \\
		                   \algorithmname{MergeInsertion}~\cite{Knuth1998} & $-1.3999$\tnotex{tn:new} & $[-1.43,-1.41] $                            & $-1.32$                   & $\Oh(n)$       & $\Oh(n^2)$\tnotex{tn:moreplace}      \\
		                    \algorithmname{MI+IS}~\cite{IwamaTeruyama2017} & $-1.4106$                &                                             &                           & $\Oh(n)$       & $\Oh(n^2)$\tnotex{tn:moreplace}      \\
		                     \algorithmname{BottomUpHeapsort} \cite{Weg93} & \notknown                & $[0.35,0.39]    $                           & $\omega(1)$               & $\Oh(1)$       & $\Oh(n \log n)$ \\
		                    \algorithmname{WeakHeapsort} \cite{Dut93,EW00} & \notknown                & $[-0.46,-0.42]  $                           & $0.09$                    & $\Oh(n)$ bits  & $\Oh(n \log n)$ \\
		      \algorithmname{RelaxedWeakHeapsort} \cite{edelkampstiegeler} & $-0.91      $            & $-0.91          $                           & $-0.91$                   & $\Oh(n)$       & $\Oh(n \log n)$ \\
		             \algorithmname{InPlaceMergesort} \cite{Reinhardt1992} & \notknown                &                                             & $-1.32$ & $\Oh(1)$        & $\Oh(n \log n)$ \\
		          \algorithmname{QuickHeapsort}~\cite{CantoneCincotti2002} & $-0.03$\tnotex{tn:ub}    & $\approx 0.20$                              & $\omega(1)$               & $\Oh(1)$       & $\Oh(n \log n)$ \\
		      Improved \algorithmname{QuickHeapsort}~\cite{DiekertWeiss2016} & $-0.99$\tnotex{tn:ub}    & $\approx -1.24$                             & $\omega(1)$               & $\Oh(n)$ bits  & $\Oh(n \log n)$ \\
		      		\algorithmname{UltimateHeapsort}~\cite{Katajainen1998} & $~\Oh(1)$    				&   $\approx 6$~\cite{DiekertWeiss2016}                           				 & $\Oh(1)$               & $\Oh(1)$  & $\Oh(n \log n)$ \\[1ex]
		                     \algorithmname{QuickMergesort}\tnotex{tn:new} & $-1.24   $               & $[-1.29,-1.27] $                            & $-0.32$\tnotex{tn:qms-wc} & $\Oh(1)$       & $\Oh(n \log n)$ \\
		\algorithmname{QuickMergesort} (IS)\tnotex{tn:new}~~\tnotex{tn:rs} & $-1.38   $               &                                             & $-0.32$\tnotex{tn:qms-wc} & $\Oh(\log n)$  & $\Oh(n \log n)$ \\
		\algorithmname{QuickMergesort} (MI)\tnotex{tn:new}~~\tnotex{tn:rs} & $-1.3999 $               & $[-1.41,-1.40] $                            & $-0.32$\tnotex{tn:qms-wc} & $\Oh(\log n)$  & $\Oh(n \log n)$ \\
     \algorithmname{QuickMergesort} (MI+IS)\tnotex{tn:new}~~\tnotex{tn:rs} & $-1.4106 $               &                                             & $-0.32$\tnotex{tn:qms-wc} & $\Oh(\log n)$  & $\Oh(n \log n)$ \\
                
	\bottomrule
	\end{tabular}
	\begin{tablenotes}
	\item[\#] \label{tn:new} in this paper
	\item[$\le$] \label{tn:ub} only upper bound proven in cited source
	\item[\dag] \label{tn:qms-wc}  assuming \algorithmname{InPlaceMergesort} as a worst-case stopper; with median-of-medians fallback pivot selection: $\Oh(1)$, without worst-case stopper: $\omega(1)$
	\item[$\bot$] \label{tn:rs} using given method for small subproblems;
				MI = \algorithmname{MergeInsertion}, 
				IS = \algorithmname{Insertionsort}. 
	\item[$\diamond$] \label{tn:moreplace} using a rope data structure and allowing additional $\Oh(n)$ space in $\Oh(n \log^2 n)$.		
	\end{tablenotes}
	\end{threeparttable}
\end{table}

\algorithmname{MergeInsertion} has a severe
drawback that renders the algorithm completely impractical, though: in a naive
the number of element moves is quadratic in $n$. Its running time can be improved to $\Oh(n \log^2 n)$ by using a rope data structure \cite{BoehmAP95} (or a similar data structure which allows random access and insertions in $\Oh(\log n)$ time) for insertion of elements (which, of course, induces additional constant-factor overhead). 
The same is true for \algorithmname{Insertionsort}, which, 
unless explicitly indicated otherwise, refers to the algorithm that inserts elements
successively into a sorted prefix by finding the insertion position by \emph{binary search}~-- 
as opposed to linear/sequential search in \algorithmname{StraightInsertionsort}.
Note that \algorithmname{MergeInsertion} or \algorithmname{Insertionsort} can still be used 
as comparison-efficient subroutines 
to sort base cases for \Mergesort (and \QuickMergesort) of size $\Oh( \log n)$
without affecting the overall running-time complexity of $\Oh(n \log n)$.

Reinhardt~\cite{Reinhardt1992} used this
trick (and others) to design an internal \algorithmname{Mergesort} variant that needs 
$n \lg n - 1.329n \pm \Oh(\log n)$ comparisons in the worst case.
Unfortunately, implementations of this \algorithmname{InPlaceMergesort} algorithm have not been
documented. 
Katajainen et al.'s~\cite{KatajainenPasanenJukka1996,GeffertKP00,ElmasryKS12} work
inspired by Reinhardt is practical, but the
number of comparisons is larger. 

Improvements over \algorithmname{MergeInsertion} have been obtained for the \emph{average} 
number of comparisons. A combination of \algorithmname{MergeInsertion} with a variant of \algorithmname{Insertionsort} (inserting two elements simultaneously) by Iwama and Teruyama uses $\le n \lg n -1.41064n$ 
comparisons on average~\cite{IwamaTeruyama2017}; as for \Mergeinsertion the overall complexity of remains quadratic (resp.\ $\Theta(n\log^2 n)$), though. 
Notice that the analysis in \cite{IwamaTeruyama2017} is based on 
 our bound on \algorithmname{MergeInsertion} in \wref{sec:mi}.

\paragraph{Previous work on QuickXsort}

Cantone and Cincotti~\cite{CantoneCincotti2002} were the first to explicitly 
naming the mixture of Quicksort with another sorting method;
they proposed \algorithmname{QuickHeapsort}.
However, the concept of \QuickXsort (without calling it like that) was 
first used in \algorithmname{UltimateHeapsort} by Katajainen~\cite{Katajainen1998}.
Both versions use an external \algorithmname{Heapsort} variant in which a heap containing $m$ elements
is not stored compactly in the first $m$ cells of the array, but may be spread out over
the whole array.
This allows to restore the heap property with $\lceil \lg n\rceil$ comparisons after extracting some element by introducing a new gap (we can think of it as an element of infinite weight) and letting it sink down to the bottom of the heap. The extracted elements are stored in an output buffer.

In \algorithmname{UltimateHeapsort}, we first find the
exact median of the array (using a linear time algorithm) and then partition the array into subarrays of equal size;
this ensures that with the above external \algorithmname{Heapsort} variant, the first half of the array (on which the heap is built)
does not contain gaps (Katajainen calls this a two-level heap); the other half of the array is used as the output buffer. 
\algorithmname{QuickHeapsort} avoids the significant additional effort for 
exact median computations by choosing the pivot as median of some smaller sample. In our terminology, it applies \QuickXsort where X is 
external Heapsort.
\algorithmname{UltimateHeapsort} is inferior to \algorithmname{QuickHeapsort} in terms of the 
average case number of comparisons, although, unlike \algorithmname{QuickHeapsort}, 
it allows an $n\lg n + \Oh(n)$ bound for the worst case number of comparisons. 
Diekert and Weiß~\cite{DiekertWeiss2016} analyzed \algorithmname{QuickHeapsort} more
thoroughly and described some improvements requiring less than $n\lg n -0.99 n
+\oh(n)$ comparisons on average (choosing the pivot as median of $\sqrt{n}$ elements).
However, both the original analysis of Cantone and Cincotti and the improved analysis 
could not give tight bounds for the average case of median-of-$k$ \QuickMergesort.

In \cite{ElmasryKS12} Elmasry, Katajainen and Stenmark proposed
\algorithmname{InSituMergesort}, following the same principle as
\algorithmname{UltimateHeapsort} but with \Mergesort replacing
\algorithmname{ExternalHeapsort}. Also \algorithmname{InSituMergesort} only uses an
expected linear algorithm for the median computation.

In the conference paper \cite{EdelkampWeiss2014}, the first and second author introduced
the name \QuickXsort and first considered \QuickMergesort as an application (including
weaker forms of the results in \wref{sec:analysis-growing-k} and \wref{sec:basecase}
without proofs). 
In \cite{Wild2018a}, the third author analyzed \QuickMergesort with constant-size pivot
sampling (see \wref{sec:analysis-fixed-k}). A weaker upper bound for the median-of-3 case
was also given by the first two authors in the preprint \cite{EdelkampWeiss2018}. The
present work is a full version of \cite{EdelkampWeiss2014} and \cite{Wild2018a}; 
it unifies and strengthens these results (including all proofs)
and it complements the theoretical findings with extensive running-time experiments.

\subsection{Contributions}

In this work, we introduce \QuickXsort as a general template for transforming an external
algorithm into an internal algorithm. As examples we consider \QuickHeapsort and
\QuickMergesort. 
For the readers convenience, we collect our results here 
(with references to the corresponding sections).

\begin{itemize}
	\item 	If X is some sorting algorithm requiring 
	$x(n) = n \lg n + b n \pm o(n)$ comparisons on expectation and $k(n) \in \omega(1) \cap o(n)$. 
	Then, median-of-$k(n)$ \QuickXsort needs 
	$x(n) \pm o(n)$
	comparisons in the average case (\wref{thm:quickXsort}).
	
	\item Under reasonable assumptions, sample sizes of $\sqrt{n}$ are optimal among all polynomial size sample sizes.

	\item The probability that median-of-$\sqrt{n}$ \QuickXsort{} needs more than
	$x_{\mathrm{wc}}(n) + 6n$ comparisons decreases exponentially in $\sqrt[4]{n}$ (\wref{pro:worstunlikely}).
	\item  We introduce \emph{median-of-medians fallback pivot selection} 
	(a trick similar to \algorithmname{Intro\-sort}~\cite{Mus97})
	which guarantees $n\lg n + \Oh(n)$ comparisons in the worst case while altering the average case only by $o(n)$-terms (\wref{thm:QuickXYsort}).
	
	\item 
Let $k$ be fixed and let X be a sorting method that needs a buffer of $\lfloor \alpha n\rfloor$ elements 
	for some constant $\alpha\in [0,1]$ to sort $n$ elements
	and requires on average $x(n)=n\lg n +bn \pm o(n)$ comparisons to do so. Then median-of-$k$ \QuickXsort{} needs
	\begin{align*}
	c(n)
	&\wwrel=
	n \lg n + (P(k,\alpha) +b ) \cdot n
	\wbin\pm o(n),
	\end{align*}	
	comparisons on average where $P(k,\alpha)$ is some constant depending on $k$ and $\alpha$ (\wref{thm:cn}). 
	We have $P(1,1) = 0.5070$ (for median-of-3 \algorithmname{QuickHeapsort} or \algorithmname{QuickMergesort}) and  $P(1,1/2) = 0.4050$ (for median-of-3 \algorithmname{QuickMergesort}).

	\item 
	We compute the standard deviation of the number of comparisons of median-of-$k$
	\QuickMergesort for some small values of $k$. 
	For $k=3$ and $\alpha = \frac12$, the standard deviation
	is $0.3268 n$ (\wref{sec:variance}).

	\item 
	When sorting small subarrays of size $\Oh(\log n)$ in \QuickMergesort with some sorting algorithm $Z$ using
	$z(n) = n \lg n + (b \pm \epsilon) n + o(n)$ comparisons on average and other operations
	taking at most $\Oh(n^2)$ time, then \QuickMergesort needs $z(n) + o(n)$ comparisons on average (\wref{cor:QMSbase}). In order to apply this result, we prove that	
	\begin{itemize}
		\item (Binary) \algorithmname{Insertionsort} needs $n \lg n - (1.3863 \pm 0.005)n + o(n)$ comparisons on average (\wref{pro:avgIns}).
		\item (A simplified version of) \algorithmname{MergeInsertion} \cite{FordJohnson1959} needs at most $n \lg n - 1.3999n+ \oh(n)$ on average (\wref{thm:mi}). 
	\end{itemize}
	Moreover, with Iwama and Teruyama's algorithm \cite{IwamaTeruyama2017} this can be improved sightly to $n \lg n	- 1.4106n + \oh(n)$ comparisons (\wref{cor:QMSbaseTI}).
	
	\item We run experiments confirming our theoretical (and heuristic) estimates for the average number of comparisons of \QuickMergesort and its standard deviation and verifying that the sublinear terms are indeed negligible (\wref{sec:experiments}).

	\item From running-time studies comparing \QuickMergesort with various other sorting methods,
	we conclude that our \QuickMergesort implementation is among the fastest internal general-purpose 
	sorting methods for both the regime of cheap and expensive comparisons (\wref{sec:experiments}).
	
\end{itemize}

To simplify the arguments, in all our analyses we assume that 
all elements in the input are distinct. 
This is no severe restriction since duplicate elements can be handled 
well using fat-pivot partitioning 
(which excludes elements equal to the pivot from recursive calls and calls to X).

\section{QuickXsort}
\label{sec:quickxsort}

In this section we give a more precise description of \algorithmname{QuickXsort}.
Let X be a sorting method that requires buffer space for storing
at most $\lfloor \alpha n\rfloor$ elements (for $\alpha \in [0,1]$) 
to sort $n$ elements. The buffer may only be accessed by swaps 
so that once X has finished its work, the buffer contains the same elements as before, 
albeit (in general) in a different order than before.

\begin{figure}[tbhp]
	\begin{captionbeside}{
			Schematic steps of \QuickXsort.
			The pictures show a sequence, where the vertical height
			corresponds to key values.
			We start with an unsorted sequence (top),
			and partition it around a pivot value (second from top).
			Then one part is sorted by X (second from bottom)
			using the other segment as buffer area (grey shaded area).
			Note that this in general permutes the elements there.
			Sorting is completed by applying the same procedure recursively
			to the buffer (bottom).
		}
		\begin{tikzpicture}[xscale=.4,yscale=.35]
		\def\ysep{5}
		\draw (13,0) -| (1,2.9)
		plot[smooth] coordinates { (1,2.9) (2,1.7) (3,0.7) (4,0.9) (5,1.1) (6,2.1) (7,1.3) (8,2.7) (9,1.9) (10,0.5) (11,2.5) (12,2.3) (13,1.5) }
		(13,1.5)
		|- (13,0) -- cycle;
		
		\begin{scope}[shift={(0,-\ysep)}]
			\fill[black!20] (13,0) -| (8.1,2)
			-- plot[smooth] coordinates { (8.1,2) (9,2.7) (10,2.1) (11,2.5) (12,2.3) (13,2.9)  }
			(13,2.9)
			|- (13,0) -- cycle;
			\draw[thick,black!50,densely dashed] (1,2) -- (13,2);
			\draw (13,0) -| (1,1.5)
			plot[smooth] coordinates { (1,1.5) (2,1.7) (3,0.7) (4,0.9) (5,1.1) (6,0.5) (7,1.3) (8,1.9) (9,2.7) (10,2.1) (11,2.5) (12,2.3) (13,2.9)  }
			(13,2.9)
			|- (13,0) -- cycle;
			\draw (8.1,0) -- ++(0,2) ;
			\draw[decoration=brace,decorate] (8.1,-.3) -- node[below=2pt] {\smaller sort by X} ++(-7.1,0);
		\end{scope}
		
		\begin{scope}[shift={(0,-2*\ysep)}]
			\draw (13,0) -| (1,0.5)
			-- (8.1,2) -- cycle;
			\draw[fill=black!20]
			(13,0) -| (8.1,2) --
			plot[smooth] coordinates { (8.1,2) (9,2.5) (10,2.1) (11,2.9) (12,2.7) (13,2.3)  }
			(13,2.3) |- (13,0) ;
			\draw (8.1,0) -- ++(0,2) ;
			\draw[decoration=brace,decorate] (13,-.3) -- node[below=2pt] {\smaller sort recursively} ++(-4.9,0);
		\end{scope}
		
		\begin{scope}[shift={(0,-3*\ysep)}]
			\draw (13,0) -| (1,0.5) -- (13,2.9) -- cycle;
			\draw (8.1,0) -- ++(0,1.9) ;
		\end{scope}
		\end{tikzpicture}
	\end{captionbeside}
	\label{fig:quickxsort-schematic}
\end{figure}

\algorithmname{QuickXsort} now works as follows:
First, we choose a pivot element; typically we use the median of a random sample of the input.
Next, we partition the array according to this
pivot element, \ie, we rearrange the array so that all elements left of the pivot are less or 
equal and all elements on the right are greater or equal than the pivot element. 
This results in two contiguous segments of $J_1$ resp.\ $J_2$ elements;
we exclude the pivot here (since it will have reached its final position), so $J_1+J_2 = n-1$.
Note that the (one-based) rank $R$ of the pivot is random, and so are the segment sizes $J_1$ and $J_2$.
We have $R=J_1+1$ for the rank.

We then sort one segment by X \emph{using the other segment as a buffer.}
To guarantee a sufficiently large buffer for X when it sorts $J_r$ ($r=1$ or $2$), 
we must make sure that $J_{3-r} \ge \alpha J_r$. 
In case both segments could be sorted by X, we use the larger of the two.
After one part of the array has been
sorted with X, we move the pivot element to its correct position 
(right after/before the already sorted part) and recurse on the other
segment of the array.
The process is illustrated in \wref{fig:quickxsort-schematic}.

The main advantage of this procedure is that the part of the array that  
is not currently being sorted can be used as temporary buffer area for 
algorithm~X. This yields fast \emph{internal} variants for
various \emph{external} sorting algorithms such as \algorithmname{Mergesort}. 
We have to make sure, however, that the contents of the buffer is not lost.
A simple sufficient condition is to require that X to maintains a permutation
of the elements in the input and buffer:
whenever a data element should be moved to the external storage, 
it is \emph{swapped} with the data element occupying that respective position 
in the buffer area.
For \Mergesort, using swaps in the merge (see \wref{sec:quickmergesort}) is sufficient.
For other methods, we need further modifications.

\begin{remark}[Avoiding unnecessary copying]
For some X, it is convenient to have the sorted sequence reside in the buffer area
instead of the input area.
We can avoid unnecessary swaps for such X
by partitioning ``in reverse order'', \ie, so that large elements are left of the pivot
and small elements right of the pivot.
\end{remark}

\paragraph{Pivot sampling}
\label{sec:pivot-sampling}

It is a standard strategy for \Quicksort to choose pivots as the median of some sample.
This optimization is also effective for \QuickXsort and we will study its effect in detail.
We assume that in each recursive call, we choose a sample of $k$ elements,
where $k=2t+1$, $t\in\N_0$ is an odd number. The sample can either be selected deterministically 
(\eg some fixed positions) or at random. Usually for the analysis we do not need random selection; 
only if the algorithm X does not preserve randomness of the buffer element, 
we have to assume randomness (see \wref{sec:recurrence}). 
However, notice that in any case random selection might be beneficial as it protects 
against against a potential adversary who provides a worst-case input permutation.

Unlike for \Quicksort, in \QuickXsort pivot selection contributes only a 
minor term to the overall running time (at least in the usual case that $k \ll n$). 
The reason is that \QuickXsort only makes a \emph{logarithmic} number of 
partitioning rounds in expectation (while \Quicksort always makes a linear 
number of partitioning rounds) since in expectation after each partitioning 
round constant fraction of the input is excluded from further consideration 
(after sorting it with X).
Therefore, we do not care about details of how pivots are selected, 
but simply assume that selecting the median of $k$ elements needs $s(k) =\Theta(k)$ 
comparisons on average (\eg\ using \algorithmname{Quickselect} \cite{Hoare61_find}).

We consider both the case where $k$ is a fixed constant and where $k = k(n)$ is an increasing function
of the (sub)problem size.
Previous results in \cite{DiekertWeiss2016,MartinezRoura2001} for \Quicksort suggest that 
sample sizes $k(n) = \Theta(\sqrt{n})$ are likely to be optimal asymptotically,
but most of the relative savings for the expected case are already realized for $k\le 10$.
It is quite natural to expect similar behavior in QuickXsort,
and it will be one goal of this article to precisely quantify these statements.

\subsection{QuickMergesort}
\label{sec:quickmergesort}

A natural candidate for X is Mergesort: it is comparison-optimal up to the linear term
(and quite close to optimal in the linear term),
and needs a $\Theta(n)$-element-size buffer for practical implementations of merging.%
\footnote{%
	Merging can be done in place using more advanced tricks (see, \eg, \cite{GeffertKP00,MannilaUkkonen1984}), 
	but those tend not to be competitive
	in terms of running time with other sorting methods.
	By changing the global structure, a ``pure'' internal Mergesort variant~\cite{KatajainenPasanenJukka1996} 
	can be achieved using part of the input as a buffer (as in QuickMergesort) 
	at the expense of occasionally having to merge runs of very different lengths.
}

\begin{algorithm}[tbhp]
{
	\def\;{\hspace{.5em}}
	\begin{codebox}
	\CLRSProcname{$\proc{SimpleMergeBySwaps}(A[\ell..r],m, B[b..e])$}
	\zi \CLRSComment Merges runs $A[\ell,m-1]$ and $A[m..r]$ in-place into $A[l..r]$ using scratch space $B[b..e]$
	\li $n_1\gets r-\ell+1$; \; $n_2\gets r-\ell+1$
	\zi \CLRSComment Assumes $A[\ell,m-1]$ and $A[m..r]$ are sorted, $n_1 \le n_2$ and $n_1 \le e-b+1$.
	\li \kw{for} $i = 0,\ldots,n_1-1$ 
		\CLRSDo
		\li		$\proc{Swap}(A[\ell+i], B[b+i])$ 
		\CLRSEnd
	\li	\kw{end for}
	\li $i_1 \gets b$; \; $i_2 \gets m$; $o \gets \ell$
	\li \CLRSWhile $i_1 < b + n_1$ and $i_2 \le r$
	\CLRSDo
		\li \CLRSIf $B[i_1] \le A[i_2]$
		\CLRSDo
			\li $\proc{Swap}(A[o],B[i_2])$; \; $o\gets o+1$; \; $i_1\gets i_1+1$
		\li \CLRSElse
			\li $\proc{Swap}(A[o],A[i_1])$; \; $o\gets o+1$; \; $i_2\gets i_2+1$
		\CLRSEnd
		\li \kw{end if}
	\CLRSEnd
	\li \kw{end while}
	\li \CLRSWhile $i_1 < b + n_1$ 
		\CLRSDo
	\li		$\proc{Swap}(A[o],B[i_2])$; \; $o\gets o+1$; \; $i_1\gets i_1+1$ 
		\CLRSEnd
	\li \kw{end while}
	\end{codebox}
}
	\caption{\strut%
		Simple merging procedure that uses the buffer only by swaps.
		We move the first run $A[\ell..m-1]$ into the buffer $B[b..b+n_1-1]$ 
		and then merge it with the second run $A[m..r]$ (still 
		in the original array) into the empty slot left by the first run. 
		By the time this first half is filled, we either have consumed enough of the second run 
		to have space to grow the merged result, or the merging was trivial, \ie, 
		all elements in the first run were smaller.%
	\strut}
	\label{alg:merge}
\end{algorithm}

\begin{figure}[thbp]
	\small
	
	\plaincenter{
		\mbox{\makeboxlike[l]{\textbf{\textsf{Step 2:}}}{\textbf{\textsf{Step 1:}}}}%
		\qquad
			\begin{scriptsize}
				\begin{tikzpicture}[scale = 0.55]
				
				\draw(0,0) -- (9,0);
				\draw(0,0) -- (3,1);
				\draw(6,1) -- (9,1);
				\draw(3,0) -- (6,1);

				\draw[thick](6,0) -- (6,1);
				
				\draw[->](1.5,0.5) ..controls(1.5,2) and (7.5, 2 ).. node[above] {swap} (7.5,1);
				
				\draw(3,0) -- (3,1);
				\draw(9,0) -- (9,1);
				\end{tikzpicture}
			\end{scriptsize}		
	}
	
	\plaincenter{
		\mbox{\makeboxlike[l]{\textbf{\textsf{Step 2:}}}{\textbf{\textsf{Step 2:}}}}%
		\qquad
			\begin{scriptsize}
				\begin{tikzpicture}[scale = 0.55]
				
				\draw(0,0) -- (9,0);
				\draw(0,1) -- (3,1);
				\draw(6,0) -- (9,1);
				\draw(3,0) -- (6,1);
				\draw(0,0) -- (0,1);

				\draw[thick](6,0) -- (6,1);
				
				\draw[<-](1.5,1) ..controls(1.5,2) and (7.5, 2.2 ).. node[above] {merge} (7.5,.5);
				\draw[<-](1.5,1) ..controls(1.5,1.6) and (4.5, 2.2 )..   (4.5,.5);
				
				\draw(3,0) -- (3,1);
				\draw(9,0) -- (9,1);
				\end{tikzpicture}
			\end{scriptsize}
	}
	
	\bigskip
	\medskip
	
	\plaincenter{
		\mbox{\makeboxlike[l]{\textbf{\textsf{Step 2:}}}{\textbf{\textsf{Result:}}}}%
		\qquad
			\begin{scriptsize}
			\begin{tikzpicture}[scale = 0.55]
			
			\draw(0,0) -- (9,0);
			\draw(0,0) -- (6,1);
			\draw(6,1) -- (9,1);

			\draw[thick](6,0) -- (6,1);
						
			\draw(9,0) -- (9,1);
			\end{tikzpicture}
		\end{scriptsize}
	}

	\caption{%
		Usual merging procedure where one of the two runs fits into the buffer.
	}
	\label{fig:merge-simple}
\end{figure}
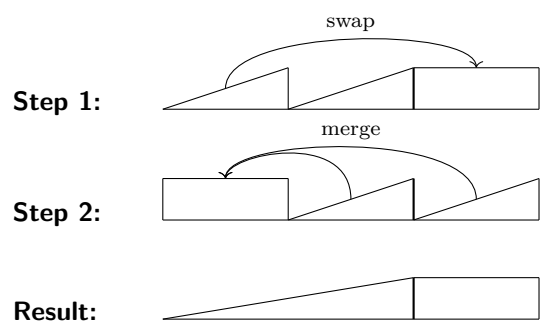

\paragraph{Simple swap-based merge}
To be usable in QuickXsort, we use a swap-based merge procedure as given in \wref{alg:merge}.
Note that it suffices to move the \emph{smaller} of the two runs to a buffer (see \wref{fig:merge-simple});
we use a symmetric version of \wref{alg:merge} when the second run is shorter.
Using classical top-down or bottom-up Mergesort as described in 
any algorithms textbook (\eg~\cite{SedgewickWayne2011}),
we thus get along with $\alpha=\frac12$.

The code in \wref{alg:merge} illustrates that very simple adaptations
suffice for \QuickMergesort.
This merge procedure leaves the merged result in the range
previously occupied by the two input runs.
This ``in-place''-style interface comes at the price of copying one run.

\paragraph{``Ping-pong'' merge}
Copying one run can be avoided if we instead write the merge result into
an output buffer (and leave it there).
This saves element moves, but uses buffer space for all $n$ elements, so we have $\alpha=1$ here.
The \Mergesort scaffold has to take care to correctly orchestrate
the merges, using the two arrays alternatingly;
this alternating pattern resembled the ping-pong game.

\paragraph{``Ping-pong'' merge with smaller buffer}

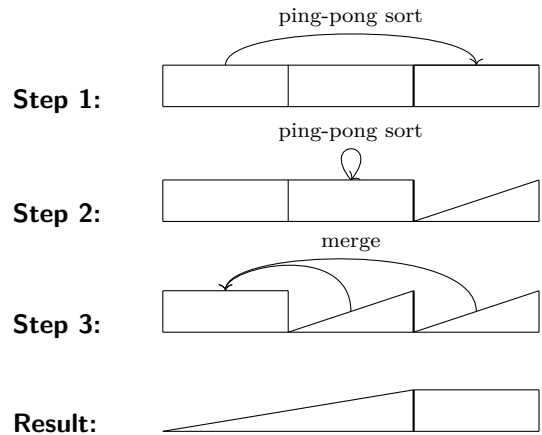
\begin{figure}[thbp]
	\small
	
	\plaincenter{
		\mbox{\makeboxlike[l]{\textbf{\textsf{Step 2:}}}{\textbf{\textsf{Step 1:}}}}%
		\qquad
		\begin{scriptsize}
			\begin{tikzpicture}[scale = 0.55]
			
			\draw(0,0) -- (9,0);
			\draw(0,1) -- (9,1);
			\draw(6,1) -- (9,1);
			\draw(0,0) -- (0,1);

			\draw[thick](6,0) -- (6,1);
			
			\draw[->](1.5,1) ..controls(1.5,2) and (7.5, 2 ).. node[above] {ping-pong sort} (7.5,1);
			
			\draw(3,0) -- (3,1);
			\draw(9,0) -- (9,1);
			\end{tikzpicture}
		\end{scriptsize}		
	}
	
	\plaincenter{
	\mbox{\makeboxlike[l]{\textbf{\textsf{Step 2:}}}{\textbf{\textsf{Step 2:}}}}%
	\qquad
	\begin{scriptsize}
		\begin{tikzpicture}[scale = 0.55]
		
		\draw(0,0) -- (9,0);
		\draw(0,1) -- (6,1);
		\draw(6,0) -- (9,1);
		\draw(0,0) -- (0,1);

		\draw[thick](6,0) -- (6,1);
		
		\draw[->](4.5,1) ..controls(3.7,2) and (5.3, 2 ).. node[above] {ping-pong sort} (4.5,1);
		
		\draw(3,0) -- (3,1);
		\draw(9,0) -- (9,1);
		\end{tikzpicture}
	\end{scriptsize}		
}

	\plaincenter{
	\mbox{\makeboxlike[l]{\textbf{\textsf{Step 2:}}}{\textbf{\textsf{Step 3:}}}}%
	\qquad
	\begin{scriptsize}
		\begin{tikzpicture}[scale = 0.55]
		
		\draw(0,0) -- (9,0);
		\draw(0,1) -- (3,1);
		\draw(6,0) -- (9,1);
		\draw(3,0) -- (6,1);
		\draw(0,0) -- (0,1);

		\draw[thick](6,0) -- (6,1);
		
		\draw[<-](1.5,1) ..controls(1.5,2) and (7.5, 2.2 ).. node[above] {merge} (7.5,.5);
		\draw[<-](1.5,1) ..controls(1.5,1.6) and (4.5, 2.2 )..   (4.5,.5);
		
		\draw(3,0) -- (3,1);
		\draw(9,0) -- (9,1);
		\end{tikzpicture}
	\end{scriptsize}
}
	
	\bigskip
	\medskip
	
	\plaincenter{
		\mbox{\makeboxlike[l]{\textbf{\textsf{Step 2:}}}{\textbf{\textsf{Result:}}}}%
		\qquad
		\begin{scriptsize}
			\begin{tikzpicture}[scale = 0.55]
			
			\draw(0,0) -- (9,0);
			\draw(0,0) -- (6,1);
			\draw(6,1) -- (9,1);

			\draw[thick](6,0) -- (6,1);

			\draw(9,0) -- (9,1);
			\end{tikzpicture}
		\end{scriptsize}
	}

	\caption{%
		Mergesort with $\alpha=1/2$ using ping-pong merges.
	}
	\label{fig:merge-ping-pong-alpha-0.5}
\end{figure}

It is also possible to implement the ``ping-pong'' merge with $\alpha= \frac{1}{2}$.
Indeed, the copying in \wref{alg:merge} can be avoided by sorting the first run with the
``ping-pong'' merge. This will automatically move it to the desired position in the buffer
and the merging can proceed as in \wref{alg:merge}. 
\wref{fig:merge-ping-pong-alpha-0.5} illustrates this idea, 
which is easily realized with a recursive procedure.
Our implementation of \QuickMergesort uses this variant.

\begin{figure}[thbp]
	\small
	
	\plaincenter{
	\mbox{\makeboxlike[l]{\textbf{\textsf{Step 2:}}}{\textbf{\textsf{Step 1:}}}}%
	\qquad
	\begin{scriptsize}
		\begin{tikzpicture}[scale = 0.55]
		
		\draw(0,0) -- (10,0);
		\draw(0,0.2) -- (2,0.2);
		\draw(2,0.2) -- (6,0.8);
		\draw(0,0) -- (0,0.2);
		\draw(6,0.2) -- (10,0.8);

		\draw[->](6.3,.3) ..controls(6.2,2) and (1, 2 ).. (0,.3);
		\draw[->](2.1,.3) ..controls(1.5,1) and (.5, 1 ).. (0,.3);
		
		\draw(2,0) -- (2,0.2);
		\draw(6,0) -- (6,0.8);
		\draw(10,0) -- (10,0.8);
		\end{tikzpicture}
	\end{scriptsize}
	}
	
	\plaincenter{
	\mbox{\makeboxlike[l]{\textbf{\textsf{Step 2:}}}{\textbf{\textsf{Step 2:}}}}%
	\qquad
	\begin{scriptsize}
		\begin{tikzpicture}[scale = 0.55]
		
		\draw(0,0) -- (10,0);
		\draw(0,0.2) -- (4,0.5);
		\draw(4,0.5) -- (6,0.8);
		\draw(6,0.2) -- (8,0.2);
		\draw(0,0) -- (0,0.2);
		\draw(8,0.5) -- (10,0.8);

		\draw[->](5.9,.9) ..controls(6.2,1.5) and (7.6, 1.9 ).. (7.9,.3);
		\draw[->](9.9,.9) ..controls(9.6,1.5) and (8.2, 1.9 ).. (7.9,.3);
		
		\draw(4,0) -- (4,0.5);
		\draw(6,0) -- (6,0.8);
		\draw(8,0) -- (8,0.5);
		\draw(10,0) -- (10,0.8);
		\end{tikzpicture}
	\end{scriptsize}
	}
	
	\bigskip
	\medskip
	
	\plaincenter{
	\mbox{\makeboxlike[l]{\textbf{\textsf{Step 2:}}}{\textbf{\textsf{Result:}}}}%
	\qquad
	\begin{scriptsize}
		\begin{tikzpicture}[scale = 0.55]
		
		\draw(0,0) -- (10,0);
		\draw(0,0.2) -- (8,0.8);
		\draw(8,0.2) -- (10,0.2);

		\draw(0,0) -- (0,0.2);
		\draw(8,0) -- (8,0.8);
		\draw(10,0) -- (10,0.2);
		\end{tikzpicture}
	\end{scriptsize}
	}

	\caption{%
		Reinhardt's merging procedure that needs only buffer space for \emph{half} of the smaller run.
		In the first step the two sequences are merged starting with the smallest 
		elements until the empty space is filled. 
		Then there is enough empty space to merge the sequences from the 
		right into the final position.
	}
	\label{fig:merge5}
\end{figure}
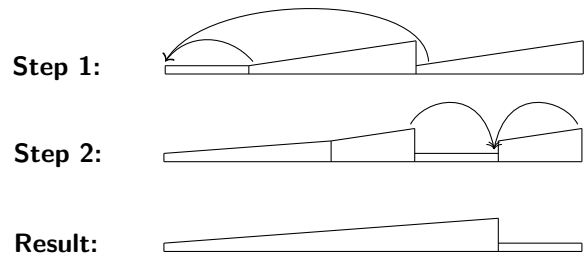

\paragraph{Reinhardt's merge}
A third, less obvious alternative was proposed by Reinhardt~\cite{Reinhardt1992},
which allows to use an even smaller $\alpha$ for merges where 
input and buffer area form a contiguous region;
see \wref{fig:merge5}.
Assume we are given an array $A$ with positions $A[1, \dots , t]$ being empty or containing dummy elements 
(to simplify the description, we assume the first case), 
$A[t+1, \dots, t + \ell ]$ and $A[t + \ell+1,\dots, t + \ell + r ]$ containing two sorted sequences. 
We wish to merge the two sequences into the space $A[1, \dots, \ell +r]$ 
(so that $A[\ell+ r+1, \dots, t + \ell + r]$ becomes empty).
We require that $r/2 \leq t < r $.
First we start from the left merging the two sequences into the empty space until 
there is no space left between the last element of the already merged part and the 
first element of the left sequence (first step in \prettyref{fig:merge5}). 
At this point, we know that at least $t$ elements of the right sequence have been 
introduced into the merged part; so, the positions $t + \ell  + 1$ through $\ell +2t$ are empty now. 
Since $\ell + t + 1 \leq \ell + r \leq \ell + 2t$, in particular, 
$A[\ell + r]$ is empty now and we can start merging the two sequences 
right-to-left into the now empty space 
(where the right-most element is moved to position $A[\ell + r]$~-- 
see the second step in \prettyref{fig:merge5}).

In order to have a balanced merge, we need $\ell = r$ and so $t \geq (\ell + r)/4$. 
Therefore, when applying this method in \QuickMergesort, we have $\alpha=\frac14$.

\begin{remark}[Even less buffer space?]
Reinhardt goes even further: even with $\epsilon n$ space, we can merge in linear time
when $\epsilon$ is fixed by moving one run whenever we run out of space.
Even though not more comparisons are needed, this method is quickly dominated by the 
additional data movements when $\epsilon < \frac14$, so we do not discuss it in this
article.

Another approach for dealing with less buffer space is to allow imbalanced merges:
for both Reinhardt's merge and the simple swap-based merge, we need only additional space
for (half) the size of the \emph{smaller} run. Hence, we can merge a short run into a long run
with a relatively small buffer. The price of this method is that the number of comparisons
increases, while the number of additional moves is better than with the previous method.
We shed some more light on this approach in \cite{EdelkampWeiss2018MQMS}.
\end{remark}

\paragraph{Avoiding Stack Space}
The standard version of \Mergesort uses a top-down recursive formulation.
It requires a stack of logarithmic height, which is usually deemed
acceptable since it is dwarfed by the buffer space for merging.
Since \QuickMergesort removes the need for the latter,
one might prefer to also avoid the logarithmic stack space.

An elementary solution is bottom-up \Mergesort, where we form pairs
of runs and merge them, except for, potentially, a lonely rightmost run.
This variant occasionally merges two runs of very different sizes,
which affects the overall performance (see \wref{sec:preliminaries-mergesort}).

A simple (but less well-known) modification that we call \emph{boustrophedonic%
\footnote{%
	after boustrophedon, a type of bi-directional text seen in ancient manuscripts 
	where lines alternate between left-to-right and right-to-left order;
	literally ``turning like oxen in ploughing''.
} 
\Mergesort} allows us to get the best of both 
worlds~\cite{GolinSedgewick1993}:
instead of leaving a lonely rightmost run unmerged (and starting again at the beginning
with the next round of merges), we start the next merging round at the same end,
moving backwards through the array. 
We hence begin by merging the lonely run, and so avoid ever having a
two runs that differ by more than a factor of two in length.
The logic for handling odd and even numbers of runs correctly 
is more involved, but constant extra space can be achieved without
a loss in the number of comparisons.

\subsection{QuickHeapsort}
\label{sec:quickheapsort}

Another good option~-- and indeed the historically first one~-- for X is Heapsort. 

\oldparagraph{Why Heapsort?}
In light of the fact that 
Heapsort is the only textbook method with reasonable overall performance 
that already sorts with \emph{constant extra space,}
this suggestion might be surprising.
Heapsort rather appears to be the candidate \emph{least} likely to profit from QuickXsort.
Indeed, it is a refined variant of Heapsort that is an interesting candidate for X.

To work in place, standard Heapsort has 
to maintain the heap in a very rigid shape 
to store it in a contiguous region of the array.
And this rigid structure comes at the price of extra comparisons.
Standard Heapsort requires up to $2 (h-1)$ comparisons to extract the maximum
from a heap of height $h$, for an overall $2 n \lg n \pm \Oh(n)$ comparisons in the worst case.

Comparisons can be saved by first finding the cascade of promotions
(\aka the \emph{special path}), \ie, the path from the root to a leaf,
always choosing to the larger of the two children.
Then, in a second step, we find the correct insertion position along this line 
of the element currently occupying the last position of the heap area.
The standard procedure corresponds to sequential search from the root.
Floyd's optimization (\aka bottom-up Heapsort \cite{Weg93}) instead uses sequential search from the leaf.
It has a substantially higher chance to succeed early (in the second phase), 
and is probably optimal in that respect for the average case.
If a better worst case is desired, one can use binary search on the special path,
or even more sophisticated methods~\cite{GonnetMunro1986}.

\paragraph{External Heapsort}
In \algorithmname{ExternalHeapsort}, we
avoid any such extra comparisons by relaxing the heap's shape.
Extracted elements go to an output buffer and we only promote the elements
along the special path into the gap left by the maximum.
This leaves a gap at the leaf level, that we fill with a sentinel value smaller 
than any element's value (in the case of a max-heap).
\algorithmname{ExternalHeapsort} uses $n\lg n\pm\Oh(n)$ comparisons in the worst case, 
but requires a buffer to hold $n$ elements. 
By using it as our X in QuickXsort, we can avoid the extra space requirement.

When using \algorithmname{ExternalHeapsort} as X, we cannot simply overwrite gaps with
sentinel values, though: we have to keep the buffer elements intact!
Fortunately, the buffer elements themselves happen to work as sentinel values.
If we sort the segment of large elements with \algorithmname{ExternalHeapsort},
we swap the max from the heap with a buffer element, which automatically is smaller
than any remaining heap element and will thus never be promoted as long as 
any actual elements remain in the heap.
We know when to stop since we know the segment sizes; after that many extractions,
the right segment is sorted and the heap area contains only buffer elements.

We use a symmetric variant (with a min-oriented heap) if the left segment shall
be sorted by X.
For detailed code for the above procedure, 
we refer to~\cite{CantoneCincotti2002} or~\cite{DiekertWeiss2016}.

\paragraph{Trading space for comparisons}

Many options to further reduce the number of comparisons have been explored.
Since these options demand extra space beyond an output buffer
and cannot restore the contents of that extra space, using them in \QuickXsort 
does not yield an internal sorting method, but
we briefly mention these variants here.

One option is to remember outcomes of sibling comparisons
to avoid redundant comparisons in following steps~\cite{McDiarmidReed}. 
In \cite[Thm.\ 4]{DiekertWeiss2016}, this is applied to \algorithmname{QuickHeapsort} 
together with some further improvements using extra space.

Another option is to modify the heap property itself.
In a weak heap, the root of a subtree is only larger than one of the subtrees,
and we use an extra bit to store (and modify) which one it is.
The more liberal structure makes construction of weak heaps more efficient: 
indeed, they can be constructed using $n-1$ comparisons. 
\algorithmname{WeakHeapsort} has been introduced by Dutton~\cite{Dut93} 
and applied to \algorithmname{QuickWeakHeapsort} in~\cite{edelkampstiegeler}.
We introduced a refined version of \algorithmname{ExternalWeakHeapsort} in~\cite{EdelkampWeiss2014} 
that works by the same principle as \algorithmname{ExternalHeapsort};
more details on this algorithm, its application in \algorithmname{QuickWeakHeapsort}, and the relation to 
\Mergesort can be found in our preprint~\cite{EdelkampW13QuickArxiv}. 

Due to the additional bit-array, which is not only space-consuming, 
but also costs time to access, \algorithmname{WeakHeapsort} and \algorithmname{QuickWeakHeapsort} are 
considerably slower than ordinary \algorithmname{Heapsort}, \algorithmname{Mergesort}, 
or \algorithmname{Quicksort}; 
see the experiments in \cite{edelkampstiegeler,EdelkampWeiss2014}. 
Therefore, we do not consider these variants here in more detail.

\section{Preliminaries}
\label{sec:prelims}

In this section, we introduce some important notation and collect known results for reference. 
The reader who is only interested in the main results may skip this section.
A comprehensive list of notation is given in \wref{app:notation}.

We use Iverson's bracket $[\mathit{stmt}]$ to mean $1$ if $\mathit{stmt}$ is true and $0$ otherwise.
$\Prob{E}$ denotes the probability of event $E$, $\E{X}$ the expectation
of random variable $X$. We write $X\eqdist Y$ to denote equality in distribution.

With $f(n) = g(n) \pm h(n)$ we mean that $\abs{f(n) - g(n)} \leq h(n)$ for all $n$,
and we use similar notation $f(n) = g(n) \pm \Oh(h(n))$ to state asymptotic bounds on 
the difference $\abs{f(n) - g(n)} = \Oh(h(n))$.
We remark that both use cases are examples of ``one-way equalities'' that are in common use for 
notational convenience, even though $\subseteq$ instead of $=$ would be formally more appropriate.
Moreover, $f(n) \sim g(n)$ means $f(n) = g(n) \pm o(g(n))$.

Throughout, $\lg$ refers to the logarithm to base $2$ while, while $\ln$ is the natural logarithm. 
Moreover, $\log$ is used for the logarithm with unspecified base (for use in $\Oh$-notation).

We write $a^{\underline b}$ (resp.\ $a^{\overline b}$)
for the  falling (resp.\ rising)  factorial power $a(a-1) \cdots(a-b+1)$ (resp.\ $a(a+1) \cdots(a+b-1)$).

\subsection{Hölder continuity}
\label{sec:hölder-continuity}

A function $f:I\to R$ defined on a bounded interval $I$ is 
\emph{Hölder-continuous} with exponent $\holdA\in(0,1]$
if 
\[
	\exists C\;
	\forall x,y\in I\wrel:
		\bigl| f(x) - f(y) \bigr|
		\wrel\le 
		C |x-y|^\holdA.
\]
Hölder-continuity is a notion of smoothness 
that is stricter than (uniform) continuity but slightly more liberal
than Lipschitz-continuity (which corresponds to $\holdA=1$).
$f:[0,1]\to\R$ with $f(z) = z \ln(1/z)$ is a stereotypical function
that is Hölder-continuous (for any $\holdA\in(0,1)$) but not Lipschitz
(see \wref{lem:x-log-x} below).

One useful consequence of Hölder-continuity is given by the following lemma:
an error bound on the difference between an integral and the Riemann sum
(\cite[\href{https://www.wild-inter.net/publications/html/wild-2016.pdf.html\#pf4a}{Proposition \mbox{2.12--(b)}}]{Wild2016}).

\begin{lemma}[{Hölder integral bound}]
\label{lem:hölder-intergral-bound}
	Let $f:[0,1] \to \R$ be Hölder-continuous with exponent $\holdA$.
	Then
	\begin{align*}
			\int_{x=0}^1 f(x) \, dx
		&\wwrel=
			\frac1{n}  \sum_{i = 0}^{n-1} f(i / n)
			\wwbin\pm
			\Oh(n^{-\holdA}),
			\qquad(n\to\infty).
	\end{align*}
\ifthenelse{\boolean{citeproofs}}{\qed}{}\end{lemma}
\ifthenelse{\boolean{citeproofs}}{}{
\begin{proof}
The proof is a simple computation.
Let $C$ be the Hölder-constant of $f$.
We split the integral into small integrals over intervals of width $\frac1n$
and use Hölder-continuity to bound the difference to the corresponding summand:
\begin{align*}
	&
	\mkern-50mu
		\Biggl|
		\int_{x=0}^1 f(x) \, dx
		\wbin-
		\frac1{n} \sum_{i=0}^{n-1} f(i / n)
		\Biggr|
\\	&\wwrel=
		\sum_{i=0}^{n-1}
			\Biggl|
				\int_{i/n}^{(i+1)/n} f(x) \, dx
				\bin- \frac{f(i /n)}{n}
			\Biggr|
\\	&\wwrel=
		\sum_{i=0}^{n-1}
				\int_{i/n}^{(i+1)/n}
				\bigl| f(x) - f(i /n)\bigr|
				\, d x
\\	&\wwrel\le
		\sum_{i=0}^{n-1}
				\int_{i/n}^{(i+1)/n}
				C \bigl| x - \tfrac{i}n\bigr|^\holdA
				\, d x
\\	&\wwrel\le
		C \sum_{i=0}^{n-1}
				\int_{i/n}^{(i+1)/n}
				\bigl(\tfrac1n\bigr)^\holdA
				\, d x
\\	&\wwrel=
		Cn^{-\holdA} \int_{0}^1 1 \, dx
\\	&\wwrel=
		\Oh(n^{-\holdA}).
\end{align*}
\end{proof}
}

\begin{remark}[Properties of Hölder-continuity]
We considered only the unit interval 
as the domain of functions, but this is no restriction:
Hölder-continuity (on bounded domains) is preserved by addition,
subtraction, multiplication and composition
(see, \eg, \cite[Section\,4.6]{Sohrab2014} for details).
Since any linear function is Lipschitz, the result above holds for 
Hölder-continuous functions $f:[a,b] \to \R$.

If our functions are defined on a bounded domain,
Lipschitz-continuity implies Hölder-continuity
and Hölder-continuity with exponent $\holdA$ implies
Hölder-continuity with exponent $\holdA' < \holdA$.
A real-valued function is Lipschitz if its derivative is bounded.
\end{remark}

\subsection{Concentration results}

We write $X\eqdist \binomial(n,p)$ if $X$ is has a binomial distribution 
with $n\in\N_0$ trials and success probability $p\in[0,1]$.
Since $X$ is a sum of independent random variables with bounded influence on the result,
Chernoff bounds imply strong concentration results for $X$.
We will only need a very basic variant given in the following lemma.

\begin{lemma}[Chernoff Bound, Theorem~2.1 of \cite{McDiarmid1998}]
\label{lem:chernoff-bound-binomial}
	Let $X\eqdist\binomial(n,p)$ and $\delta\ge0$. Then
	\begin{align}
	\label{eq:chernoff-bound-binomial}
			\Prob[\Bigg]{ \biggl|\frac{X}n - p\biggr| \ge \delta }
		&\wwrel\le
			2\exp (-2 \delta^2 n ).
	\end{align}
\qed\end{lemma}

A consequence of this bound is that we can bound expectations of the form $\E{f(\frac Xn)}$, 
by $f(p)$ plus a small error term
if $f$ is ``sufficiently smooth''.
Hölder-continuous (introduced above) is an example for such a criterion:

\begin{lemma}[Expectation via Chernoff]
\label{lem:E-f-X-concentration}
	Let \(p\in(0,1)\) 
	and \(X \eqdist \binomial(n, p)\), and let
	$f:[0,1]\to\R$ be a function that is bounded by $|f(x)| \le A$ and Hölder-continuous with 
	exponent $\holdA \in(0,1]$ and constant $C$.
	Then it holds that
	\begin{align*}
			\E[\bigg]{ f\biggl(\frac{X}n\biggr) }
		&\wwrel=
			f(p) \wbin\pm \rho,
	\end{align*}
	where we have for any \(\delta\ge0\) that
	\begin{align*}		
			\rho
		&\wwrel\le 
			\frac{C}{\ln 2} \cdot \delta^\holdA \bigl(1 - 2 e^ {-2 \delta^2 n}\bigr)
			\bin+
			4A e^{-2 \delta^2 n}
	\end{align*}
	For any fixed \(\varepsilon>\frac{1-\holdA}2\), we obtain
	$\rho = o(n^{-1/2+\varepsilon})$ as $n\to\infty$ for a suitable choice of $\delta$.
\end{lemma}

\begin{proof}[\wref{lem:E-f-X-concentration}]
By the Chernoff bound
we have
\begin{align*}
		\Prob[\Bigg]{ \biggl| \frac{X}n - p \biggr| \ge \delta }
	&\wwrel\le
		2u \exp (-2 \delta^2 n) .
\numberthis\label{eq:chernoff-X-by-n-minus-p-greater-delta}
\end{align*}
To use this on 
$\E[\big]{ \bigl| f\bigl(\frac{X}n\bigr) - f(p) \bigr| }$, 
we divide the domain $[0,1]$ of $\frac{X}n$ into 
the region of values with distance 
at most $\delta$ from $p$, and all others.
This
yields
\begin{align*}
		\E[\bigg]{ \biggl| 
			f\biggl(\frac{X}n\biggr) - f(p) 
		\biggr| } 
	&\wwrel{\relwithref{eq:chernoff-X-by-n-minus-p-greater-delta}\le}
		\sup_{\xi \rel: |\xi| < \delta} \bigl| 
			f(p + \xi) - f(p) 
		\bigr| 
		\cdot \Bigl(1 - 2 e^{-2 \delta^2 n}\Bigr)
\\*	&\wwrel\ppe\quad{}	
		\bin+
		\sup_{x} \,\bigl| f(x) - f(p)\bigr|
		{}\cdot 2 e ^{-2 \delta^2 n}
\\	&\wwrel{\relwithref[r]{lem:x-log-x}\le}
		C\cdot \delta^\holdA \cdot \Bigl(1 - 2 e^{-2 \delta^2 n}\Bigr)
		\bin+
		2A\cdot 2e^{-2 \delta^2 n}.
	\end{align*}
	This proves the first part of the claim.
	
	For the second part, we assume
	$\varepsilon>\frac{1-\holdA}2$ is given, so we can write
	$\holdA = 1-2\varepsilon + 4\beta$ for a constant $\beta>0$, and
	$\holdA = (1-2\varepsilon) / (1-2\beta')$ for another constant $\beta'>0$.
	We may further assume $\varepsilon<\frac12$; for larger values the claim is vacuous.
	We then choose $\delta = n^c$ with 
	$c=\bigl(-\frac12 - \frac{1/2-\varepsilon}{\holdA}\bigr)/2 
	= -\frac14-\frac{1-2\varepsilon}{4\holdA}$.
	For large $n$ we thus have
	\begin{align*}
			\rho \cdot n^{1/2-\epsilon}
		&\wwrel\le
			C\delta^\holdA n^{1/2-\epsilon} \bigl(1-2\exp(-2\delta^2 n)\bigr)
			\bin+
			4A n^{1/2-\epsilon} \exp(-2\delta^2 n)
	\\	&\wwrel=
			\underbrace{ C n^{-\beta} }_{{}\to 0} 
				\cdot \bigl(1 - \underbrace{2 \exp (-2 n^{\beta'})}_{{} \to 0} \bigr)
			\bin+
			4A \underbrace{  \exp \Bigl(-2 n^{\beta'} + (\tfrac12-\varepsilon) \ln(n)\Bigr) }_{{} \to 0}
	\\[-1ex]	&\wwrel\to
			0
	\end{align*}
	for $n\to\infty$, which implies the claim.
	\end{proof}

\subsection{Beta distribution}
\label{sec:beta-dist}

The analysis in \wref{sec:analysis-fixed-k} makes frequent use of the 
\emph{beta distribution:} For $\betL,\betR \in \R_{>0}$,
$X\eqdist\betadist(\betL,\betR)$ if $X$ admits the density 
$f_X(z) = z^{\betL-1}(1-z)^{\betR-1} / \BetaFun(\betL,\betR)$
where $\BetaFun(\betL,\betR) = \int_0^1 z^{\betL-1}(1-z)^{\betR-1} \, dz$ is the beta function. 
It is a standard fact that for $\betL, \betR \in \N_{\ge1}$ we have 
\begin{align}
\label{eq:betaFun} 
		\BetaFun(\betL,\betR) 
	\wrel= 
		\frac{(\betL - 1)! (\betR-1)!}{(\betL + \betR - 1)!};
\end{align}
a generalization of this identity using the gamma function holds for any $\betL,\betR > 0$
\cite[{\href{https://dlmf.nist.gov/5.12.E1}{Eq.\,(5.12.1)}}]{DLMF}.
We will also use the \emph{regularized incomplete beta function}
\begin{align*}
\numberthis\label{eq:regularized-incomplete-beta}
		I_{x,y}(\betL,\betR)
	&\wwrel=
		\int_x^y \frac{z^{\betL-1}(1-z)^{\betR-1}}{\BetaFun(\betL,\betR)} \, dz
		,\qquad (\betL,\betR\in\R_+, 0\le x\le y\le 1).
\end{align*}
Clearly $I_{0,1}(\betL,\betR) = 1$.

Let us denote by $h$ the function $h:[0,1]\to \R_\ge0$  
with $h(x) = -x \lg x$.
We have for a beta-distributed random variable
$X\eqdist \betadist(\betL,\betR)$ for $\betL,\betR\in\N_{\ge1}$ that
\begin{align}
\label{eq:E-h-X}
		\E{ h(X) }
	&\wwrel=
		\BetaFun(\betL,\betR) \bigl( \harm{\betL+\betR} - \harm{\betL} \bigr) .
\end{align}
This follows directly from a well-known closed form 
a ``logarithmic beta integral''
(see, \eg,~\cite[\href{https://www.wild-inter.net/publications/html/wild-2016.pdf.html\#pf46}{Eq.~(2.30)}]{Wild2016}).
\begin{align*}
		\int_0^1 \ln (z) \cdot z^{\betL-1} (1-z)^{\betR-1} \, dz
	&\wwrel=
		\BetaFun(\betL,\betR) \bigl( \harm{\betL-1} - \harm{\betL+\betR-1} \bigr)
\end{align*}
We will make use of the following elementary properties of $h$ later
(towards applying \wref{lem:E-f-X-concentration}).

\begin{lemma}[Elementary Properties of $h$]
\label{lem:x-log-x}
	Let \(h: [0,1]^u \to \R_{\ge0}\) with 
	\(h(x) = -x \lg(x)\).
	\begin{thmenumerate}{lem:x-log-x}
	\item \label{lem:x-log-x-bounds} 
		$h$ is bounded by $0 \le h(x) \le \frac{\lg e}e \le  0.54$ for $x\in[0,1]$.
	\item \label{lem:x-log-x-hölder}
		\(g(x) \ce -x \ln x = \ln(2) h(x)\) is Hölder-continuous in 
		\([0,1]\) for any exponent \(\holdA\in(0,1)\),
		\ie, there is a constant \(C=C_\holdA\) such that 
		\(|g(y) - g(x)| \le C_\holdA |y - x|^\holdA\) for all 
		\(x, y \in [0,1] \).
		A possible choice for \(C_\holdA\) is given by 
		\begin{align*}
		\numberthis\label{eq:C-alpha}
				C_\holdA 
			&\wwrel= 
				\biggl( \int_0^1 \bigl|\ln(t)+1\bigr|^{\frac1{1-\holdA}} \biggr)^{1-\holdA}
		\end{align*}
		For example, \(\holdA=0.99\) yields \(C_\holdA \approx 37.61\).
	\end{thmenumerate}
\qed\end{lemma}
A detailed proof for the second claim appears in 
\cite[\href{https://www.wild-inter.net/publications/html/wild-2016.pdf.html\#pf4b}{Lemma 2.13}]{Wild2016}.
Hence, $h$ is sufficiently smooth to be used in \wref{lem:E-f-X-concentration}.

\subsection{Beta-binomial distribution}
\label{sec:beta-binomial-dist}
Moreover, we use the \emph{beta-binomial distribution,} which is a conditional binomial distribution 
with the success probability being a beta-distributed random variable.
If $X\eqdist \betaBinomial(n,\betL,\betR)$ then
\[
		\Prob{X=i} 
	\wrel= 
		\binom ni \frac{\BetaFun(\betL+i,\betR+(n-i)) }{ \BetaFun(\betL,\betR)}.
\]
Beta-binomial distributions are precisely the distribution of subproblem sizes after partitioning in 
\algorithmname{Quicksort}.
We detail this in \wref{sec:distribution}.

A property that we repeatedly use here is a local limit law showing that the 
normalized beta-binomial distribution converges to the beta distribution. 
Using Chernoff bounds after conditioning on the beta distributed
success probability shows that $\betaBinomial(n,\betL,\betR)/n$ converges to $\betadist(\betL,\betR)$
(in a specific sense);
but we obtain stronger error bounds for fixed $\betL$ and $\betR$ by directly comparing 
the probability density functions (PDFs). 
This yields the following result;
(a detailed proof appears in \cite[{\href{https://www.wild-inter.net/publications/html/wild-2016.pdf.html\#pf66}{Lemma~2.38}}]{Wild2016}).

\begin{lemma}[{Local Limit Law for Beta-Binomial, \cite{Wild2016}}]
\label{lem:beta-binomial-convergence-to-beta}
	Let \((\ui In)_{n\in\N_{\ge1}}\) 
	be a family of random variables with beta-binomial distribution,
	\(\ui In \eqdist \betaBinomial(n, \betL,\betR)\) where \(\betL,\betR\in\{1\}\cup\R_{\ge2}\), 
	and let \(f_B(z) = z^{\betL-1}(1-z)^{\betR-1} / \BetaFun(\betL,\betR)\) be the density of the \(\betadist(\betL,\betR)\) distribution.
	Then we have uniformly in \(z\in(0,1)\) that 
	\begin{align*}
				n \cdot \Prob[\big]{ I = \lfloor z(n+1)\rfloor } 
			\wwrel= 
				f_B(z) \bin\pm \Oh(n^{-1})
				,\qquad (n\to\infty).
	\end{align*}
	That is, \(\ui In/n\) converges to \(\betadist(\betL,\betR)\) in distribution, 
	and the probability weights converge uniformly to the limiting density at rate \(\Oh(n^{-1})\).
\end{lemma}

\subsection{Continuous Master Theorem}
\label{sec:cmt}

For solving recurrences, we build upon Roura's master theorems~\cite{Roura2001}.
The relevant \textsl{continuous master theorem} is restated here for convenience:

\begin{theorem}[Roura's Continuous Master Theorem (CMT)]
\label{thm:CMT}
	\oneline{Let \(F_n\) be recursively} defined~by
	\begin{align}
	\label{eq:CMT-recurrence}
		F_n \wwrel= \begin{dcases*}
			b_n\:,	
				& for \( 0 \le n < N \); \\
			\vphantom{\bigg|}
				t_n \bin+ \smash{\sum_{j=0}^{n-1} w_{n,j} \, F_j}, 
				& for  \(n \ge N\)\,, 
		\end{dcases*}
	\end{align}
	where \(t_n\), the toll function, satisfies \(t_n \sim K n^\cmtA \log^\cmtB(n)\) as
	\(n\to\infty\) for constants \(K\ne0\), \(\cmtA\ge0\) and \(\cmtB > -1\).
	Assume there exists a function \(w:[0,1]\to \R_{\ge0}\), the \textit{shape function,}
	with \(\int_0^1 w(z) dz \ge 1 \) and
	\begin{align}
	\label{eq:CMT-shape-function-condition}
		\sum_{j=0}^{n-1} \,\biggl|
			w_{n,j} \bin- \! \int_{j/n}^{(j+1)/n} \mkern-15mu w(z) \: dz
		\biggr|
		\wwrel= \Oh(n^{-d}),
		\qquad(n\to\infty),
	\end{align}
	for a constant \(d>0\).
	With \(\displaystyle H \ce 1 - \int_0^1 \!z^\cmtA w(z) \, dz\), we
	have the following cases:
	\begin{enumerate}[itemsep=0ex]
		\item If \(H > 0\), then \(\displaystyle F_n \sim \frac{t_n}{H}\).
			\(\vphantom{\displaystyle\int_0^1}\)
		\item \label{case:CMT-H0} 
		If \(H = 0\), then 
		\(\displaystyle
		F_n \sim \frac{t_n \ln n}{\widetilde H}\) with 
		\(\displaystyle \widetilde H = -(\cmtB+1)\int_0^1 \!z^\cmtA \ln(z) \, w(z) \, dz\).
		\item \label{case:CMT-theta-nc}
		If \(H < 0\), then \(F_n = \Oh(n^c)\) for the unique
		\(c\in\R\) with \(\displaystyle\int_0^1 \!z^c w(z) \, dz = 1\).
	\end{enumerate}
\qed\end{theorem}

\noindent
\wref{thm:CMT} is the ``reduced form'' of the CMT,
which appears as Theorem~1.3.2 in Roura's doctoral thesis~\cite{Roura1997},
and as Theorem~18 of \cite{MartinezRoura2001}.
The full version (Theorem~3.3 in~\cite{Roura2001})
allows us to handle sublogarithmic factors in the toll function, as well, 
which we do not need here.

\subsection{Average costs of Mergesort}
\label{sec:preliminaries-mergesort}

We recapitulate some known facts about standard mergesort.
The average number of comparisons for Mergesort has the same~-- optimal~-- leading
term $n \lg n$ in the worst and best case;
this is true for both the top-down and bottom-up variants.
The coefficient of the \emph{linear} term of the asymptotic expansion, though, is not a constant,
but a bounded periodic function with period $\lg n$, and the functions differ for best, worst,
and average case and the variants of 
Mergesort~\cite{SedgewickFlajolet2013,FlajoletGolin1994,PannyProdinger1995,Hwang1996,Hwang1998}.

For this paper, we confine ourselves to upper and lower \emph{bounds} for the average case 
of the form
$x(n) = a n \lg n + b n \wbin\pm \Oh(n^{1-\epsilon}) $ with \emph{constant} $b$ valid for all $n$.
Setting $b$ to the infimum resp.\ supremum of the periodic function,
we obtain the following lower resp.\ upper bounds for \underline top-\underline down~\cite{Hwang1998} 
and \underline bottom-\underline up~\cite{PannyProdinger1995} Mergesort
\begin{align*}
\numberthis\label{eq:xtd}
		x_\mathrm{td}(n)
&	\wwrel=
		n \lg n - \biggl\{\mkern-5mu\begin{array}{c}
			1.2645 n \\
			1.2408 n
		\end{array} \mkern-5mu + 2 \bin\pm \Oh(n^{-1})
\\	&\wwrel=
		n \lg n -(1.25265 \pm 0.01185)n + 2  \bin\pm \Oh(n^{-1}) \qquad \text{and}
\\
		x_\mathrm{bu}(n)
&	\wwrel=
		n \lg n - \biggl\{\mkern-5mu\begin{array}{c}
			1.2645n \\
			0.2645n
		\end{array} \mkern-5mu \wbin\pm \Oh(1)
\\	&\wwrel=
		n \lg n -(0.7645 \pm 0.5)n \wbin\pm \Oh(1).
\end{align*}

\section{The QuickXsort recurrence}
\label{sec:recurrence}

In this section, we set up a recurrence equation for the costs of \QuickXsort.
This recurrence will be the basis for our analyses below.
We start with some prerequisites and assumptions about X.

\subsection{Prerequisites}

For simplicity we will assume that below a constant subproblem size~$w$ 
(with $w\geq k$ in the case of constant size-$k$ samples for pivot selection) 
are sorted with X (using a constant amount of extra space). Nevertheless, we could use any other algorithm for that as this only influences the \emph{constant term} of costs.
A common choice in practice is replace X by \algorithmname{StraightInsertionsort} to sort the small cases.

We further assume that selecting the pivot from a sample of size $k$
costs $s(k)$ comparisons, where we usually assume $s(k) = \Theta(k)$,
\ie, a (expected-case) linear selection method is used.

Now, let $c(n)$ be the expected number of comparisons in QuickXsort on arrays of size $n$,
where the expectation is over the random choices for selecting the pivots
for partitioning.

\oldparagraph{Preservation of randomness?}

Our goal is to set up a recurrence equation for $c(n)$.
We will justify here that such a recursive relation exists.

For the Quicksort part of QuickXsort, only the ranks of the chosen pivot
elements has an influence on the costs; partitioning itself
always needs precisely one comparison per element.%
\footnote{%
	We remark that this is no longer true for multiway partitioning methods
	where the number of comparisons per element is not necessarily the same
	for all possible outcomes.
	Similarly, the number of swaps in the standard partitioning method
	depends not only on the rank of the pivot, but also on how ``displaced''
	the elements in the input are.
}
Since we choose the pivot elements randomly (from a random sample),
the order of the input does not influence the costs of the Quicksort part
of QuickXsort.

For general X, the sorting costs do depend on the order of the input,
and we would like to use the average-case bounds for X, when it is applied
on a random permutation.
We may assume that our \emph{initial} input is indeed a random permutation of the elements,%
\footnote{
	It is a reasonable option to \emph{enforce} this assumption in an implementation
	by an explicit \emph{random shuffle} of the input before we start sorting.
	Sedgewick and Wayne, for example, do this
	for the implementation of Quicksort in their textbook~\cite{SedgewickWayne2011}.
}
but this is not sufficient!
We also have to guarantee that the inputs for recursive calls are again random
permutations of their elements.

A simple sufficient condition for this ``randomness-preserving'' property
is that X may not compare buffer contents.
This is a natural requirement, \eg, for our \algorithmname{Mergesort} variants.
If no buffer elements are compared to each other and
the original input is a random permutation of its elements, 
so are the segments after partitioning,
and so will be the buffer after X has terminated.
Then we can set up a recurrence equation for $c(n)$ using the average-case cost
for X.
We may also replace the random sampling of pivots by choosing any 
\emph{fixed} positions without affecting the expected costs $c(n)$.

However, not all candidates for X meet this requirement. 
(Basic) \algorithmname{QuickHeapsort} does compare buffer elements to each other
(see \wref{sec:quickheapsort})
and, indeed, the buffer elements are \emph{not} in random order when the \algorithmname{Heapsort} part has finished.
For such X, we assume that genuinely random samples for pivot selection are used.
Moreover, and we will have to use conservative bounds for the number of comparisons incurred by X,
\eg, worst or best case results, 
as the input of X is not random anymore.
This only allows to derive upper or lower bounds for $c(n)$, whereas for randomness preserving methods,
the expected costs can be characterized precisely by the recurrence.

In both cases, we use $x(n)$ as (a bound for) the number of comparisons needed by X to sort $n$ elements,
and we will assume that 
\begin{align*}
x(n)
&\wwrel=
a n \lg n +b n \wbin\pm \Oh(n^{1-\epsilon}) 
,\qquad(n\to\infty),
\end{align*}
for constants $a$, $b$ and $\epsilon\in(0,1]$. 

\subsection{The recurrence for the expected costs}

We can now proceed to the recursive description of the expected costs $c(n)$ of \QuickXsort.
The description follows the recursive nature of the algorithm.
Recall that \QuickXsort tries to sort the largest segment with X for which the other
segment gives sufficient buffer space.
We first consider the case $\alpha=1$, in which this largest segment is always the smaller of
the two segments created.

\paragraph{\boldmath Case $\alpha=1$}
Let us consider the recurrence for $c(n)$ (which holds for both constant and growing size~$k = k(n)$). 
We distinguish two cases: 
first, let $\alpha= 1 $. We obtain the recurrence
\begin{align*}
		c(n) 
	&\wwrel=
		x(n) \ge 0,
		\qquad(\text{for }n\le w)
\\
c(n)
&\wwrel=
\begin{aligned}[t]
		\smash{\underbrace{\vphantom{\big(}n-k(n)}_{\mathclap{\text{partitioning}}}{} 
		\bin+ \underbrace{s\bigl(k(n)\bigr)}_{\mathclap{\text{pivot sampling}}}{}}
		\quad
		&\bin+ \E[\big]{[J_1 > J_2](x(J_1) + c(J_2))}\\
		&\bin+ \E[\big]{[J_1 \le J_2](x(J_2) + c(J_1))}
&\qquad (\text{for }n > w)
\end{aligned}
\\	&\wwrel=
		\sum_{r=1}^2\E{A_r(J_r) c(J_r)}  + t(n)
		\shortintertext{where}
		A_1(J)
	&\wwrel= 
		[J \le J'], 
	\qquad
		A_2(J)
	\wwrel= 
		[J < J'] 
		\qquad\text{with }J' = (n-1)-J,
\\
		t(n)
	&\wwrel=
		n-k + s(k)
		\bin+ \E*{A_2(J_2) x(J_1)}
		\bin+ \E*{A_1(J_1) x(J_2)}.
\end{align*}
The expectation here is taken over the choice for the random pivot, \ie, over the
segment sizes $J_1$ resp.\ $J_2$.
Note that we use both $J_1$ and $J_2$ to express the conditions in a convenient form,
but actually either one is fully determined by the other via $J_1 + J_2 = n-1$. We call $t(n)$ the \emph{toll function}.
Note how $A_1$ and $A_2$ change roles in recursive calls and toll functions, 
since we always sort one segment recursively and the other segment by X.

\paragraph{\boldmath General $\alpha$}
For $\alpha<1$, we obtain two cases:
When the split induced by the pivot is ``uneven''~-- 
namely when $\min\{J_1,J_2\} < \alpha\max\{J_1,J_2\}$, \ie, 
$\max\{J_1,J_2\} > \frac{n-1}{1+\alpha}$~-- 
the smaller
segment is not large enough to be used as buffer.
Then we can only assign the large segment as a buffer and run X on the \emph{smaller} segment.
If however the split is ``about even'', \ie, both segments are $\le \tfrac{n-1}{1+\alpha}$
we can sort the \emph{larger} of the two segments by X.
These cases also show up in the recurrence of costs.
\begin{align*}
		c(n) 
	&\wwrel=
		x(n) \ge 0,
		\qquad(\text{for }n\le w)
\\
		c(n)
	&\wwrel=
		\begin{aligned}[t]
		(n-k) + s(k)  
		&\bin+ \E*{\big[J_1,J_2 \le \tfrac1{1+\alpha} (n-1) \big]\cdot[J_1 > J_2]\cdot \big(x(J_1) + c(J_2)\big)}\\
		&\bin+ \E*{\big[J_1,J_2 \le \tfrac1{1+\alpha} (n-1) \big]\cdot[J_1 \le J_2]\cdot \big(x(J_2) + c(J_1)\big)}\\
		&\bin+ \E*{\big[J_2 > \tfrac1{1+\alpha}(n-1)\big]\cdot\big(x(J_1) + c(J_2)\big)}\\
		&\bin+ \E*{\big[J_1 > \tfrac1{1+\alpha}(n-1)\big]\cdot\big(x(J_2) + c(J_1)\big)}
		&\mkern-50mu (\text{for }n > w)
		\end{aligned}
\\	&\wwrel=
		\sum_{r=1}^2\E{A_r(J_r) c(J_r)}  + t(n)
\shortintertext{where}
		A_1(J)
	&\wwrel=
		\left[J,J' \le \tfrac1{1+\alpha} (n-1) \right]\cdot [J \le J'] \bin+ \left[J > \tfrac1{1+\alpha}(n-1)\right]
		\qquad\text{with }J' = (n-1)-J
\\
		A_2(J)
	&\wwrel=
		\left[J,J' \le \tfrac1{1+\alpha} (n-1) \right]\cdot [J < J'] \bin+ \left[J > \tfrac1{1+\alpha}(n-1)\right]
\\
		t(n)
	&\wwrel=
		n-k + s(k)
		\bin+ \E*{A_2(J_2) x(J_1)}
		\bin+ \E*{A_1(J_1) x(J_2)}
\end{align*}
The above formulation actually covers $\alpha=1$ as a special case, so
in both cases we have 
\begin{align*}
c(n)&\wwrel=
		\sum_{r=1}^2\E{A_r(J_r) c(J_r)}  + t(n)
		\numberthis\label{eq:recurrence-cn-E}
		\intertext{where $A_1$ (resp.\ $A_2$)  is the indicator random variable for the event ``left (resp.\ right) segment sorted recursively''  and}
t(n) &\wwrel=
		n-k + s(k)
		\bin+ \sum_{r=1}^2 \E{A_r \, x(J_{3-r})}.
		\numberthis\label{eq:definition-toll}
\end{align*}
We note that the expected number of partitioning rounds is only $\Theta(\log n)$
and hence also 
the expected overall number of comparisons used in all pivot sampling rounds combined is only 
$\Oh(\log n)$
when $k$ is constant.

\paragraph{Recursion indicator variables}
It will be convenient to rewrite $A_1(J_1)$ and $A_2(J_2)$ 
in terms of the \emph{relative subproblem size}:
\begin{align*}
A_1(J_1)
&\wwrel=
\biggl[\frac{J_1}{n-1} \rel\in \Bigl[\frac\alpha{1+\alpha},\frac12\like{\Bigr)}{\Bigr]}
\cup \Bigl(\frac1{1+\alpha},1\Bigr]\biggr] ,
\\
A_2(J_2)
&\wwrel=
\biggl[\frac{J_2}{n-1} \rel\in \Bigl[\frac\alpha{1+\alpha},\frac12\Bigr)
\cup \Bigl(\frac1{1+\alpha},1\Bigr]\biggr] .
\end{align*}
Graphically, if we view $J_1/(n-1)$ as a point in the unit interval,
the following picture shows which subproblem is sorted recursively for typical values of $\alpha$;
(the other subproblem is sorted by X).

\begin{center}\small
	\begin{tikzpicture}[yscale=7,xscale=10]
	\foreach \a/\y/\label in {
		0.5/0/\frac12,%
		0.25/1/\frac14,%
		1/2/1%
	} {
		\begin{scope}[shift={(0,-.25*\y)}]
		\draw (0,0) -- (1,0);
		\ifthenelse{\equal{\a}{1}}{
			\draw[{Bracket[]}-{Bracket[]},thick] (0,0) -- node[below] {$A_2=1$} ({\a/(1+\a)},0) ;
			\draw[{Parenthesis[]}-{Bracket[]},thick] (1/2,0) -- node[above] {$A_1=1$} (1,0) ;
			\foreach \x/\l [evaluate=\x] in {
				0/0,%
				0.5/{\frac12},%
				1/1%
			} {
				\node at (\x,-0.075) {$\l$} ;
			}
		}{
			\draw[{Bracket[]}-{Parenthesis[]},thick] (0,0) -- node[below] {$A_2=1$} ({\a/(1+\a)},0) ;
			\draw[{Parenthesis[]}-{Bracket[]},thick] ({1/(1+\a)},0) -- node[above] {$A_1=1$} (1,0) ;
			\draw[{Parenthesis[]}-{Bracket[]},thick] (1/2,0) -- node[below] {$A_2=1$} ({1/(1+\a)},0) ;
			\draw[{Bracket[]}-{Bracket[]},thick] ({\a/(1+\a)},0) -- node[above] {$A_1=1$} (1/2,0) ;
			\foreach \x/\l [evaluate=\x] in {
				0/0,%
				{\a/(1+\a)}/{\frac{\alpha}{1+\alpha}},%
				0.5/{\frac12},%
				{1/(1+\a)}/{\frac{1}{1+\alpha}},%
				1/1%
			} {
				\node at (\x,-0.075) {$\l$} ;
			}
		}
		\node at (1.1,0) {$\alpha = \label$} ;
		\end{scope}
	}
	\end{tikzpicture}
\end{center}
Obviously, we have $A_1 + A_2 = 1$ for any choice of $J_1$, 
which corresponds to having exactly one recursive call
in QuickXsort.

\subsection{Distribution of subproblem sizes}\label{sec:distribution}

A vital ingredient to our analyses below is to characterize the distribution
of the subproblem sizes $J_1$ and $J_2$.

Without pivot sampling, we have $J_1 \eqdist \uniform[0..n-1]$, 
a discrete uniform distribution.
In this paper, though, we assume throughout that pivots
are chosen as the me the \emph{median} of a random sample of $k=2t+1$,
elements, where $t\in\N_0$.
$k$ may or may not depend on $n$; we write $k=k(n)$ to emphasize a potential dependency.

By symmetry, the two subproblem sizes always have the same distribution, $J_1\eqdist J_2$.
We will therefore in the following simply write $J$ instead of $J_1$
when the distinction between left and right subproblem is not important.

\paragraph{Combinatorial model}
What is the probability $\Prob{J=j}$ to obtain a certain subproblem size~$j$?
An elementary counting argument yields the result. 
For selecting the $j+1$-st element as pivot, the sample needs to contain $t$ elements smaller 
than the pivot and $t$ elements large than the pivot. 
There are $\binom{n}{k}$ possible choices for the sample in total, 
and $\binom{j}{t}\cdot \binom{n-1-j}{t}$  of which will select the $j+1$-st element as pivot.
Thus,
\begin{align*}
		\Prob{J= j } 	
	&\wwrel=
		\frac{\binom{j}{t}\binom{n-1-j}{t}}{\binom{n}{k}}
\end{align*}
Note that this probability is $0$ for $j<t$ or $j > n-1-t$,
so we can always write $J = I+t$ for a random variable $I\in[0..n-k]$
with $\Prob{I=i} = \Prob{J = i+t}$.

The following lemma can be derived by direct elementary calculations,
showing that $J$ is concentrated around its expected value $\frac{n-1}2$.
\begin{lemma}[{{\cite[Lemma\,2]{DiekertWeiss2016}}}]\label{lm:prob_bound}
	Let $0<\delta < \frac{1}{2}$. 
	If we choose the pivot as median of a random sample of $k = 2t +1$ elements where
	$k \leq\frac{n}{2}$, 
	then the rank of the pivot $R = J_1 + 1$ satisfies
	\begin{align*}
		\Prob*{ R \leq \tfrac{n}{2} - \delta n} &\wwrel< k \rho^t
	\qquad\text{and}\qquad
		\Prob*{ R \ge \tfrac{n}{2} + \delta n} \wwrel< k \rho^t
	\end{align*}
	where $\rho = 1 - 4\delta^2 < 1$.
\ifthenelse{\boolean{citeproofs}}{\qed}{}\end{lemma}

\ifthenelse{\boolean{citeproofs}}{}{
\begin{proof}
	First note that the probability for choosing the $r$-th element as pivot satisfies 
	\[\binom{n}{k}\cdot \Prob{R = r } \wwrel=  \binom{r-1}{t}\binom{n-r}{t}{} .\]
	We use the notation of \emph{falling factorial} $x^{\underline \ell}
	= x \cdots (x-\ell +1)$. Thus, $\binom x \ell = {x^{\underline \ell}} / {\ell!}$. 
	
	\begin{align*}
			\Prob{R = r } 
		&\wwrel= 
			\dfrac{k!\cdot(r-1)^{\underline t}\cdot(n-r)^{\underline t}}{(t!)^2 \cdot n^{\underline {k}}}
	\\	&\wwrel= 
			\binom{2t}{t} \frac{k}{(n-k-1)} \prod_{i=0}^{t-1}\frac{(r-1-i)(n-r-i)}{(n-2i-1) (n-2i)} .
	\end{align*}
	For  $r \leq t$ we have  $\Prob{R = r} =0$. So, let  $t < r \leq  \frac{n}{2} - \delta n$ and 
	let us consider an index $i$ in the product with  $0\leq i< t$:
	
	\begin{align*}
			\frac{(r-1-i)(n-r-i)}{(n-2i-1) (n-2i)} 
		&\wwrel\leq 
			\frac{(r-i)(n-r-i)}{(n-2i) (n-2i)}
	\\	&\wwrel= 
			\frac{\left(\left(\frac{n}{2}-i\right)-\left(\frac{n}{2}-r\right)\right) 
			\cdot \left(\left(\frac{n}{2}-i\right)+\left(\frac{n}{2}-r\right)\right) }{\left(n-2i\right)^2}
	\\	&\wwrel= 
			\frac{\left(\frac{n}{2}-i\right)^2-\left(\frac{n}{2}-r\right)^2}{\left(n-2i\right)^2}
	\\	&\wwrel\leq 
			\frac{1}{4}- \frac{\left(\frac{n}{2}-\left(\frac{n}{2}-\delta n\right)\right)^2}{n^2}  
		\wwrel= 
			\frac{1}{4}-\delta^2 .
	\end{align*}
	We have  $\binom{2t}{t}\leq 4^t$. Since $k \leq\frac{n}{2}$, we obtain: 
	\begin{align*}
			\Prob{R = r} 
		&\wwrel\leq 
			4^t \frac{k}{(n-2t)}\left(\frac{1}{4}-\delta^2\right)^t
		\wwrel< 
			k \frac{2}{n}\rho^t .
	\end{align*}
	Now, we obtain the desired result:
	\begin{align*}
			\Prob{R \leq \frac{n}{2} - \delta n}
		&\wwrel< 
			\sum_{k=0}^{\floor{\frac{n}{2} - \delta n}} k \frac{2}{n}\rho^t 
		\wwrel\leq
			k \rho^t. 
	\end{align*}
\end{proof}
}

\paragraph{Uniform model}
There is a second view on the distribution of $J$
that will turn out convenient for our analysis.
Suppose our input consists of $n$ real numbers drawn \iid uniformly from $(0,1)$.
Since our algorithms are comparison based and the ranks of these numbers
form a random permutation almost surely, this assumption is without loss of generality
for expected-case considerations.

The vital aspect of this uniform model is that we can separate 
the \emph{value} $P\in(0,1)$ of the (first) pivot from its \emph{rank} $R\in[1..n]$.
In particular, $P$ only depends on the values in the random sample, whereas
$R$ necessarily depends on the values of all elements in the input.
It is a well-known result that the median of a sample of $\uniform(0,1)$ random variates 
has a \emph{beta distribution:} $P\eqdist \betadist(t+1,t+1)$.
Indeed, the density of the beta distribution is proportional to $x^t(1-x)^t$, 
which is the probability to have $t$ of the $\uniform(0,1)$ elements $\le x$ and
$t$ elements $\ge x$ (for a given value $x$ of the sample median).

Now suppose the pivot value $P$ is fixed.
Then, conditional on $P$, all further (non-sample) elements fall into the categories
``smaller than $P$'' resp.\ ``larger than $P$'' \emph{independently} and with probability
$P$ resp.\ $1-P$ (almost surely there are no duplicates).
Apart from the $t$ small elements from the sample, $J_1$ precisely 
counts how many elements are less than $P$,
so we can write $J_1 = I_1 + t$ where $I_1$ is the number of elements that turned out to be
smaller than the pivot during partitioning.

Since each of the $n-k$ non-sample elements is smaller than $P$ with probability $P$ independent of 
all other elements, we have conditional on $P$ that $I_1 \eqdist \binomial(n-k,P)$.
$I_1$ is said to have \emph{mixed} binomial distribution, with a beta-distributed \emph{mixer} $P$.
If we drop the conditioning on $P$, we obtain the so-called \emph{beta-binomial distribution}:
$I_1\eqdist \betaBinomial(n-k,t+1,t+1)$. 
We can express the probability weights by ``integrating $P$ out'':
\begin{align*}
		\Prob{I = i}
	&\wwrel=
		\Eover* P{\binom{n-k}i P^i (1-P)^i}
\\	&\wwrel=
		\int_{x=0}^1{\binom{n-k}i x^i (1-x)^{n-k-i}} \cdot \frac{x^t(1-x)^t}{\BetaFun(t+1,t+1)} \, dx
\\	&\wwrel=
		\frac{\binom{n-k}i}{\BetaFun(t+1,t+1)} \int_{x=0}^1 x^{t+i} (1-x)^{t+n-k-i}\, dx
\\[.1em]	&\wwrel=
		\binom{n-k}i \frac{\BetaFun(t+1+i,t+1+n-k-i)}{\BetaFun(t+1,t+1)} ,
\end{align*}
which yields the expression given in \wref{sec:beta-binomial-dist}.
(The last step above uses the definition of the beta function.)
Note that for $t=0$, \ie, no sampling, we have 
$t+\betaBinomial(n-k,t+1,t+1) = \betaBinomial(n-1,1,1) = \uniform[0..n-1]$,
so we recover the uniform case.

The uniform model is convenient since it allows to compute expectations involving $J$
by first conditioning on $P$, and then in a second step also taking expectations
\wrt $P$, formally using the law of total expectation.
In the first step, we can make use of the simple Chernoff bounds for the binomial
distribution (\wref{lem:chernoff-bound-binomial}) instead of \wref{lm:prob_bound}.
The second step is often much easier than the original problem and can use known 
formulas for integrals, such as the ones given in \wref{sec:beta-dist}.
For a larger collection of such properties and connections to other stochastic processes
see \cite[\href{https://www.wild-inter.net/publications/html/wild-2016.pdf.html\#pf64}{Section 2.4.7}]{Wild2016}.

\paragraph{Connection between models}
We obtained two expressions for $\Prob{J=j}$ from the two points of view above;
the reader might find it reassuring that they can indeed be proven equal by elementary 
term rewriting 
(see also \cite[\href{https://www.wild-inter.net/publications/html/wild-2016.pdf.html\#pfde}{Lemma 6.3}]{Wild2016}):
\begin{align*}
		\Prob{J= j } 	
	&\wwrel=
		\frac{\binom{j}{t}\binom{n-1-j}{t}}{\binom{n}{k}}
	\wwrel=
		\frac{k!\,(n-k)!}{n!}\cdot \frac{j!}{t!\,(j-t)!}\cdot \frac{(n-1-j)!}{t!\,(n-1-j-t)!}
\intertext{%
	setting $j = i+t$ and using $k=2t+1$, we obtain%
}
	&\wwrel= 
		\frac{k!\,(n-k)!}{n!}\cdot \frac{(i+t)!}{t!\,i!}\cdot \frac{(n-k-i+t)!}{t!\,(n-k-i)!}
\\	&\wwrel= 
		\frac{(n-k)!}{i!\,(n-k-i)!}\cdot 
		\frac{(t+i)!\,(t+n-k-i)!}{n!}
		\bin{\bigg/}
		\frac{t!\, t!} {k!}
\\	&\wwrel{\eqwithref{eq:betaFun}}
		\binom{n-k}{i} \frac{\BetaFun(i+t+1, n - i - t)}{\BetaFun(t+1, t+1)}.
\end{align*}

\section{Analysis for growing sample sizes}
\label{sec:analysis-growing-k}

In this and the following section, we derive general transfer theorems 
that allow us to express the total cost of \QuickXsort
in terms of the costs of X (as if used in isolation).
We can then directly use known results about X from the literature.

As in plain \Quicksort, the performance of \QuickXsort is heavily influenced by
the method for choosing pivots (though the influence is only on the linear term of the number of comparisons).
We distinguish two regimes here.
The first considers the case that the median of a large sample is used; more precisely,
the sample size is chosen as a growing but sublinear function in the subproblem size.
This method yields optimal asymptotic results and allows a rather clean analysis.
This case is covered in \wref{sec:analysis-growing-k}.

It is known for \Quicksort that
increasing the sample size yields rapidly diminishing marginal returns~\cite{Sedgewick1977},
and it is natural to assume that \QuickXsort behaves similarly.
Asymptotically, a growing sample size will eventually be better,
but the evidence in \wref{sec:experiments} shows
that a small, fixed sample size gives the best practical performance on realistic input sizes,
so these variants deserve further study.
This will be the purpose of \wref{sec:analysis-fixed-k}.

We mainly focus on the number of key comparisons as our cost model;
the transfer theorems derived here are, however, oblivious to this.

In this section, we derive general results which hold for a wide class of algorithms X.
As we will show, the average number of comparisons of X and of median-of-$k(n)$ \QuickXsort differ only by an
$o(n)$-term (if $k(n)$ grows as $n$ grows and under some natural assumptions).

\subsection{Expected costs}

Throughout this section, we assume
that the pivot is selected as the median of $k=k(n)$
elements where $k(n)$ grows when $n$ grows.
The following theorem allows to transfer an asymptotic approximation for the costs of X
to an asymptotic approximation of the costs of \QuickXsort.
We will apply this theorem to concrete methods X in \wref{sec:analysis-for-concrete-X}.

\begin{theorem} [Transfer theorem (expected costs, growing $k$)]
\label{thm:quickXsort}
	Let $c(n)$ be defined by \weqref{eq:recurrence-cn-E} 
	(the recurrence for the expected costs of QuickXsort) and assume
	$x(n)$ (the costs of X) and $k = k(n)$ (the sample size) fulfill
	$x(n) = a n \lg n + b n \pm o(n)$ for constants $a\ge1$ and $b$,
	and $k = k(n) \in \omega(1) \cap o(n)$ as $n\to\infty$ with $1\leq k(n) \leq n$ for all $n$.
	
	Then, $c(n) \le x(n) + o(n)$.
	For $a=1$, the above holds with equality,
	\ie, 
	$c(n) = x(n) + c'(n)$ with $c'(n) = o(n)$.
	Moreover, in the typical case with $k(n) = \Theta(n^\kappa)$ for $\kappa \in (0,1)$ and 
	$x(n) = a n \lg n + b  n \pm \Oh(n^{\delta})$ with $\delta \in [0,1)$,
	we have for any fixed $\epsilon > 0$ that
	\[
			c'(n) 
		\wwrel= 
			\Theta( n^{\max\{\kappa,1-\kappa\}}) 
			\bin\pm 
			\Oh(n^{\max\{\delta,1/2 + \epsilon\}})
	.\]
\end{theorem}

We note that this result considerably strengthens the error term
from $o(n)$ in versions of this theorem in earlier work to $\Oh(n^{1/2+\epsilon})$
(for $k(n) = \sqrt n$).
Since this error term is the only difference between the costs of \QuickXsort and X
(for $a=1$), we feel that this improved bound is not merely a technical contribution,
but significant strengthens our confidence in the utility and practicality of \QuickXsort
as an algorithmic template.

\begin{remark}[Optimal sample sizes]
	The experiments in \cite{DiekertWeiss2016} and  the results for \algorithmname{Quickselect} in \cite{MartinezRoura2001} suggest that 
	sample sizes $k(n) = \Theta(\sqrt{n})$ are likely to be optimal \wrt 
	balancing costs of pivot selection and benefit of better-quality pivots 
	within the lower order terms.
	
	\wref{thm:quickXsort} gives a proof for this in a special situation: assume that $a=1$,
	the error term $\xi(n) \in \Oh(n^{\delta})$ for some $\delta \in [0,\frac12]$ and that we are
	restricted to sample sizes $k(n) = \Theta(n^\kappa)$, for $\kappa\in(0,1)$. In this case
	\wref{thm:quickXsort} shows that $\kappa = \frac12$ is the optimal choice, \ie, 
	$k(n) = \sqrt{n}$ has the ``best polynomial growth'' among all feasible polynomial sample sizes.
\end{remark}

\begin{proof}[{{\prettyref{thm:quickXsort}}}]
	Let $c(n)$ denote the average number of comparisons performed by  \QuickXsort{}  
	on an input array of length $n$ and let $x (n) = a n \lg n + b n \pm \xi(n)$ with 
	$\xi(n) \in o(n)$ be (upper and lower) bounds for the average number of comparisons performed 
	by the algorithm~X on an input array of length $n$. 
	Without loss of generality we may assume that $\xi(n)$ is monotone.

	Let $A_1$ be the indicator random variable for the event ``left segment sorted recursively''
	and $A_2 = 1-A_1$ similarly for the right segment.
	Recall that $c(n)$ fulfills the recurrence
	\begin{align*}
			c(n)
		&\wwrel=
			\sum_{r=1}^2 \E{A_r \,c(J_r)} \bin+ t(n), \qquad\text{where}
	\\		t(n)
		&\wwrel=
			n-k(n) \bin+ s\bigl(k(n)\bigr) \bin+ \sum_{r=1}^2 \E{A_r \, x(J_{3-r})}
	\end{align*}
	and $J_1$ and $J_2$ are the sizes for the left resp.\ right segment
	created in the first partitioning step and $s(k) \in \Theta(k)$ is the expected number of comparisons to find the median of the sample of $k$ elements.
	
	\paragraph{Recurrence for the difference}
	To prove our claim, we will bound the difference $c'(n) = c(n) - x(n)$;
	it satisfies a recurrence very similar to the one for $c(n)$:
	\begin{align*}
			c'(n)
		&\wwrel=
			n
			-k(n)\bin+ s\bigl(k(n)\bigr)
			\bin+ \E[\Big]{ A_1 \cdot \bigl( c'(J_1) + x(J_1) + x(J_2) \bigr)} 
	\\*	&\wwrel\ppe{}
			\bin+ \E[\Big]{ A_2 \cdot \bigl( c'(J_2) + x(J_2) + x(J_1) \bigr)}
			\bin- x(n) 
	\\	&\wwrel=
			\E[\big]{ A_1 \, c'(J_1) }
			+\E[\big]{ A_2 \, c'(J_2) }
			\bin+ {\underbrace {
				n
				-k(n)+ s\bigl(k(n)\bigr)
				+ \E[\big]{x(J_1)} + \E[\big]{x(J_2)}
				- x(n) 
			}_ {t'(n)}}.
	\numberthis\label{eq:recurrence-c'n}
	\end{align*}
	(Note how taking the difference here turns the complicated
	terms $\E{A_r x(J_{3-r})}$ from $t(n)$ into the simpler $\E{x(J_r)}$ terms in $t'(n)$.)
	
	\paragraph{Approximating the toll function}
	We will eventually bound $c'(n)$; the first step is to 
	study the (asymptotic) behavior of the residual toll function $t'(n)$.
	
	\begin{lemma}[Approximating $t'(n)$]
	\label{lem:apx-t'}
	\label{lem:t'n-asymptotic}
		Let $t'(n)$ as in \weqref{eq:recurrence-c'n}. 
		Then for $\epsilon > 0$, we have
		\begin{align*}
				t'(n)	
			&\wwrel=
			(1 - a)n
			\bin+\Theta\Bigl(k(n) + \tfrac n{k(n)}\Bigr)
			\wbin\pm \Oh\Bigl(
				  \xi(n) 
				+ n^{1/2+\epsilon}
			\Bigr).
		\end{align*}
	Moreover, if $a=1$, $k(n) = \Theta(n^\kappa)$ for $\kappa \in (0,1)$ and 
		$\xi(n) = \Oh(n^{\delta})$ for $\delta \in [0,1)$, we have
	\[
		t'(n)
	\wwrel=
		\Theta\left(n^{\max\{\kappa,1-\kappa\}}\right)
		\wbin\pm \Oh\Bigl(n^{\max\{\delta,1/2 + \epsilon\}} \Bigr).
	\]
	\end{lemma}

\begin{proof}[\wref{lem:apx-t'}]
	
	We start with the simple observation that
	\begin{align*}
			J \lg J
		&\wwrel=
			J \bigl(\lg (\tfrac Jn) + \lg n\bigr)
		\wwrel=
			n \cdot \Bigl( \tfrac Jn \lg \tfrac Jn + \tfrac Jn \lg n  \Bigr)
		\wwrel=
			\tfrac J n \, n \lg n\bin+ \tfrac Jn \lg \bigl(\tfrac Jn\bigr)\, n.
		\numberthis\label{eq:JlnJ}
	\end{align*}
	With that, we can simplify $t'(n)$ to (recall $s(k) \in \Theta(k)$)
	\begin{align*}
			t'(n)
		&\wwrel=
			n -k(n)+ s\bigl(k(n)\bigr)
			\bin+ \sum_{r=1}^2 \E[\big]{a J_r\lg J_r + bJ_r \pm \xi(J_r)}
			- x(n)
	\\	&\wwrel=
			n
			\bin+ \sum_{r=1}^2 \Biggl(
				a \E{J_r} \lg n + a \E{\tfrac{J_r}n \lg(\tfrac{J_r}n)} n + b\E{J_r} \pm \xi(n)
			\Biggr)
			- x(n) 
			\wbin+ \Theta(k(n)) 
	\\	&\wwrel=
			n
			\bin+ \bigl(a n \lg n \pm \Oh(\log n)\bigr)
			+ 2a \E*{\tfrac{J_1}n \lg(\tfrac{J_1}n)} n
			+ (bn \pm \Oh(1))
			\pm 2\xi(n)
	\\*	&\wwrel\ppe{}
			- \bigl(a n\lg n + bn \pm \xi(n)\bigr)
			\wbin+ \Theta(k(n))
	\\	&\wwrel=
			\biggl(1
			+ 2a \E*{\tfrac{J_1} n \lg(\tfrac{J_1}n)} \biggr) n
			\wbin+ \Theta\bigl(k(n)\bigr) \pm \Oh(\xi(n))
	\numberthis\label{eq:t'n-rewritten}
	\end{align*}
	The expectation $\E*{\tfrac{J_1} n \lg(\tfrac{J_1}n)} = - \E{h(J_1/n)}$ is almost of the form addressed in
	\wref{lem:E-f-X-concentration} when we write the beta-binomial distribution of
	$J_1$ as the mixed distribution $J_1 = t(n) + I_1$, where $I_1\eqdist \betaBinomial(n-k,t+1,t+1)$: 
	we only have to change the argument from $- \E{h(J_1/n)}$ to $- \E{h(I_1/(n-k))}$. 
	The first step is to show that this can be done with a sufficiently small error. 
	For brevity we write $J$ (resp.\ $I$) instead of $J_1$ (resp.\ $I_1$).
	
	Let $\delta = \delta (n) = 1/\sqrt[4]{k(n)}$. Thus, by \wref{lm:prob_bound} and $1+x \le \exp(x)$, we obtain
	\begin{align*}
			\Prob*{J\leq (1/2-\delta)n}
		&\wwrel\leq 
			k(n) \cdot \left( 1 - 4 \cdot \frac1{\sqrt{k(n)}}\, \right)^{\!{(k(n)-1)/2}} 
	\\	&\wwrel\leq 
			k(n) \cdot \exp\left(-\frac{2(k(n)-1)}{\sqrt{k(n)}}\right)
	\\	&\wwrel\leq 
			k(n) \cdot \exp\bigl({\textstyle-\sqrt{k(n)}}\,\bigr)
	\\	&\wwrel= 
			\Oh\bigl(k(n)^{-2}\bigr).
	\numberthis\label{eq:prob_small}
	\end{align*}
	Notice that better bounds are easily possible, but do not affect the result.
	We need to change the argument in the expectation from $J/n$ to $I/(n-k)$ where $J = I+t$. 
	The idea is that we split the expectation into two ranges: 
	one for $J\in\left[\floor{(1/2-\delta)n .. \ceil{(1/2+\delta)n}}  \right] $ and one outside. 
	By \weqref{eq:prob_small}, the outer part has negligible contribution.
	For the inner part, we will now show that the difference between $J/n$ and $I/(n-k)$ is very small.
	So let $j \in\big[\ceil{(1/2+\delta)n}..\floor{(1/2-\delta)n}  \big] $ and write $j= i+t $.
	Then it holds that	
	\begin{align*}
			\frac{j}{n} - \frac{i}{n-k} 	
		&\wwrel= 
			\frac{j(n-k) - (j-t) n}{n(n-k)}
		\wwrel= 
			\frac{tn - jk}{n(n-k)}
	\\	&\wwrel=  
			\frac{t- k\cdot\bigl(\frac 12 \pm (\delta + 1/n)\bigr)}{n-k}
				\tag{because $j= n/2 \pm (\delta n + 1)$}
	\\	&\wwrel=  
			\frac{-\frac 12 \pm k(\delta + 1/n)}{n-k}  
		\wwrel=
			\Oh\left(\tfrac{k^{3/4}}{n}\right)\tag{because $k=2t+1$}
	\end{align*}
	(Note that this difference is $\Omega(k/n)$ for unrestricted values of $j$;
	only for the region close to $n/2$, the above bound holds.)
	
	Now, recall from \wref{lem:x-log-x} that $h$ is Hölder-continuous for any
	exponent $\holdA\in(0,1)$ with Hölder constant $C_\holdA/\ln 2$. 
	Thus, $\abs{h(y) - h(z)} = \Oh\Bigl( \Bigl(\tfrac{k(n)^{3/4}}{n}\Bigr)^{\!\holdA} \Bigr)$ for $y,z \in [0,1]$ with 
	$\abs{y-z} = \Oh\bigl(\frac{k(n)^{3/4}}{n}\bigr)$. 
	We use this observation to show:
	\begin{align*}
			\E*{- h(J/n)} &\wwrel= -\sum_{j=0}^{n} \Prob*{J=j} h(j/n)
	\\	&\wwrel{\eqwithref[r]{lem:x-log-x-bounds}}
			-\mkern-20mu \sum_{j=\floor{(1/2-\delta)n}}^{\ceil{(1/2+\delta)n}}  \mkern-20mu
				\Prob*{J=j} \cdot h\left(\tfrac{j}{n}\right) 
				\bin\pm 2\Prob[\big]{J \leq (1/2-\delta)n }\cdot \frac{\lg e}e
	\\	&\wwrel{\relwithtext[r]{Hölder-cont.}=}
			-\mkern-20mu \sum_{j=\floor{(1/2-\delta)n}}^{\ceil{(1/2+\delta)n}}  \mkern-20mu
				\Prob*{J=j} \cdot h\left(\tfrac{j-t}{n-k}\right) 
				\bin\pm 2\Prob[\big]{J \leq (1/2-\delta)n }\cdot \frac{\lg e} e
				\bin\pm \Oh\Bigl( \Bigl(\tfrac{k(n)^{3/4}}{n}\Bigr)^{\!\holdA} \Bigr)
	\\	&\wwrel{\eqwithref[r]{lem:x-log-x-bounds}}
			-\sum_{j=0}^{n} \Prob*{J=j} \cdot h\left(\tfrac{j-t}{n-k}\right) 
				\wbin\pm 4\Prob[\big]{J \leq (1/2-\delta)n }\cdot \frac{\lg e}{e}
				\bin\pm \Oh\Bigl( \Bigl(\tfrac{k(n)^{3/4}}{n}\Bigr)^{\!\holdA} \Bigr)
	\\	&\wwrel{\eqwithref{eq:prob_small}}
			\E*{\tfrac{I}{n-k} \lg(\tfrac{I}{n-k})} 
				\wwbin\pm  \Oh\biggl(
					\tfrac{1}{k(n)^2} + \Bigl(\tfrac{k(n)^{3/4}}{n}\Bigr)^{\!\holdA} 
				\biggr).
	\numberthis\label{eq:J-to-I}
	\end{align*}

	Thus, it remains to examine $\E*{- h(I/(n-k))}$ further. 
	By the definition of the beta binomial distribution, we have $I \eqdist \binomial(n-k(n),P)$
	conditional on the \emph{value} of the pivot $P\eqdist\betadist(t(n)+1,t(n)+1)$ 
	(see \wref{sec:distribution}).
	So we apply \wref{lem:E-f-X-concentration} on the \emph{conditional} expectation 
	to get for any $\zeta \ge 0$:
	\begin{gather*}
			\E*{h\bigl(\tfrac{I}{n-k}\bigr) \given P}
		\wwrel=
			h(P) \pm \rho
	\shortintertext{where}
			\rho
		\wwrel=
			\frac{C_\holdA}{\ln 2} \cdot 
				\zeta^\holdA \Bigl(1 - 2 e^ {-2 \zeta^2 (n-k(n))}\Bigr)
			+
			4 \frac{\lg e}{e} e^{-2 \zeta^2 (n-k(n))}.
	\end{gather*}
	By \weqref{eq:E-h-X} and the asymptotic expansion of the harmonic numbers
	(see, \eg, \cite[Eq.\,(9.89)]{GrahamKnuthPatashnik1994}),
%	(or \cite[{\href{https://www.wild-inter.net/publications/html/wild-2016.pdf.html\#pf71}{Proposition 2.55}}]{Wild2016}),
	we find
	\begin{align*}
			\E*{-h\left(\tfrac{I}{n-k}\right)}
		&\wwrel=
			-\E{h(P)} \pm \rho
	\\	&\wwrel{\eqwithref{eq:E-h-X}}
			-\frac{ \frac12 (\harm{k(n)+1} - \harm{(k(n)+1)/2}) }{ \ln 2 } \wwbin\pm \rho
	\\	&\wwrel=
			-\frac12 \cdot \frac{ \ln 2 - \Theta(1/k(n)) }{ \ln 2 } \wwbin\pm \rho
	\intertext{and using the choice for $\zeta$ from \wref{lem:E-f-X-concentration}}
		&\wwrel=
			-\frac12 + \Theta\bigl( k(n)^{-1} \bigr) \wwbin\pm \Oh\bigl( 
				 n^{-1/2+\epsilon} 
			\bigr)
	\end{align*}
	for any fixed $\epsilon \in (0,\frac12)$ with \(\varepsilon>\frac{1-\holdA}2\) 	(recall that still $\epsilon$ can be an arbitrarily small constant). Together with \wref{eq:t'n-rewritten} and \wref{eq:J-to-I} this allows us to estimate 
$t'(n)$. Here, we set $\epsilon' = 1-\holdA$:
	\begin{align*}
			t'(n)
		&\wwrel=
			(1 - a)n
			\bin+\Theta\Bigl(k(n) + \tfrac n{k(n)}\Bigr)
			\wbin\pm \Oh\Bigl(
				\xi(n) 
				+ \sqrt n \cdot n^{\epsilon} 
				+ \tfrac n{k(n)^2}
				+ k(n)^{3/4}\cdot \bigl(\tfrac n{{k(n)}^{3/4}}\bigr)^{\epsilon'}  
			\Bigr)
	\intertext{%
		replacing $\epsilon$ and $\epsilon'$ by their maximum, 
		we obtain for any small enough $\epsilon > 0$, that%
	}
			t'(n)	
		&\wwrel=
			(1 - a)n
			\bin+\Theta\Bigl(k(n) + \tfrac n{k(n)}\Bigr)
			\wbin\pm \Oh\Bigl(
				\xi(n) 
				+ n^{\epsilon}\cdot\left(\sqrt n +k(n)^{3/4}\right)  
				+ \tfrac n{k(n)^2} 
			\Bigr)
	\\	&\wwrel=
			(1 - a)n
			\bin+\Theta\Bigl(k(n) + \tfrac n{k(n)}\Bigr)
			\wbin\pm \Oh\Bigl(
				\xi(n)  
				+ n^{1/2+\epsilon} 
			\Bigr).
	\end{align*}
To see the last step, let us verify that $n^{\epsilon} k(n)^{3/4} = \Oh( n^{1/2 + \epsilon} ) + o(k(n))$: we write $\N = N_1 \cup N_2$ with $N_1 = \set{n\in \N}{k(n) \leq \sqrt{n}}$ and $N_1 = \set{n\in \N}{k(n) \geq \sqrt{n}}$. For $n \in N_1$ clearly we have $n^{\epsilon} k(n)^{3/4}\leq n^{1/2 + \epsilon}$. For $n \in N_2$, we have $\sqrt[4]{k(n)} \geq \sqrt[8]{n}\geq n^{\epsilon + \epsilon''}$ for some small $\epsilon''>0$ (here we need that $\epsilon$ is small); thus, $n^{\epsilon} k(n)^{3/4}\leq k(n) n^{-\epsilon''}$. Altogether, we obtain $n^{\epsilon} k(n)^{3/4} = \Oh( n^{1/2 + \epsilon} ) + o(k(n))$.	

In the case that $a=1$, $k(n) = \Theta(n^\kappa)$ for $\kappa \in (0,1)$ and 
	$\xi(n) = \Oh(n^{\delta})$ for $\delta \in [0,1)$, we have
	\begin{align*}
			t'(n)	
		&\wwrel=
			\Theta\left(n^{\max\{\kappa,1-\kappa\}}\right)
				\wbin\pm \Oh\Bigl(n^{\max\{\delta,\tfrac 12 + \epsilon\}} \Bigr) 
	\end{align*}
\end{proof}
	Note that $t'(n)$ can be positive or negative (depending on $x(n)$),
	but the $\Theta$-bound is definitively a positive term,
	and it will be minimal for $k(n) \sim \sqrt{n}$.
	Now that we know the order of growth of $t'(n)$, we can proceed to 
	our recurrence for the difference $c'(n)$.

	\paragraph{Bounding the difference} 
	The final step is to bound $c'(n)$ from above. 
	Recall that by \wref{eq:recurrence-c'n}, we have $ c'(n) = \E[\big]{ A_1 \,  c(J_1) }
	+\E[\big]{ A_2 \,  c'(J_2) }+ t'(n)	$. 
	For the case $a>1$, \wref{lem:apx-t'} tells us that $t'(n)$ is eventually negative
	and asymptotic to $(1-a)n$. Thus $c'(n)$ is eventually negative, as well,
	\ie, $c(n) \le x(n)$ for large enough $n$. The claim follows.
	
	We therefore are left with the case $a=1$.
	\wref{lem:apx-t'} only gives us a bound in that case and certainly $t'(n) = o(n)$.
	The fact that $t'(n)$ can in general be positive or negative and 
	need not be monotonic, makes solving the recurrence for $c'(n)$ a formidable problem,
	but the following simpler problem can easily be solved.

	\begin{lemma}\label{lem:difference-bound}
		Let $\hat t: \R_{\ge0}\to \R_{\ge0}$ be monotonically increasing and consider the recurrence 
		\[ 
				\hat c(n) 
			\wwrel= 
				  \E[\big]{ A_1 \, \hat c(J_1) }
				+ \E[\big]{ A_2 \, \hat c(J_2) } 
				+ \hat t(n)
		\] 
		with $\hat c(n) = c_0$ for $n\leq 1$. 
		Then for any constant $\beta \in (\frac12,1)$ there is a constant $C = C(\beta)>0$ such that
		\[
				\hat c(n) 
			\wwrel\leq 
				C \sum_{i=0}^{\lceil\log_{1/\beta} (n)\rceil} \hat t(n \beta^i)
		\]
\end{lemma}
	\begin{proof}[\wref{lem:difference-bound}]
		Since $\hat t(n)$ is non-negative and monotonically increasing, 
		so is $\hat c(n)$ and we can bound
	\[	\hat c(n)
		\wwrel\le
		\E[\big]{ \hat c(\max\{J_1,J_2\}) } \bin+ \hat t(n).\]
		Let us abbreviate $\hat J = \max\{J_1,J_2\}$.
		For given constant $\beta\in(\frac12,1)$,
		we have by the law of total expectation and monotonicity of $\hat c$ that
		\begin{align*}
		\hat c(n)
		&\wwrel\le
		\E[\big]{ \hat c(\hat J) } \bin+ \hat t(n)
		\\	&\wwrel=
		\E[\big]{ \hat c(\hat J) \given \hat J \le \beta n } \cdot \Prob[\big]{ \hat J \le \beta n }
		\bin+
		\E[\big]{ \hat c(\hat J) \given \hat J > \beta n } \cdot \Prob[\big]{ \hat J > \beta n }
		\bin+ \hat t(n)
		\\	&\wwrel\le
		\Prob[\big]{ \hat J \le \beta n } \cdot \hat c(\beta n) 
		+ \Prob[\big]{ \hat J > \beta n } \cdot \hat c(n) \bin+ \hat t(n).
		\intertext{%
			For fixed  $\beta>\frac12$, we can bound 
			$\Prob[\big]{ \hat J \le \beta n } \ge C > 0$ for a constant $C$
			and all large enough $n$. Hence for $n$ large enough,%
		}
		\hat c(n)
		&\wwrel\le
		\frac{\Prob[\big]{ \hat J \le \beta n }}{1-\Prob[\big]{ \hat J > \beta n }} \cdot \hat c(\beta n) \bin+ 
		\frac1{1-\Prob[\big]{ \hat J > \beta n }} \cdot \hat t(n)
		\\	&\wwrel\le
		\hat c(\beta n) \bin+ 
		\frac1{C} \cdot \hat t(n).
		\end{align*}
		Iterating the last inequality $\lceil{\log_{1/\beta}(n)}\rceil$ times,
		we find 
		$\hat c(n) \leq \frac1{C} \sum_{i=0}^{\lceil{\log_{1/\beta} (n)}\rceil} \hat t(n \beta^i)$.
	\end{proof}
	We can apply this lemma if we replace $t'(n)$ by $\hat t(n) \ce \max_{m\le n} |t'(m)|$,
	which is both non-negative and monotone.
	We clearly have $t'(n) \le \hat t(n)$ by definition. 
	Moreover, if $t'(n) = \Oh(g(n))$ for a monotonically increasing function $g$, then
	also $\hat t(n) = \Oh(g(n))$, and the same statement holds with $\Oh$ replaced by $o$.

	Now let $\hat c(n)$ be defined by the recurrence 
	$\hat c(n) = \E[\big]{ A_1 \, \hat c(J_1) }
				+\E[\big]{ A_2 \, \hat c(J_2) }
				+\hat t(n)
	$. 
	Then, we have $|c'(n)|\le \hat c(n)$.
	We will now bound $\hat c(n)$.

	\paragraph{\boldmath $o(n)$ bound}
	We first show that $ \hat c(n) = o(n)$.
	By \wref{lem:t'n-asymptotic} we have $t'(n) = o(n)$, and by the above argument also $\hat t(n) = o(n)$.
	Since $\hat{t}(n) \in o(n)$, we know that for every $ \epsilon>0$, 
	there is some $N_\epsilon \in \N$ such that for  $n \geq N_\epsilon$,
	we have  $\hat t(n) \leq n \epsilon$. 
	Let $D_\epsilon = \sum_{i=0}^{N_\epsilon} \hat t(i)$.	
	Then, for any $\beta \in (\frac12,1)$ by \wref{lem:difference-bound} 
	there is some constant $C$ such that for all $n$ we have
	\[
			|c'(n)|
		\wwrel\le 
			\hat c(n) 
		\wwrel\leq 
			C \sum_{i=0}^{\ceil{\log_{1/\beta} (n)}} \hat t(n \beta^i) 
		\wwrel\leq 
			D_\epsilon + C\sum_{i=0}^{\ceil{\log_{1/\beta} (n)}} \epsilon\beta^i n 
		\wwrel\leq 
			CD_\epsilon + \epsilon' n
	\]
	for 
	$
			\epsilon' 
		\ce 
			\frac{C}{1-\beta}\cdot \epsilon 
		\ge 
			C\epsilon \sum_{i=0}^{\lceil{\log_{1/\beta} (n)}\rceil} \beta^i 
	$. 
	Since we can hence find a suitable 
	$\epsilon = \epsilon(\epsilon')>0$ for any given $\epsilon' > 0$,
	the above inequality holds for all $\epsilon' > 0$, and therefore $\hat c(n) = o(n)$ holds. 
	This proves the first part of \wref{thm:quickXsort}.
			
	\paragraph{Refined bound}
	Now, consider the case that $k(n) = \Theta(n^\kappa)$ for $\kappa \in (0,1)$ and 
	$\xi(n) \in  \Oh(n^{\delta})$ with $\delta \in [0,1)$. 
	Then, by \wref{lem:t'n-asymptotic}, we have $t'(n) = \Oh(n^\gamma)$ for some $\gamma \in (0,1)$, 
	\ie, there is some constant $C_\gamma$ such that $t'(n) \leq C_\gamma n^\gamma + \Oh(1)$. 
	By \wref{lem:difference-bound}, we obtain
	\[
			c'(n) 
		\wwrel\leq 
			\sum_{i=0}^{\lceil{\log_{1/\beta} (n)}\rceil} C_\gamma (n \beta^i)^\gamma + \Oh(1) 
		\wwrel\leq 
			C_\gamma n^\gamma\sum_{i\geq 0} (\beta^{\gamma})^i + \Oh(1) 
		\wwrel= \Oh(n^\gamma).
	\]
	Moreover, if $t'(n) \in \Theta(n^\gamma)$ 
	(the case that $\max\{\kappa, 1-\kappa\} > \max\{\delta, \frac{1}{2} + \epsilon \}$ in \wref{lem:t'n-asymptotic}), 
	then also $c'(n) \in \Theta(n^\gamma)$ as $c'(n) \ge t'(n)$. 
	This concludes the proof of the last part of \wref{thm:quickXsort}.
\end{proof}

\wref{thm:quickXsort} shows that for methods X that have optimal costs up to linear terms ($a=1$), 
also median-of-$k(n)$ \QuickXsort with $k(n) = \Theta(n^\kappa)$ and $\kappa \in (0,1)$ as $n\to\infty$
is optimal up to linear terms.
We obtain the best lower-order terms with median-of-$\sqrt n$ \QuickXsort, namely
$c(n) = x(n) \pm \Oh(n^{1/2+\epsilon})$, and we will in the following focus on this case.

Note that our proof actually gives slightly more information than stated in the theorem
for the case that the cost of X are not optimal in the leading-term coefficient ($a>1$).
Then \QuickXsort uses asymptotically \emph{fewer} comparisons than X, 
whereas for X with optimal leading-term costs,
\QuickXsort uses slightly more comparisons.

\subsection{Large-deviation bounds}

Does \QuickXsort{} provide a
good bound for the worst case? The obvious answer is ``no''. If always
the $\sqrt{n}$ smallest elements are chosen for pivot selection, a
running time of $\Theta(n^{3/2})$ is obtained. However, we can
prove that such a worst case is very unlikely. In fact, let
$x_{\mathrm{wc}}(n)$ be the worst case number of comparisons of the algorithm X.
\prettyref{pro:worstunlikely} states that the probability that
\QuickXsort needs more than $x_{\mathrm{wc}}(n) + 6n$ comparisons decreases
exponentially in $n$. 
 (This bound is not tight, but since we do not aim for exact
 probabilities, \prettyref{pro:worstunlikely} is enough for us.)

\begin{proposition}\label{pro:worstunlikely}
Let $\epsilon > 0$. The probability that median-of-$\sqrt{n}$ \QuickXsort{} needs more than
$x_{\mathrm{wc}}(n) + 6n$ comparisons is less than $(3/4+ \epsilon)^{\sqrt[4]{n}}$
for $n$ large enough.
\end{proposition}

\begin{proof}%[\prettyref{pro:worstunlikely}]
	Let $n$ be the size of the input. We say that we are in a \emph{good} case if an
	array of size $m$ is partitioned in the interval $[m/4,\, 3m/4]$, \ie, if the
	pivot rank is chosen in that interval. We can obtain a bound for the desired
	probability by estimating the probability that we always are in such a good case
	until the array contains only $\sqrt{n}$ elements. For smaller arrays, we can
	assume an upper bound of $\sqrt{n}^2 = n$ comparisons for the worst case.
	If we are always in a good case,
	all partitioning steps sums up to less than $n\cdot\sum_{i \geq
		0}(3/4)^i =
	4n$ comparisons. We also have to consider the
	number of comparisons required to find the pivot element. At any stage the
	pivot is chosen as median of at most $\sqrt{n}$ elements. Since the median can
	be determined in linear time, for all stages together this sums up to less than
	$n$ comparisons if we are always in a good case and $n$ is large enough.
	Finally, for all the sorting phases with
	X we need at most $x_{\mathrm{wc}}(n)$ comparisons in total (that is only a rough upper bound
	which can be improved). Hence, we
	need at most $x_{\mathrm{wc}}(n) + 6n$ comparisons if always a good case occurs.
	
	Now, we only have to estimate the probability that always a good case occurs.
	By \prettyref{lm:prob_bound}, the probability for a good case in the first partitioning step is at least $1-d\cdot\sqrt{n}\cdot\left(3/4\right)^{\sqrt{n}}$ for some constant $d$. 
	We have to choose $\log_{3/4}(\sqrt n / n) < 1.21\lg n$ times a pivot in the interval $[m/4,\, 3m/4]$, 
	then the array has size less than $\sqrt{n}$. 
	We  only have to consider partitioning steps where the array has size greater than 
	$\sqrt{n}$ (if the size of the array is already less than $\sqrt{n}$ we define the 
	probability of a good case as $1$). 
	Hence, for each of these partitioning steps we obtain that the probability for a 
	good case is greater than $1-d\cdot\sqrt[4]{n}\cdot\left(3/4\right)^{\sqrt[4]{n}}$. 
	Therefore, we obtain
	\begin{align*}
			\Prob{\text{always good case}}
		&\wwrel\geq
			\left(1-d\cdot\sqrt[4]{n}\cdot\left(3/4\right)^{\sqrt[4]{n}}\right)^{1.21\lg(n)}
	\\	&\wwrel\geq 
			1 - 1.21\lg(n) \cdot d\cdot\sqrt[4]{n}\cdot\left(3/4\right)^{\sqrt[4]{n}}
	\end{align*}
	by Bernoulli's inequality. For $n$ large enough we have $1.21\lg(n) \cdot
	d\cdot\sqrt[4]{n}\cdot\left(3/4\right)^{\sqrt[4]{n}}\leq (3/4+
	\epsilon)^{\sqrt[4] {n}}$.
\end{proof}

\subsection{Worst-case guarantees}

In order to obtain a provable bound for the worst case complexity we
apply a simple trick similar to the one used in \algorithmname{Introsort} \cite{Mus97}. 
We choose some $\delta \in (0,1/2)$. 
Now, whenever the pivot is more than $\delta n$
off from the median (\ie, if $J_1 \leq (1/2-\delta)n$ or $J_2 \leq (1/2-\delta)n$), 
we choose the next pivot as median of the whole array using the median-of-medians 
algorithm~\cite{BFPRT73} (or some other selection algorithm with a linear worst case). 
Afterwards we continue with the usual sampling strategy. 
We call this median-of-medians fallback pivot selection. 

\begin{remark}
	Notice that instead of choosing the next pivot as median, we can also switch to an
	entirely different sorting algorithm as it is done in \algorithmname{Introsort}~-- as we
	proposed in \cite{EdelkampWeiss2014}. The advantage in \cite{EdelkampWeiss2014} is that
	theoretically a better worst-case can be achieved: indeed, we showed that the worst-case
	is only $n + o(n)$ comparisons above the worst case of the fallback algorithm. Thus, using
	Reinhardt's \algorithmname{Mergesort} \cite{Reinhardt1992}, we obtain a worst case of
	$n\log n - 0.25n + o(n)$. However, here we follow a different approach for two reasons:
	first, we want to give a (almost) self-contained description of the algorithm; second, we
	are not aware of a fallback algorithm which in practice performs better than our approach:
	\algorithmname{Heapsort} and most internal \algorithmname{Mergesort} variants are
	considerably slower. Moreover, we are even not aware of an implementation of Reinhardt's
	\algorithmname{Mergesort}.
\end{remark}

\begin{theorem}[\QuickXsort{} Worst-Case]\label{thm:QuickXYsort}
	Let X be a sorting algorithm with at most $x(n) = n \lg n + bn +o(n)$
	comparisons in the average case and $x_{\mathrm{wc}}(n) = n \lg n + \Oh(n)$
	comparisons in the worst case and let $k(n) \in \omega(1) \cap o(n)$ with $1 \leq k(n) \leq n$ for all $n$. 
	If $k(n) = \omega(\sqrt{n})$, we additionally require that always some worst-case linear time algorithm is used for pivot selection (\eg\ using \algorithmname{IntroSelect} or the median-of-medians algorithm); otherwise, the worst-case is allowed to be at most quadratic (\eg\ using \algorithmname{Quickselect}).
	
	Then, median-of-$k(n)$
	\QuickXsort{} with median-of-medians fallback pivot selection is a sorting algorithm that performs $x(n) + o(n)$ comparisons in the average case and $n \lg n + \Oh(n)$ comparisons in the worst case.	
\end{theorem} 

Thus, by applying the median-of-medians fallback pivot selection, the average case changes only in the $o(n)$-terms. Notice that the $\Oh(n)$-term for the worst case of \QuickMergesort is rather large because of the median-of-medians algorithm. 
Nevertheless, in \cite{EdelkampWeiss2018MQMS}, we elaborate the technique of median-of-medians pivot selection in more detail. In particular, we show how to reduce the $\Oh(n)$-term for the worst case down $3.58n$ for \QuickMergesort.

\begin{proof}
It is clear that the worst case is $n \lg n + \Oh(n)$ comparisons since there can be at
most $\max\{2\lg n, \log_{1/2+\delta} n\}$ rounds of partitioning (by the additional
requirement, pivot selection takes at most linear time).
Thus, it remains to consider the average case~-- for which we follow the proof of
\wref{thm:quickXsort}. We say a pivot choice is ``bad'' if the next pivot is selected as
median of the whole array (\ie, if $J_1 \leq (1/2-\delta)n$ or $J_2 \leq (1/2-\delta)n$),
otherwise we call the pivot ``good''.
	
The difference to the situation in \wref{thm:quickXsort} is that now we have four segments
to distinguish instead of two: let $A'_1$ be the indicator random variable for the event
``left segment sorted recursively''
and $A'_2$ similarly for the right segment~-- both for the case that the pivot was good.
Likewise, let $A''_1$ be the indicator random variable for the event ``left segment sorted
recursively ''
and $A''_2 = 1-A'_1-A'_2-A''_1$ ``right segment sorted recursively'' in the case that the
pivot was bad. Then, $A_1= A'_1 + A''_1$ is the indicator random variable for the event
``left segment sorted recursively'' and $A_2 =A'_2 + A''_2$  the same for the right
segment.
Let $c(n)$ denote the average number of comparisons of median-of-$k(n)$
\QuickXsort{} with median-of-medians fallback pivot selection and $\tilde c(n)$ the same
but in the case that the first pivot is selected with the median-of-medians algorithms. We
obtain the following recurrence
\begin{align*}
c(n)
&\wwrel=
\underbrace{n-k(n)}_{\mathclap{\text{partitioning}}}{} 
\bin+ \underbrace{s\bigl(k(n)\bigr)}_{\mathclap{\text{pivot sampling}}}{}
\bin+\E[\Big]{ A'_1 \cdot \bigl( c(J_1) + x(J_2) \bigr) 
	\bin+ A'_2 \cdot \bigl( c(J_2) + x(J_1) \bigr)}  \\*
&\wwrel\ppe{}\bin+ \E[\Big]{ A''_1 \cdot \bigl( \tilde c(J_1) + x(J_2) \bigr) 
	\bin+ A''_2 \cdot \bigl( \tilde c(J_2) + x(J_1) \bigr)}
\\	&\wwrel=
\sum_{r=1}^2 \E{A_r  \,c(J_r)} \bin+ t(n), \qquad\text{where}
\\		t(n)
&\wwrel=
n-k(n) \bin+ s\bigl(k(n)\bigr) + \sum_{r=1}^2 \E{A_r  \,x(J_r)} + \sum_{r=1}^2 \E{A''_{r+2} \,(\tilde c(J_r) - c(J_r))}.
\end{align*}
As before $s(k)$ is the number of comparisons to select the median from the $k$ sample elements and $J_1$ and $J_2$ are the sizes for the left resp.\ right segment
created in the first partitioning step. Since $n \log n- \Oh(n) \leq  \tilde c(n)  \leq c_{\mathrm{wc}}(n)$ and  $c_{\mathrm{wc}}(n) = n \log n + \Oh(n)$, it follows that 
$\tilde c(n) - c(n) \in \Oh(n)$. By \wref{lm:prob_bound} we have $\Prob{A''_1}, \Prob{A''_2} \in o(1)$. Thus,
\[\zeta(n) \wwrel\ce \sum_{r=1}^2 \E{A''_{r} \,(\tilde c(J_r) - c(J_r))} \wwrel\in o(n).\]
As for \wref{thm:quickXsort} we now consider $c'(n) = c(n) - x(n)$ yielding
\begin{align*}
		c'(n)
	&\wwrel=
		n - k(n)\bin+ s\bigl(k(n)\bigr) \bin+ \E[\Big]{ A_1 \cdot \bigl( c'(J_1) + x(J_1) + x(J_2) \bigr)} 
\\*	&\wwrel\ppe{}
		\bin+ \E[\Big]{ A_2 \cdot \bigl( c'(J_2) + x(J_2) + x(J_1) \bigr)} \bin+  \zeta(n) 
		\bin- x(n)
\\	&\wwrel=
		\E[\big]{ A_1 \, c'(J_1) }
		+\E[\big]{ A_2 \, c'(J_2) } + t'(n)
\end{align*}
for $t'(n) =	n
-k(n)+ s\bigl(k(n)\bigr)
+ \E[\big]{x(J_1)} + \E[\big]{x(J_2)} +  \zeta(n)
- x(n)$. Now the proof proceeds exactly as for \wref{thm:quickXsort}.
\end{proof}

\section{Analysis for fixed sample sizes}
\label{sec:analysis-fixed-k}

In this section, we consider the practically relevant version of QuickXsort,
where we choose pivots as the median of a sample of fixed size~$k$.
We think of $k$ as a design parameter of the algorithm that we have to choose.
Setting $k=1$ corresponds to selecting pivots uniformly at random;
good practical performance is often achieved for moderate values, say, $k=3,\ldots,9$. 

For very small subproblems, when $n\le w$ for a constant $w\ge k$, 
we switch to another sorting method (for simplicity we can assume that they are sortied directly with X).
Clearly this only influences the \emph{constant term} of costs
in \QuickXsort. Moreover the costs of sampling pivots is $\Oh(\log n)$
in expectation (for constant $k$ and $w$), so we how the median of the $k$ sample elements
is found is immaterial.

\subsection{Transfer theorem for fixed k}

We now state the main result of this section, the transfer theorem for
median-of-$k$ \QuickXsort when $k$ is fixed.
Instantiations for actual X  are deferred to \wref{sec:analysis-for-concrete-X}.
Recall that $I_{x,y}(\betL,\betR)$ denotes the regularized incomplete beta function,
see \wpeqref{eq:regularized-incomplete-beta}.

\begin{theorem}[Transfer theorem (expected costs, fixed $k$)]
\label{thm:cn}
	Let $c(n)$ be defined by \weqref{eq:recurrence-cn-E} 
	(the recurrence for the expected costs of QuickXsort) and assume
	$x(n)$ (the costs of X) fulfills
	$x(n)=a n\lg n +bn \pm \Oh(n^{1-\epsilon})$ for constants $\epsilon\in(0,1]$, $a\ge 1$ and $b$.
	Assume further that $k$ (the sample size) is a fixed odd constant $k=2t+1$, $t\in\N_0$.
	Then it holds that
	\begin{align*}
			c(n)
		&\wwrel=
			x(n) \bin+ q \cdot n
			\wbin\pm\Oh(n^{1-\epsilon} + \log n),
	\shortintertext{where}
			q
		&\wwrel=
			\frac 1H - a\cdot\frac {\harm{k+1}-\harm{t+1}}{H \ln 2}
	\\[.2em]
			H
		&\wwrel=  I_{0,\frac\alpha{1+\alpha}}(t+2,t+1),
			\bin+ I_{\frac12,\frac1{1+\alpha}}(t+2,t+1).
	\end{align*}
\end{theorem}

Before we prove \wref{thm:cn}, let us look at the consequences for the number of comparisons of \QuickXsort.

\paragraph{The QuickXsort penalty}
\label{sec:quicksort-penalty}

Since all our choices for X are optimal up to linear terms, so will \QuickXsort be. 
We thus have $a=1$ in \wref{thm:cn}; $b$ (and the allowable $\alpha$) still depend on X.
We then find that going from X to \QuickXsort basically adds a ``penalty'' $q$
in the linear term that depends on the sampling size (and $\alpha$) but not on X.
\wref{tab:q} shows that this penalty is $\approx n$ without sampling, but can be reduced
drastically when choosing pivots from a sample of $3$ or $5$ elements.

\begin{table}[bhtp]
	\plaincenter{
		\begin{tabular}{ r *{6}{c} }
		\toprule
			                  & $k=1$    &  $k=3$   &  $k=5$   &  $k=7$   &  $k=21$   & $t\to\infty$ \\
		\midrule
			$\alpha=1$        & $1.1146$ & $0.5070$ & $0.3210$ & $0.2328$ & $0.07705$ &     $0$      \\
			$\alpha=\sfrac12$ & $0.9120$ & $0.4050$ & $0.2526$ & $0.1815$ & $0.05956$ &     $0$      \\
			$\alpha=\sfrac14$ & $0.6480$ & $0.2967$ & $0.1921$ & $0.1431$ & $0.05498$ &     $0$      \\
		\bottomrule
		\end{tabular}
	}
	\caption{%
		QuickXsort penalty.
		QuickXsort with $x(n) = n\lg n + bn$ yields
		$c(n) = n\lg n + (q +b) n$, where $q$, the QuickXsort penalty, is given in the table.
	}
	\label{tab:q}
\end{table}

As we increase the sample size, we converge to the situation for growing sample sizes
where no linear-term penalty is left (\wref{sec:analysis-growing-k}).
That $q$ is less than $0.08$ already for a sample of $21$ elements indicates most
benefits from pivots sampling are achieved for moderate sample sizes.
It is noteworthy that the improvement from no sampling to median-of-3 yields a 
reduction of $q$ by more than $50\%$, which is much more than
its effect on Quicksort itself 
(where it reduces the leading term of costs by 15\,\% from $2n\ln n$ to $\frac{12}7 n\ln n$).

\begin{proof}[\wref{thm:cn}]
The proof of \wref{thm:cn} fills the remainder of this section.
We start with \wpeqref{eq:recurrence-c'n}, the recurrence for $c'(n) = c(n) - x(n)$.%
\footnote{%
	Although the statement of the theorem is the same, our proof here
	is significantly shorter than the one given in 
	\cite[\href{https://www.wild-inter.net/publications/html/wild-2018a.pdf.html\#pf5}{Theorem~5.1}]{Wild2018a}.
	First taking the difference $c(n) - x(n)$ turns the much more complicated
	terms $\E{A_r x(J_{3-r})}$ from $t(n)$ into the simpler $\E{x(J_r)}$ in $t'(n)$,
	which allowed us to omit
	\cite[\href{https://www.wild-inter.net/publications/html/wild-2018a.pdf.html\#pf11}{Lemma E.1}]{Wild2018a}.
}
Recall that $c(n)$ denotes the expected number of comparisons performed by \QuickXsort{}.
With $x(n) = a n \lg n + bn \pm \xi(n)$
for a monotonic function $\xi(n) = \Oh(n^{1-\epsilon})$, the same arguments
as in the proof of \wref{thm:quickXsort} lead to 
\begin{align*}
		t'(n)
	&\wwrel=
		\biggl(1
		+ 2a \E*{\tfrac{J_1} n \lg(\tfrac{J_1}n)} \biggr) n
		\wbin+ \Theta\bigl(s(k(n))\bigr) \pm \Oh(\xi(n))
\tagref{eq:t'n-rewritten}
\\	&\wwrel=
		\biggl(1
		+ 2a \E*{\tfrac{J} n \lg(\tfrac{J}n)} \biggr) n
		\wbin\pm \Oh(n^{1-\epsilon}).
\numberthis\label{eq:t'n-fixed-k}
\end{align*}

The main complication for fixed $k$ is that~-- 
unlike for the median-of-$\sqrt n$ case, 
where the pivot was very close to the overall median with high probability~--
$\frac Jn$ here has significant variance.
We will thus have to compute $\E[\big]{\tfrac{J} n \lg(\tfrac{J}n)}$
more precisely and also solve the recurrence for $c'(n)$ precisely.
As a consequence, 
we need additional techniques over what we used in the previous section;
these are established below.
In terms of the result,
more details of the algorithm have significant influence on the overall cost,
in particular $\alpha$ and the choice which subproblem is sorted recursively
will influence the linear term of costs.

\subsection{Approximation by beta integrals}
\label{sec:approx-by-beta-integrals}

In this section, we compute certain expectations that arise, \eg, 
in the toll function of our recurrence.
The idea is to approximate $\frac Jn$ by a 
beta distributed variable, relying on the local limit law \wref{lem:beta-binomial-convergence-to-beta}.
the conditionals translate to bounds of an integral. 
Carefully tracing the error of this approximation yields the following result.

\begin{lemma}[Beta-integral approximation]
\label{lem:E-Jn-ln-Jn}
	Let $J \eqdist \betaBinomial(n-c_1,\betL,\betR) + c_2$ be a random variable
	that differs by fixed constants $c_1$ and $c_2$ from a beta-binomial variable 
	with parameters $n\in \N$ and $\betL, \betR\in\N_{\ge1}$.
	
	Then for any $\holdA\in(0,1)$ holds
	\[
			\E*{\tfrac Jn \ln \tfrac Jn} 
		\wwrel= 
			\frac{\betL}{\betL+\betR} (\harm{\betL}-\harm{\betL+\betR}) 
			\bin\pm \Oh(n^{-\holdA})
			,\qquad(n\to\infty)
		.
	\]
\end{lemma}

\begin{proof}[\wref{lem:E-Jn-ln-Jn}]
	By the local limit law for beta binomials (\wref{lem:beta-binomial-convergence-to-beta}) 
	it is plausible to expect a reasonably small error when
	we replace $\E[\big]{J \lg J}$ by
	$\E[\big]{ (P n) \lg (P n) }$
	where $P\eqdist \betadist(\betL,\betR)$ is beta distributed.
	We bound the error in the following.
	
	We first replace $J$ by $I\eqdist\betaBinomial(n,\betL,\betR)$ and argue later that this 
	results in a sufficiently small error.
	\begin{align*}
			\E[\big]{\tfrac In \ln \bigl(\tfrac In \bigr)}
		&\wwrel=
			\sum_{i = 0} ^ {n}
				\tfrac in \ln \bigl(\tfrac in \bigr) \cdot \Prob{I = i}
	\\	&\wwrel=
			\frac 1n \sum_{i = 0} ^ {n}
				\tfrac in \ln \bigl(\tfrac in \bigr) \cdot n \Prob{I = i}
	\\	&\wwrel{\relwithref[r]{lem:beta-binomial-convergence-to-beta}=}
			\frac 1n \sum_{i = 0} ^ {n}
				\tfrac in \ln \tfrac in \cdot 
				\biggl(\frac{(i/n)^{\betL-1}(1-(i/n))^{\betR-1}}{\BetaFun(\betL,\betR)} \bin\pm \Oh(n^{-1})\biggr)
	\\	&\wwrel=
			- \frac1 {\BetaFun(\betL,\betR)} \cdot \frac 1n \sum_{i = 0} ^ {n}
				f(i/n) \wwbin\pm \Oh(n^{-1}),
	\end{align*}
	where 
	$f(z) = \ln (1/z) \cdot z^{\betL} (1-z)^{\betR-1}$.
	Since the derivative is $\infty$ for $z=0$, $f$ cannot be Lipschitz-continuous,
	but it is Hölder-continuous
	on $[0,1]$ for any exponent $\holdA \in (0,1)$. 
	This is because $z\mapsto z\ln(1/z)$ is Hölder-continuous (\wref{lem:x-log-x-hölder}),
	products of Hölder-continuous function remain so on bounded intervals
	and the remaining factor of $f$ is a polynomial in $z$, which is Lipschitz- and
	hence Hölder-continuous.
	By \wref{lem:hölder-intergral-bound} we then have
	\begin{align*}
			\frac 1n \sum_{i = 0} ^ {n}
						f(i/n)
		&\wwrel=
			\int_0^1 f(z) \,dz  \wbin\pm \Oh(n^{-\holdA}).
	\end{align*}
	Recall that we can choose $\holdA$ as close to $1$ as we wish; this will
	only affect the constant inside $\Oh(n^{-\holdA})$.
	
	Changing from $I$ back to $J$ has no influence on the given approximation:
	To compensate for the difference in the number of trials ($n-c_1$ instead of $n$),
	we use the above formulas for with $n-c_1$ instead of $n$; since we let $n$ go to
	infinity anyway, this does not change the result.
	Moreover, replacing $I$ by $I+c_2$ changes the value of the argument $z=I/n$
	of $f$ by $\Oh(n^{-1})$; 
	since $z \mapsto z \ln(1/z)$ is smooth, namely Hölder-continuous,
	this also changes $z \ln(1/z)$ by at most $\Oh(n^{-\holdA})$.
	
	It remains to evaluate the beta integral; it is given in \weqref{eq:E-h-X}.
	Inserting, we find
	\begin{align*}
			\E{\tfrac Jn \ln \tfrac Jn}
		&\wwrel=
			\E{\tfrac In \ln \tfrac In} \wbin\pm\Oh(n^{-\holdA})
	\\	&\wwrel=
			\frac{\betL}{\betL+\betR} \bigl( \harm{\betL} - \harm{\betL+\betR} \bigr)
				\wbin\pm\Oh(n^{-\holdA})
	\end{align*}
	for any $\holdA \in (0,1)$.
\end{proof}

\begin{remark}[Generalization of beta-integral approximation]
	The technique above directly extends to $\E{g(\frac Jn)}$ for any Hölder-continuous function $g$.
	For computing the variance in \wref{sec:variance}, we will have to deal 
	with more complicated functions including the indicator variables $A_1(J)$ resp.\
	$A_2(J)$.
	As long as $g$ is \textit{piecewise} Hölder-continuous, the same arguments and error bounds
	apply: 
	We can break the sums resp.\ integrals into several parts and apply the above
	approximation to each individually.
	The indicator variables simply translate into restricted bounds of the integral.
	For example, we obtain for constants $0\le x\le y\le 1$ that
	\begin{align*}
			\E[\big]{[x n\le J\le y n] \cdot J \lg J}
		&\wwrel=
			\frac{\betL}{\betL+\betR} \, I_{x,y}(\betL+1,\betR) \cdot n\lg n \wbin\pm\Oh(n)
			,\qquad(n\to\infty).
	\end{align*}
\end{remark}

\subsection{The toll function}
\label{sec:fixed-k-toll-function}

Building on the preparatory work from \wref{lem:E-Jn-ln-Jn}, 
we can easily determine an asymptotic approximation for the toll function.
We find
\begin{align*}
		t'(n)
	&\wwrel=
		\biggl(1 + 2a \E*{\tfrac{J} n \lg(\tfrac{J}n)} \biggr) n
		\wbin\pm \Oh(n^{1-\epsilon})
\\	&\wwrel=
		\biggl(1 + 2a \frac{ \E*{\tfrac{J} n \ln(\tfrac{J}n)} } {\ln 2} \biggr) n
		\wbin\pm \Oh(n^{1-\epsilon})
\\	&\wwrel{\eqwithref[r]{lem:E-Jn-ln-Jn}}
		\Biggl(1 + \frac{2a}{\ln 2} \biggl(\frac{t+1}{2(t+1)}(\harm{t+1}-\harm{2t+2}) \pm \Oh(n^{-\holdA})\biggr) \Biggr) n
		\wbin\pm \Oh(n^{1-\epsilon})
\\	&\wwrel=
		\underbrace{\biggl(1 - \frac{a\bigl(\harm{k+1}-\harm{t+1}\bigr)}{\ln 2}  \biggr)}_{\hat q} n
		\wbin\pm \Oh(n^{1-\epsilon} + n^{1-\holdA} ).
\numberthis\label{eq:t'n-asymptotic-fixed-k}
\end{align*}

\subsection{The shape function}
\label{sec:shape-function}
The expectations $\E{A_r(J_r) c'(J_r)}$ in \weqref{eq:recurrence-c'n}
(and in the same way for the original costs in \weqref{eq:recurrence-cn-E})
are finite sums over the values $0,\ldots, n-1$ that $J \ce J_1$ can attain.
Recall that $J_2 = n-1 - J_1$ and
$A_1(J_1) + A_2(J_2) = 1$ for any value of $J$.
With $J = J_1 \eqdist J_2$, we find
\begin{align*}
		\sum_{r=1}^2\E{A_r(J_r) c(J_r)}
	&\wwrel=
		{}\bin\ppe
		\E[\Bigg]{\biggl[\frac{J}{n-1} \rel\in \Bigl[\frac{\alpha}{1+\alpha},\frac12\like{\Bigr)}{\Bigr]}\cup \Bigl(\frac1{1+\alpha},1\Bigr]\biggr] \cdot c(J)}
\\*	&\wwrel\ppe{}	
		\bin+ 
		\E[\Bigg]{\biggl[\frac{J}{n-1} \rel\in \Bigl[\frac{\alpha}{1+\alpha},\frac12\Bigr)\cup \Bigl(\frac1{1+\alpha},1\Bigr]\biggr] \cdot c(J)}
%}
\\	&\wwrel=
		\sum_{j=0}^{n-1} w_{n,j} \cdot c(j)
		,\qquad \text{where}
\\[1ex]			
		w_{n,j}
	&\wwrel={}
		\phantom{{}+{}}\Prob{J=j} \cdot 
			\Bigl[\tfrac{j}{n-1} \in [\tfrac{\alpha}{1+\alpha},\tfrac12\like(]\cup (\tfrac1{1+\alpha},1]\Bigr]
\\*	&\wwrel\ppe{}
		+\Prob{J=j} \cdot 
			\Bigl[\tfrac{j}{n-1} \in [\tfrac{\alpha}{1+\alpha},\tfrac12)\cup (\tfrac1{1+\alpha},1]\Bigr]
\\[.5ex]	&\wwrel= 
		\begin{dcases}
			2 \cdot \Prob{J = j} & \text{if } \tfrac{j}{n-1} \in [\tfrac{\alpha}{1+\alpha},\tfrac12)\cup (\tfrac1{1+\alpha},1] \\
			1 \cdot \Prob{J=j} & \text{if } \tfrac{j}{n-1} = \tfrac12 \\
			0	& \text{otherwise}.
		\end{dcases}
\end{align*}
We thus have a recurrence of the form required by the Roura's \textsl{continuous master theorem (CMT)}
(see \wref{thm:CMT}) with the weights $w_{n,j}$ from above.
\wref{fig:wnj} shows a specific example for how these weights look like.

\begin{figure}[htb]
	\plaincenter{
		\externalizedpicture{shape-function-t1}
		\begin{tikzpicture}
		\begin{axis}[
			width=.7\textwidth,
			height=.43\textwidth,
			title={$n \cdot w_{n,zn}$ vs. $w(z)$ \quad \smaller($n=51$, $k=3$)},
			ymax=3.3,
			xlabel=$z$,
		]
			\addplot[mark=none,black!50,draw=none,fill=black!10] table[x=z,y=w(z)] {pics/shape-function-t1-closed.tab} \closedcycle;
			\addplot[mark=none,black!50,ultra thick] table[x=z,y=w(z)] {pics/shape-function-t1.tab} ;
			\addplot[mark=none,black!50,thick,dotted] coordinates { (1/3,0) (1/3,2.667)};
			\addplot[mark=none,black!50,thick,dotted] coordinates { (1/2,0) (1/2,3)};
			\addplot[mark=none,black!50,thick,dotted] coordinates { (2/3,0) (2/3,2.667)};
			\addplot[only marks,mark=*,mark size=1.5pt,fill=white,fill opacity=.5,thin,ycomb,draw=black!60] 
				table [x=z,y={n*wnzn}] {pics/wnzn-n51-t1.tab} ;
		\end{axis}
		\end{tikzpicture}
	}
	\caption{%
		The weights $w_{n,j}$ (circles) for $n=51$, $t=1$ and $\alpha=\frac12$ 
		and the corresponding shape function $w(z)$ (fat gray line); 
		note the singular point at $j=25$.
	}
\label{fig:wnj}
\end{figure}
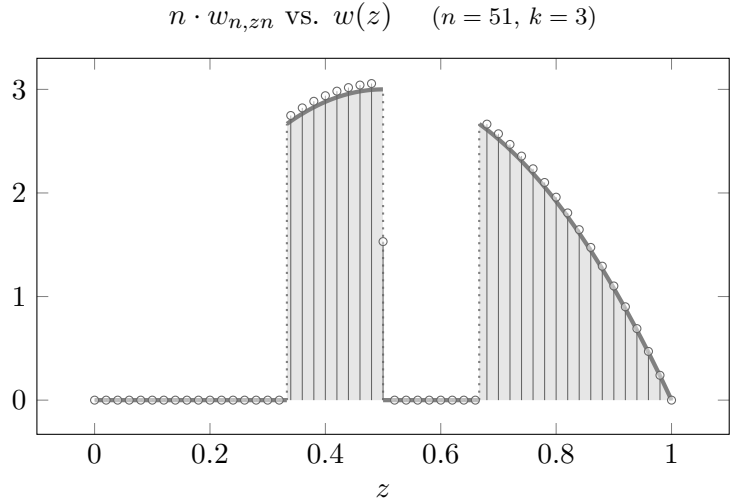

It remains to determine $\Prob{J=j}$.
Recall that we choose the pivot as the median of $k=2t+1$ elements for a fixed constant $t\in\N_0$,
and the subproblem size $J$ fulfills $J = t + I$ with $I\eqdist \betaBinomial(n-k,t+1,t+1)$.
So we have for $i\in[0,n-1-t]$ by definition
\begin{align*}
		\Prob{I=i}
	&\wwrel=
		\binom{n-k}{i} \frac{\BetaFun\bigl(i+t+1,(n-k-i)+t+1\bigr)}{\BetaFun(t+1,t+1)}
\\	&\wwrel=
		\binom{n-k}{i} \frac{ (t+1)^{\overline{i}} (t+1)^{\overline{n-k-i}} }{(k+1)^{\overline{n-k}}}
\end{align*}
The first step towards applying the CMT is to identify a shape function $w(z)$
that approximates the relative subproblem size probabilities $w(z) \approx n w_{n,\lfloor z n\rfloor}$ 
for large $n$.
Now the local limit law for beta binomials 
(\wref{lem:beta-binomial-convergence-to-beta}) says that
the normalized beta binomial $I/n$ converges to a beta variable ``in density'',
and the convergence is uniform.
With the beta density $f_P(z) = z^t(1-z)^t / \BetaFun(t+1,t+1)$, 
we thus find by \wref{lem:beta-binomial-convergence-to-beta} that
\begin{align*}
		\Prob{J=j}
	&\wwrel=
		\Prob{I=j-t}
	\wwrel=
		\frac1n f_P(j/n) \wwbin\pm \Oh(n^{-2}) 
		,\qquad (n\to\infty).
\end{align*}
The shift by the small constant $t$ from $(j-t)/n$ to $j/n$
only changes the function value by $\Oh(n^{-1})$ since $f_P$ is Lipschitz continuous on $[0,1]$
(see \wref{sec:hölder-continuity}).

With this observation, a natural candidate for the shape function of the recurrence is
\begin{align*}
\numberthis\label{eq:shape-function}
		w(z)
	&\wwrel=
		2\,\left[\tfrac{\alpha}{1+\alpha} < z < \tfrac12 \bin\vee z > \tfrac1{1+\alpha}\right] 
			\frac{z^{t}(1-z)^{t}}{\BetaFun(t+1,t+1)}.
\end{align*}

It remains to show that this is indeed a suitable shape function,
\ie, that $w(z)$ fulfills \weqref{eq:CMT-shape-function-condition},
the approximation-rate condition of the CMT.

We consider the following ranges for 
$\frac{\lfloor zn\rfloor}{n-1} = \frac j{n-1}$ separately:
\begin{itemize}
\item $\frac{\lfloor zn\rfloor}{n-1} < \frac{\alpha}{1+\alpha}$ and $\frac12<\frac{\lfloor zn\rfloor}{n-1}<\frac1{1+\alpha}$.\\
	Here $w_{n,\lfloor zn\rfloor} = 0$ and so is $w(z)$.
	So actual value and approximation are exactly the same.
\item $\frac{\alpha}{1+\alpha}<\frac{\lfloor zn\rfloor}{n-1} < \frac12$ and $\frac{\lfloor zn\rfloor}{n-1}>\frac1{1+\alpha}$.\\
	Here $w_{n,j} = 2 \Prob{J=j}$ and $w(z) = 2 f_P(z)$ where $f_P(z) = z^t(1-z)^t / \BetaFun(t+1,t+1)$
	is twice the density of the beta distribution $\betadist(t+1,t+1)$.
	Since $f_P$ is Lipschitz-continuous on the bounded interval $[0,1]$ (it is a polynomial)
	the uniform pointwise convergence from above is enough to bound the sum of
	$\bigl| w_{n,j} \bin- \! \int_{j/n}^{(j+1)/n} w(z) \: dz \bigr|$ over all $j$ in the range by
	$\Oh(n^{-1})$.
\item $\frac{\lfloor zn\rfloor}{n-1} \in \{\frac{\alpha}{1+\alpha},\frac12,\frac1{1+\alpha}\}$.\\
	At these boundary points, the difference between $w_{n,\lfloor zn\rfloor}$ and $w(z)$ does not
	vanish (in particularly $\frac12$ is a singular point for $w_{n,\lfloor zn\rfloor}$), 
	but the absolute difference is bounded.
	Since this case only concerns $3$ out of $n$ summands, the overall contribution to the error
	is $\Oh(n^{-1})$.
\end{itemize}
Together, we find that \weqref{eq:CMT-shape-function-condition} is fulfilled as claimed:
\begin{align}
	\sum_{j=0}^{n-1} \,\biggl|
		w_{n,j} \bin- \! \int_{j/n}^{(j+1)/n} \mkern-15mu w(z) \: dz
	\biggr|
	\wwrel= \Oh(n^{-1})
	\qquad(n\to\infty).
\end{align}

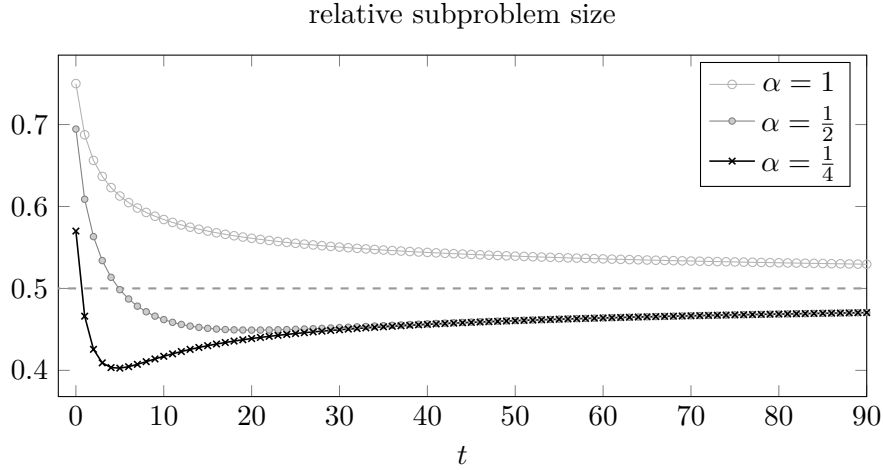
\begin{figure}[htpb]
	\plaincenter{
	\externalizedpicture{relative-subproblem-size}
		\begin{tikzpicture}
		\begin{axis}[
			width=.8\textwidth,
			height=.4\textwidth,
			title={relative subproblem size},
			xlabel=$t$,
			xmin=-2,xmax=90,
		]
			\draw[thick,dashed,black!40] (axis cs: -10,.5) -- (axis cs:130,.5) ;
			\addplot[mark=o,mark size=1.6pt,thin,draw=black!30] 
				table [x=t,y={relative-subproblem-size}] {pics/relative-subproblem-size-alpha1.tab} ;
			\addlegendentry{$\alpha=1$}
			\addplot[mark=*,mark size=1.2pt,thin,draw=black!50,mark options={fill=black!20}] 
				table [x=t,y={relative-subproblem-size}] {pics/relative-subproblem-size-alpha0.5.tab} ;
			\addlegendentry{$\alpha=\frac12$}
			\addplot[mark=x,mark size=1.75pt,semithick,draw=black] 
				table [x=t,y={relative-subproblem-size}] {pics/relative-subproblem-size-alpha0.25.tab} ;
			\addlegendentry{$\alpha=\frac14$}
		\end{axis}
		\end{tikzpicture}
	}
	\caption{%
		$\int_0^1 z w(z)\,dz$, the relative recursive subproblem size, as a function of $t$.
	}
	\label{fig:recursive-subproblem-size}
\end{figure}

\begin{remark}[Relative subproblem sizes]
The integral $\int_0^1 z w(z)\,dz$ is precisely the expected relative subproblem size for the recursive
call. 
This is of independent interest;
while it is intuitively clear that for $t\to\infty$, \ie, the case of exact medians as pivots,
we must have a relative subproblem size of exactly \smash{$\frac12$},
this convergence is not
obvious from the behavior for finite $t$: the mass of the integral $\int_0^1 z w(z)\,dz$
concentrates at $z=\frac12$, a point of discontinuity in $w(z)$.
It is also worthy of note that for, \eg, $\alpha=\frac12$, 
the expected subproblem size is initially larger than $\frac12$ 
($0.69\overline 4$ for $t=0$), then decreases to $\approx 0.449124$ around $t=20$ and then starts 
to slowly increase again
(see \wref{fig:recursive-subproblem-size}).
This effect is even more pronounced for $\alpha=\frac14$.
\end{remark}

\subsection{Which case of the CMT?}

We are now ready to apply the CMT (\wref{thm:CMT}).
Assume that $a\ne \ln 2 / (\harm{k+1}-\harm{t+1})$; the other (special) case will be addressed later.
Then by \weqref{eq:t'n-asymptotic-fixed-k} our toll function fulfills 
$t'(n) \sim \hat q n$ for 
$\hat q = \bigl(1 - {a(\harm{k+1}-\harm{t+1})}/{\ln 2}  \bigr)$.
Thus, we have $\cmtA=1$, $\cmtB=0$ and $K = \hat q\ne 0$ and we compute 
\begin{align*}
		H
	&\wwrel=
		1-\int_0^1 z\,w(z) \:dz 
\\	&\wwrel= 
		1 - \int_0^1 2\,\left[\tfrac{\alpha}{1+\alpha} < z < \tfrac12 \bin\vee z > \tfrac1{1+\alpha}\right]\, 
					\frac{z^{t+1}(1-z)^{t}}{\BetaFun(t+1,t+1)} \: dz
\\	&\wwrel= 
		1 - 2\frac{t+1}{k+1}
				\int_0^1 \left[\tfrac{\alpha}{1+\alpha} < z < \tfrac12 \bin\vee z > \tfrac1{1+\alpha}\right]\, 
				\frac{z^{t+1}(1-z)^{t}}{\BetaFun(t+2,t+1)} \: dz
\\	&\wwrel= 
		1 - \Bigl( I_{\frac{\alpha}{1+\alpha},\frac12}(t+2,t+1)
			+ I_{\frac1{1+\alpha},1}(t+2,t+1)\Bigr)
\\	&\wwrel= 
		  I_{0,\frac{\alpha}{1+\alpha}}(t+2,t+1)
		+ I_{\frac12,\frac1{1+\alpha}}(t+2,t+1)
\numberthis\label{eq:H}
\end{align*}
For any sampling parameters, we have $H>0$,
so by Case~1 of \wref{thm:CMT}, we have that
\begin{align*}%\SwapAboveDisplaySkip
		c'(n)
	&\wwrel\sim
		\frac{t'(n)}{H}
	\wwrel\sim
		\frac{\hat q n }{H}
	\wwrel=
		qn
		,\qquad (n\to\infty).
\end{align*}

\paragraph{\boldmath Special case for $a$}
If $a = \ln 2 / (\harm{k+1}-\harm{t+1})$, \ie, $\hat q = 0$,
then $t'(n) = \Oh(n^{1-\epsilon})$.
Then the claim follows from a coarser bound for $c'(n) = \Oh(n^{1-\epsilon} + \log n)$
which can be established by the same arguments as in the proof of \wref{thm:quickXsort}.

\subsection{Error bound}

Since our toll function is not given precisely, but only up to an error term
$\Oh(n^{1-\epsilon})$ for a given fixed $\epsilon\in(0,1]$,
we also have to estimate the overall influence of this term.
For that we consider the recurrence for $c(n)$ again, but replace $t(n)$ (entirely) by
$C \cdot n^{1-\epsilon}$.
If $\epsilon > 0$, $\int_0^1 z^{1-\epsilon} w(z)\, dz < \int_0^1 w(z)\, dz = 1$,
so we still find $H>0$ and apply case~1 of the CMT\@. The overall contribution of the 
error term is then $\Oh(n^{1-\epsilon})$.
For $\epsilon=1$, we have $H=0$ and case~2 applies, giving an overall error term of $\Oh(\log n)$.

\medskip\noindent
This completes the proof of \wref{thm:cn}.
\end{proof}

\section{Analysis of QuickMergesort and QuickHeapsort}
\label{sec:analysis-for-concrete-X}

We have analyzed the expected cost of the \QuickXsort scheme in great detail.
Next, we apply our transfer theorems to the concrete choices for X
discussed in \wref{sec:quickxsort}.
Besides describing how to overcome technical complications in the analysis, 
we also discuss our results.
Comparing with analyses and measured comparison counts from previous work, we find
that our exact solutions for the \QuickXsort recurrence yield more accurate
predictions for the overall number of comparisons.

\subsection{QuickMergesort}
\label{sec:analysis-quickmergesort}

We use \QuickMergesort here to mean the ``ping-pong'' variant with smaller buffer ($\alpha=\frac12$)
as illustrated in \wpref{fig:merge-ping-pong-alpha-0.5}.
Among the variations of \Mergesort (that are all usable in \QuickXsort)
we discussed in \wref{sec:quickmergesort},
this is the most promising option in terms of practical performance.
The analysis of the other variants is very similar.

We assume a variant of \Mergesort that generates an optimally balanced merges. 
Top-down mergesort is the typical choice for that, but there are also
variations of bottom-up mergesort that achieve the same result without using logarithmic
extra space for a recursion stack~\cite{GolinSedgewick1993}.

\begin{corollary} [Average Case \algorithmname{QuickMergesort}]
\label{cor:quickmergesort-average-case}
	The following results hold for the expected number of comparisons
	when sorting a random permutation of $n$ elements.
\begin{thmenumerate}{cor:quickmergesort-average-case}
\item
\label{cor:QMS}
	Median-of-$\sqrt n$ \QuickMergesort is an internal sorting algorithm 
	that performs $n \lg n - (1.25265 \pm 0.01185)n \pm \Oh(n^{1/2+\epsilon})$
	comparisons on average for any constant $\epsilon>0$.
\item
\label{cor:constQMS}
	Median-of-3 \QuickMergesort (with $\alpha = 1/2$) is an internal sorting algorithm 
	that performs $n \lg n - (0.84765 \pm 0.01185)n \pm \Oh(\log n)$ comparisons on average.
\end{thmenumerate}
\end{corollary}

\begin{proof}
	We first note that \Mergesort does never compare buffer elements to each other:
	The buffer contents are only accessed in swap operations.
	Therefore, \QuickMergesort preserves randomness: if the original input is a random permutation,
	both the calls to \Mergesort and the recursive call operate on a random permutation
	of the respective elements.
	The recurrence for $c(n)$ thus gives the exact expected costs 
	of \QuickMergesort when we insert for $x(n)$ 
	the expected number of comparisons used by \Mergesort on a random permutation of $n$ elements.
	The latter is given in \wpeqref{eq:xtd}.
	
	Note that these asymptotic approximations in \weqref{eq:xtd} are \emph{not} of 
	the form required for our transfer theorems;
	we need a \emph{constant} coefficient in the linear term.
	But since $c(n)$ is a monotonically increasing function in $x(n)$, 
	we can use upper and lower bounds on $x(n)$ to derive
	upper and lower bounds on $c(n)$.
	We thus apply \wref{thm:quickXsort} and \wref{thm:cn} separately with $x(n)$ replaced by
	\begin{align*}
			\underline x(n)
		&\wwrel=
			n \lg n - 1.2645 n - \Oh(1)
			\qquad\text{resp.}
	\\[.5ex]
			\overline x(n)
		&\wwrel=
			n \lg n - 1.2408 n + \Oh(1).
	\end{align*}
	For part (a), we find 
	$\underline x(n) \pm \Oh(n^{1/2+\epsilon}) 
	\le c(n) \le 
	\overline x(n) \pm \Oh(n^{1/2+\epsilon})$ for any fixed $\epsilon>0$.
	Comparing upper and lower bound yields the claim.
	
	For part (b) we obtain with $q = 0.4050$ the bounds
	$\underline x(n) +qn \pm \Oh(\log n) 
	\le c(n) \le 
	\overline x(n) +qn \pm \Oh(\log n)$.
\end{proof}

\begin{remark}[Randomization vs average case]
	We can also prove a bound for the \emph{expected} performance on \emph{any} input, 
	where the expectation is taken over the random choices for pivot sampling.
	By using an upper bound for the worst case of \Mergesort,
	$\overline x(n) = n \lg n - 0.91392n + 1$, we find
	that the expected number of comparisons is at most
	$n \lg n - 0.91392n \pm \Oh(n^{1/2+\epsilon})$ for median-of-$\sqrt n$ \QuickMergesort
	and at most $n \lg n - 0.50892n + \Oh(\log n)$ for median-of-3 \QuickMergesort.
\end{remark}

\begin{figure}[htbp]
	\plaincenter{\small
	\includegraphics{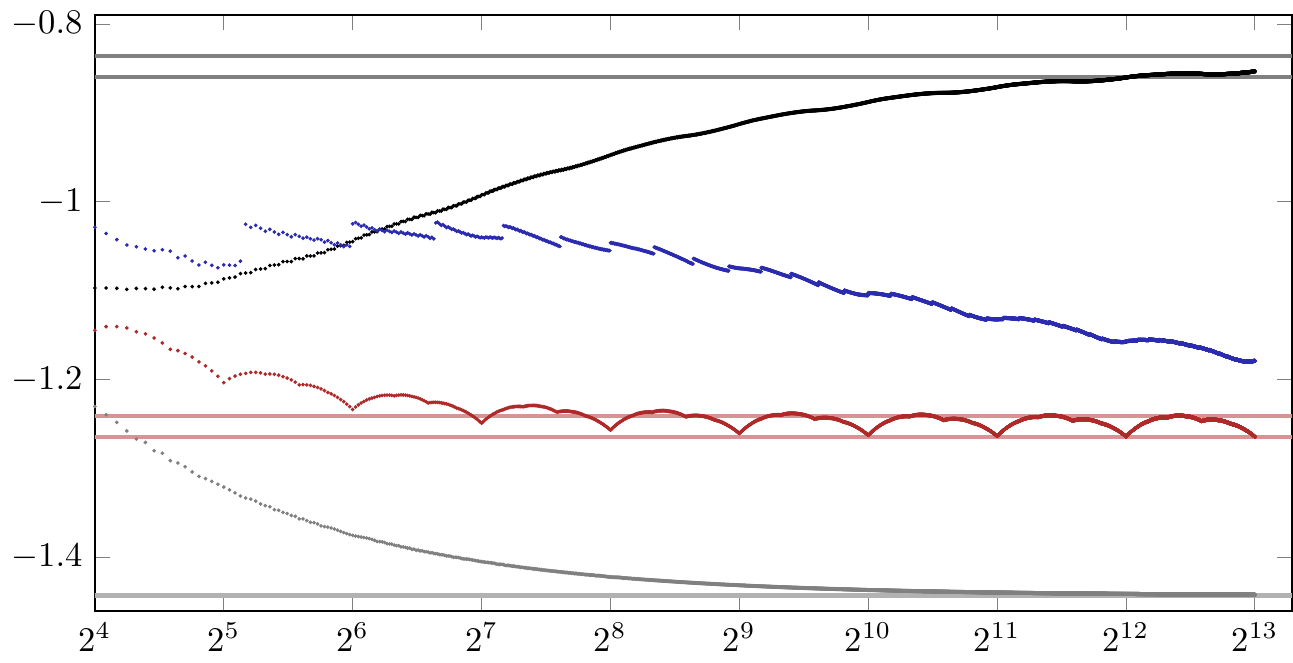}
	}
	\caption{%
		Exact comparison count of  \Mergesort (red), median-of-3 \QuickMergesort (black)
		and median-of-$\sqrt n$ \QuickMergesort (blue)
		for small input sizes, computed from the recurrence.
		The information-theoretic lower bound (for the average case) is also shown (gray).
		The $x$-axis shows $n$ (logarithmic), the $y$-axis shows $\bigl(c(n) - n \lg n\bigr) / n$.
		The horizontal lines are the supremum and infimum of the asymptotic periodic terms.
	}
	\label{fig:quickmergesort-expected-comparisons-small-n}
\end{figure}

Given that the error term of our approximation for fixed $k$ is only of logarithmic growth,
we can expect very good predictive quality for our asymptotic approximation.
This is confirmed by numbers reported on in \wref{sec:comparison-experiments} below.
If we consider the relative error between the exact value of $c(n)$ and the approximation
$n\lg n - 0.84765n$, then for $n\ge 400$, we have less than 1\% error.

\wref{fig:quickmergesort-expected-comparisons-small-n} gives a closer look for small $n$.
The numbers are computed from the exact recurrences for \Mergesort (see \wref{sec:variance-mergesort}) 
and \QuickMergesort (\weqref{eq:recurrence-cn-E})
by recursively tabulating $c(n)$ for all $n\le 2^{13} = 8192$.
For the pivot sampling costs $s(k)$, we use the average cost of finding the median with
Quickselect, which are known precisely~\cite[p.\,14]{Knuth2000}.
For the numbers for median-of-$\sqrt n$ \QuickMergesort,
we use $k(n) = 2 \lfloor \sqrt{n} /2 \rfloor + 1$.
The computations were done using Mathematica.

For standard \Mergesort, the linear coefficient reaches its asymptotic regime
rather quickly; this is due to the absence of a logarithmic term.
For median-of-3 \QuickMergesort, considerably larger inputs are needed, but for
$n\ge 2000$ we are again close to the asymptotic regime.
Median-of-$\sqrt n$ \QuickMergesort needs substantially larger inputs
than considered here to come close to \Mergesort.
It is interesting to note that for roughly $n\le 100$, the median-of-3 variant
is better, but from then onwards, the median-of-$\sqrt n$ version uses fewer comparisons.

\wref{fig:quickmergesort-expected-comparisons-small-n} 
shows the well-known periodic behavior for \Mergesort.
Oscillations are clearly visible also for \QuickMergesort, 
but compared to the rather sharp ``bumps'' in \Mergesort's
cost, \QuickMergesort's costs are smoothed out.
\wref{fig:quickmergesort-expected-comparisons-small-n}
also confirms that the amplitude of the periodic term 
is very small in \QuickMergesort.

\subsection{QuickHeapsort}
\label{sec:analysis-quickheapsort}

By \algorithmname{QuickHeapsort} we refer to \QuickXsort using 
the basic \algorithmname{ExternalHeapsort} version (as described in \wref{sec:quickheapsort}) as X.
We 

\begin{corollary} [Expected Case \algorithmname{QuickHeapsort}]
\label{cor:quickheapsort-expected-case}
	The following results hold for the expected number of comparisons
	where the expectation is taken over the random choices of the pivots.
\begin{thmenumerate}{cor:quickheapsort-expected-case}
\item
	Median-of-$\sqrt n$ \QuickHeapsort is an internal sorting algorithm 
	that performs $n \lg n + (0.54305 \pm 0.54305) n \pm \Oh(n^{1/2+\epsilon})$
%	that performs at most $n \lg n + 1.0861 n \pm \Oh(n^{1/2+\epsilon})$
%	and at least $n \lg n \pm \Oh(n^{1/2+\epsilon})$
	comparisons for any constant $\epsilon>0$.
\item
	Median-of-3 \QuickHeapsort is an internal sorting algorithm 
	that performs $n \lg n + (1.05005 \pm 0.54305) n \pm \Oh(n^{\epsilon})$
%	that performs at most $n \lg n + 1.5931 n \pm \Oh(n^{\epsilon})$
%	and at least $n \lg n + 0.5070 n \pm \Oh(n^{\epsilon})$
	comparisons for any constant $\epsilon>0$.
\end{thmenumerate}
\end{corollary}

\begin{proof}
\ExternalHeapsort
always traverses one path in the heap from root to bottom and does one comparison for each edge followed,
\ie, $\lfloor \lg n \rfloor$ or $\lfloor \lg n \rfloor-1$ many per delete-max operation.
By counting how many leaves we have on each level
one can show that we need
\[
n \bigl(\lfloor \lg n\rfloor -1\bigr) + 2 \bigl(n-2^{\lfloor \lg n\rfloor}\bigr) \bin\pm\Oh(\log n) 
\wwrel\le 
n\lg n -0.913929n \bin\pm\Oh(\log n)
\]
comparisons for the sort-down phase (both in the best and worst case)~\cite[Eq.\ 1]{DiekertWeiss2016}.
The constant of the given linear term is $1-\frac1{\ln2}-\lg(2\ln 2)$, the supremum of the
periodic function at the linear term. 
Using the classical heap construction method adds between $n-1$ and $2n$ comparisons
and $1.8813726n$ comparisons on average~\cite{Doberkat1984}.
We therefore find the following upper bounds for the average and worst case 
and lower bound for the best case
of \ExternalHeapsort:
\begin{align*}
		x_{\mathrm{ac}}(n)
	&\wwrel=
		n\lg n + 0.967444 n \wwbin\pm \Oh(n^\epsilon) 
\\
		x_{\mathrm{wc}}(n)
	&\wwrel=
		n\lg n + 1.086071 n \wwbin\pm \Oh(n^\epsilon) 
\\
		x_{\mathrm{bc}}(n)
	&\wwrel=
		n\lg n \wwbin\pm \Oh(n^\epsilon) 
\end{align*}
for any $\epsilon>0$. 

Notice that every deleteMax operation performs comparisons until the element inserted at the top of the heap (replacing the maximum) reaches the bottom of the heap.
That means when the heap is already quite empty, some of those comparisons are between
two buffer elements and these buffer elements are exchanged according to the outcome of
the comparison. Therefore, \ExternalHeapsort does \emph{not} preserve 
the randomness of the buffer elements.
Our recurrence, \weqref{eq:recurrence-cn-E}, is thus not valid for \QuickHeapsort directly.

We can, however, study a hypothetical method X that always uses $x(n) = x_{\mathrm{wc}}(n)$
comparisons on an input of size $n$, and consider the costs $c(n)$ of \QuickXsort for this method.
This is clearly an upper bound for the cost of \QuickHeapsort since $c(n)$
is a monotonically increasing function in $x(n)$.
Similarly, using $x(n) = x_{\mathrm{bc}}(n)$ yields a lower bound.
The results then follow by applying \wref{thm:quickXsort} and \wref{thm:cn}.
\end{proof}

We note that our transfer theorems are only applicable to worst resp.\ best case bounds
for \ExternalHeapsort, but nevertheless, 
using the average case $x_{\mathrm{ac}}(n)$ still might give us a better (heuristic) 
approximation of the actual numbers. 

\begin{table}[htpb]
	\plaincenter{
		\smaller
		\def\na{\multicolumn1c{---}}
		\begin{tabular}{ r *{5}{ >{$} r <{$} } }
			\toprule
			\multicolumn1c{\textbf{Instance}}                            & \multicolumn1c{\textbf{observed}} & \multicolumn1c{\textbf{estimate}} & \multicolumn1c{\textbf{upper bound}} & \multicolumn1c{\textbf{CC}} & \multicolumn1c{\textbf{DW}} \\
			\midrule
Fig.\,4~\cite{CantoneCincotti2002}, $n=10^2$, $k=1$ & 806                      & +67             & +79                & +158               & +156               \\
Fig.\,4~\cite{CantoneCincotti2002}, $n=10^2$, $k=3$ & 714                      & +98             & +110                & \na                & +168               \\
Fig.\,4~\cite{CantoneCincotti2002}, $n=10^5$, $k=1$ & 1\,869\,769              & -600            & +11\,263               & +90\,795           & +88\,795           \\
Fig.\,4~\cite{CantoneCincotti2002}, $n=10^5$, $k=3$ & 1\,799\,240              & +9\,165         & +21\,028           & \na                & +79\,324           \\
Fig.\,4~\cite{CantoneCincotti2002}, $n=10^6$, $k=1$ & 21\,891\,874             & +121\,748       & +240\,375          & +1\,035\,695       & +1\,015\,695       \\
Fig.\,4~\cite{CantoneCincotti2002}, $n=10^6$, $k=3$ & 21\,355\,988             & +49\,994        & +168\,621           & \na                & +751\,581          \\[1ex]
Tab.\,2~\cite{DiekertWeiss2016}, $n=10^4$, $k=1$    & 152\,573                 & +1\,125         & +2\,311            & +10\,264           & +10\,064           \\
Tab.\,2~\cite{DiekertWeiss2016}, $n=10^4$, $k=3$    & 146\,485                 & +1\,136         & +2\,322            & \na                & +8\,152            \\
Tab.\,2~\cite{DiekertWeiss2016}, $n=10^6$, $k=1$    & 21\,975\,912             & +37\,710        & +156\,337           & +951\,657          & +931\,657          \\
Tab.\,2~\cite{DiekertWeiss2016}, $n=10^6$, $k=3$    & 21\,327\,478             & +78\,504        & +197\,131           & \na                & +780\,091          \\
			\bottomrule
		\end{tabular}%
	}
	\caption{%
		Comparison of estimates from this paper where we use the average for \algorithmname{ExternalHeapsort} (estimate) and where we use the worst case for \algorithmname{ExternalHeapsort} (upper bound),
		Theorem~6 of~\cite{CantoneCincotti2002} (CC) and Theorem~1 of~\cite{DiekertWeiss2016} (DW);
		shown is the difference between the estimate and the observed average.
	}
	\label{tab:cmps-qhs}
	\vspace{-2ex}
\end{table}

\paragraph{Comparison with previously reported comparison counts}
Both~\cite{CantoneCincotti2002} and~\cite{DiekertWeiss2016} report averaged comparison counts
from running time experiments. We compare them in \wref{tab:cmps-qhs}
against the estimates from our results and previous analyses. 
We compare both proven upper bound from above
and the heuristic estimate using \ExternalHeapsort's average case. 

While the approximation is not very accurate for $n=100$ (for all analyses),
for larger $n$, our estimate is correct up to the first three digits, whereas 
previous upper bounds have almost one order of magnitude bigger errors. 
Our provable upper bound is somewhere in between.
Note that we can expect even our estimate to be still on the conservative side 
because we used the supremum of the periodic linear term for \ExternalHeapsort.

\section{Variance of QuickXsort}
\label{sec:variance}

If an algorithm's cost regularly exceeds its expectation by far,
good expected performance is not enough.
In this section, we approximate the \emph{variance} of the number of comparisons 
in \QuickXsort under certain restrictions.
Similar to the expected costs, 
we prove a general transfer theorem for the variance.
We then review results on the variance of the number of comparisons in \Mergesort and 
\algorithmname{ExternalHeapsort}, the two main methods of interest for \QuickXsort,
and discuss the application of the transfer theorem.

\subsection{Transfer theorem for variance}

The purpose of this section is to explore what influence the distribution
of the costs of X have on \QuickXsort.
We assume a constant sample size $k$ in this section.
Formally, our result is the following.

\begin{theorem}[Variance of QuickXsort]
\label{thm:transfer-variance}
	Assume X is a sorting method whose comparison cost have
	expectation
	$x(n) = a n\lg n + bn \pm \Oh(n^{1-\epsilon})$ and variance
	$v_X(n) = a_v n^2 + \Oh(n^{2-\epsilon})$ for a constant $a_v$ and $\epsilon>0$; 
	the case $a_v = 0$ is allowed.
	Moreover, let \QuickXsort preserve randomness.
	
	Assuming the technical conjecture $t_v(n) = \Oh(n^2)$ (see below),
	median-of-$k$ \QuickXsort is a sorting method whose
	comparison cost has variance $v(n) \sim c n^2$ for
	an explicitly computable constant $c$ that depends only on $k$, $\alpha$
	and $a_v$.
\end{theorem}

\begin{remark}
	We could confirm the conjecture mentioned above for all tried combinations
	of values for $\alpha$ and $k$, but were not able to prove it in the general setting,
	so we have to formally keep it as a prerequisite.
	We have no reason to believe it is not always fulfilled.
\end{remark}

\begin{proof}[\wref{thm:transfer-variance}]
This transfer theorem can be proven with similar techniques
as for the expected value, but the computations become lengthier.

\paragraph{Distributional recurrence}

We can precisely characterize the distribution
of the random number of comparisons, $C_n$, that we need to sort an
input of size $n$.
We will generally denote the random variables by capital letter $C_n$
and their expectations by lowercase letters $c(n)$.
We describe the distribution of $C_n$ in the form of a \emph{distributional recurrence,} \ie,
a recursive description of the distribution of the family of random variables $(C_n)_{n\in\N}$.
From these, we can mechanically derive recurrence equations for the moments
of the distribution and in particular for the variance.
We have
\begin{align*}
		C_n
	&\wwrel\eqdist
		\underbrace{ n-k + s(k) \bin+ A_1 \cdot X_{J_2} + A_2 \cdot \tilde X_{J_1} }_ {T_n}{}
		+ A_1\cdot C_{J_1} + A_2\cdot \tilde C_{J_2}
		,\qquad(n>w)
\numberthis\label{eq:Cn-distributional-rec}
\end{align*}
for $(X_n)_{n\in \N}$ the family of (random) comparisons
to sort a random permutation of $n$ elements with X.
$(\tilde C_n)_{n\in\N}$ and $(\tilde X_n)_{n\in\N}$ are independent copies of 
$(C_n)_{n\in\N}$ and $(X_n)_{n\in\N}$, respectively,
and these are also independent of $(J_1,J_2)$;
we will in the following omit the tildes for legibility; 
we implicitly define all terms in an equation from the same family 
as each coming from its own independent copy.
Base cases for small $n$ are given by the recursion-stopper method and
are immaterial for the asymptotic regime (for constant $w$).

\paragraph{Recurrence for the second moment}

We start with the elementary equation $\Var{C_n} = \E{C_n^2} - \E{C_n}^2$.
Of course, $\E{C_n}^2 = c^2(n)$, which we already know by \wref{thm:cn}.
From the distributional recurrence, we can compute the second moment
$m_2(n) = \E{C_n^2}$ as follows:
Square of both sides in \weqref{eq:Cn-distributional-rec},
and then take expectations; that leaves $m_2(n)$ on the left-hand side.
To simplify the right-hand side, we use the law of total expectation
to first take expectations conditional on $J_1$ (which also fixes $J_2 = n-1-J_1$)
and then take expectations over $J_1$.
We find
\begin{align*}
		\E{C_n^2 \given J_1}
	&\wwrel=
		\E*{\Bigl(\textstyle T_n + \sum_{r=1}^2 A_r C_{J_r}\Bigr)^{\!2} \given J_1}
\\	&\wwrel=
		\E[\big]{T_n^2 \given J_1} 
		+ \sum_{r=1}^2 \E[\Big]{\underbrace{A_r^2}_{=A_r} C_{J_r}^2 \given J_1}
\\*[-1ex]	&\wwrel\ppe{}
		+ \E[\Big]{2 \underbrace{A_1 A_2}_{=0} C_{J_1} C_{J_2} \given J_1}
		+ \sum_{r=1}^2 \E[\big]{2 T_n \cdot A_r C_{J_r} \given J_1}
\intertext{%
	since $A_1$ and $A_2$ are fully determined by $J_1$, 
	and since $T_n$ and $C_{J_r}$ are \emph{conditionally independent} given $J_1$, this is
}
	&\wwrel=
		\E[\big]{T_n^2 \given J_1} 
		+ \sum_{r=1}^2 A_r \, m_2(J_r)
		+ 2\E{T_n \given J_1} \sum_{r=1}^2 A_r \, c(J_r).
\end{align*}
We now take expected values also \wrt $J_1$ and exploit symmetries $J_1\eqdist J_2$. 
We will write $A \ce A_1$ and $J \ce J_1$;
we find
\begin{align*}
		m_2(n)
	&\wwrel=
		2 \E{ A \, m_2(J) }
		\bin+ \underbrace{
			\E[\big]{T_n^2} 
			+ 2\sum_{r=1}^2  \Eover[\Big] J { A_r\, \E{T_n \given J_1}  \, c(J_r) }
		} _ {t_{m_2}(n)}.
\end{align*}
To continue, we have to unfold $t_{m_2}(n)$ a bit more.
We start with the simplest one, the conditional expectation of $T_n$.
For constant $k$, we find
\begin{align*}
		\E{T_n \given J}
	&\wwrel=
		\E*{ \textstyle n \pm \Oh(1) + \sum_{r=1}^2 (1-A_r) X_{J_r} \given J}
\\	&\wwrel=
		n + \sum_{r=1}^2 (1-A_r) \E*{ X_{J_r} \given J}
		\wbin\pm\Oh(1)
\\	&\wwrel=
		n + \sum_{r=1}^2 (1-A_r) x(J_r)
		\wwbin\pm\Oh(1).
\end{align*}
So we find for the last term in the equation for $m_2(n)$
\begin{align*}
		&
		2\sum_{r=1}^2  \Eover[\Big] J { A_r\, \E{T_n \given J_1}  \, c(J_r) }
\\	&\wwrel=
		2\sum_{r=1}^2 \Eover[\bigg] J {  A_r \Bigl( \textstyle n + \sum_{\ell=1}^2 (1-A_\ell) x(J_\ell)
				\wwbin\pm\Oh(1) \Bigr) \, c(J_r) }
\\	&\wwrel=
		2n\sum_{r=1}^2 \E[\big]{A_r \, c(J_r)}
		+2\sum_{r=1}^2 \E[\big] {  A_r \, x(J_{3-r}) \, c(J_r) }
		\wwbin\pm\Oh(n \log n)
\\	&\wwrel=
		4n \E[\big]{A \, c(J)}
		+4 \E[\big] {  A \, c(J) \, x(n-1-J) }
		\wwbin\pm\Oh(n \log n).
\end{align*}
It remains to we compute the second moment of $T_n$:
\begin{align*}
		\E{T_n^2}
	&\wwrel=
		\E*{ \Bigl(\textstyle n\bigl(1\pm\Oh(n^{-1})\bigr) + \sum_{r=1}^2 (1-A_r) X_{J_r}  \Bigr)^{\!2}\,}
\\	&\wwrel=
		\sum_{r=1}^2\Eover[\Big] {J_r}{(1-A_r) \E{  X^2_{J_r} \given J_r}}
		+ n^2\bigl(1\pm\Oh(n^{-1})\bigr)
\\*	&\wwrel\ppe{}
		+ 2n\bigl(1\pm\Oh(n^{-1})\bigr) 2\E{(1-A) x(J)},
\intertext{
	denoting $\Var{X_n}$ by $v_X(n) = \Theta(n)$ and using $\E{X^2} = \E{X}^2 + \Var X$
}
	&\wwrel=
		2\E[\Big] {(1-A) \bigl(x^2(J) + v_X(J)\bigr)}
		+ n^2 
		\wbin\pm \Oh(n \log n)
\\	&\wwrel=
		2\E[\Big] {(1-A) x^2(J)}
		+ 2a_v \E{(1-A) J^2}
		+ 4n\E{(1-A) x(J)}
		+ n^2 
		\wbin\pm \Oh(n^{2-\epsilon}).
\end{align*}
We can see here that the variance of X only influences lower order terms
of the variance of \QuickXsort
when $v_X(n) = o(n^2)$.

\paragraph{Recurrence for the variance}
We now have all ingredients together to compute an asymptotic solution of
the recurrence for $m_2(n)$, the second moment of costs for 
\QuickMergesort. However, it is more economical to first subtract $c^2(n)$
on the level of recurrences, since many terms will cancel.
We thus now derive from the above results a direct recurrence for $v(n) = \Var{C_n}$.
\begin{align*}
		v(n)
	&\wwrel=
		m_2(n) - c^2(n)
\\	&\wwrel=
		2 \E{ A \, v(J) } 
		\bin+ \underbrace{
			2 \E{ A \, c^2(J) }
			- c^2(n) + t_{m_2}(n)
		} _ {t_v(n)}.
\numberthis\label{eq:variance-recurrence}
\end{align*}
For brevity, we write $\overline J = n-1-J$.
We compute using $c(n) = x(n) + q n \pm \Oh(n^\delta)$ for a $\delta<1$
\begin{align*}
		t_v(n)
	&\wwrel=
		2\E*{ A\, \bigl( x(J) + q J \pm \Oh(n^{1-\epsilon}) \bigr)^2 }
		\bin-\bigl(x(n) + q n \pm \Oh(n^{1-\epsilon})\bigr)^2
\\*	&\wwrel\ppe{}
		\bin+4n \E* {A \, \bigl( x(J) + q J \pm \Oh(n^{1-\epsilon}) \bigr)}
		+4 \E* {  A \, \bigl( x(J) + q J \pm \Oh(n^{1-\epsilon}) \bigr) \, x(\overline J) }
\\*	&\wwrel\ppe{}
		\bin+ 2\E[\big]{(1-A) x^2(J)}
		+ 2a_v\E{(1-A) J^2}
		+4n \E[\big] {  (1-A) \, x(J) }
		+n^2
		\wbin\pm\Oh(n^{2-\epsilon})
\\	&\wwrel=
		2\E[\big]{ A\, x^2(J)} + 4q\E[\big]{A J x(J) } + 2q^2 \E[\big]{A J^2}
		\bin- x^2(n) -2q x(n) n  - q^2 n^2
\\*	&\wwrel\ppe{}
		+ 4n \E* {A x(J) } +4qn\E{A J}
		\bin+ 4 \E* { A  x(J) x(\overline J) } + 4q \E* { A  J x(\overline J) }
\\*	&\wwrel\ppe{}
		\bin+ 2\E[\big] {(1-A) x^2(J)}
		+ 2a_v \E{(1-A) J^2}
		+4n \E[\big] {  (1-A) \, x(J) }
		+n^2
\\*	&\wwrel\ppe{}
		\bin\pm\Oh(n^{2-\epsilon} \log n)
\\	&\wwrel=
		2\E[\big]{x^2(J)} 
		+ 4 \E* { A  x(J) x(\overline J) } 
		- x^2(n) 
\\*	&\wwrel\ppe{}\bin
		+ 4n \E* {x(J) } 
		+ 4q\E[\big]{A J \bigl(x(J)+x(\overline J)\bigr) } 
		-2q x(n) n  
\\*	&\wwrel\ppe{}\bin
		+ n^2
		+ 2q^2 \E[\big]{A J^2}
		+ 2a_v \E{(1-A) J^2}
		+4qn\E{A J}
		- q^2 n^2
		\wwbin\pm\Oh(n^{2-\epsilon} \log n).
		\label{eq:variance-toll-mathematica}
\end{align*}
At this point, the only route to make progress seems to be to expand all occurrences of $x$
into $x(n) = a n \lg n + bn + \Oh(n^{1-\epsilon})$ and compute the expectations.
For that, we use the approximation by incomplete beta integrals that we introduced
in \wref{sec:approx-by-beta-integrals} to compute the expectations of the form $\E{g(J)}$,
where $g$ only depends on $J$.
Writing $z = \frac Jn$ and $\overline z = 1-z = \frac{\overline J}n$, we can expand
all occurring functions $g$ as follows:
\begin{align*}
		J^2 \lg^2(J)
	&\wwrel=
		z^2 \cdot n^2 \lg^2 n + 2z^2 \lg z\cdot n^2 \lg n + z^2 \lg^2 z \cdot n^2
\\	
		J^2 \lg J
	&\wwrel=
		z^2 \cdot n^2 \lg n + z^2 \lg z \cdot n^2
\\	
		J \overline J \lg (J) \lg (\overline J)
	&\wwrel=
		z\overline z \cdot n^2 \lg^2 n
		+ z \overline z(\lg z + \lg \overline z) \cdot n^2 \lg n
		+ z \overline z \lg (z) \lg (\overline z) \cdot n^2
\\		J \overline J \lg J
	&\wwrel=
		z \overline z \cdot n^2 \lg n + z \overline z \lg z \cdot n^2.
\end{align*}
The right hand sides are all Hölder-continuous functions in $z\in[0,1]$,
and so the same arguments and error bounds as in \wref{lem:E-Jn-ln-Jn} apply here.
The actual computation is laborious and the expression for $t_v(n)$ 
is too big to state here in full,
but it can easily be found and evaluated for fixed values of $t$ by computer algebra.
We provide a Mathematica notebook for this step as 
supplementary material~\cite{varianceNoteook}.

The incomplete beta integrals resulting form the rewritten expectations
are principally solvable symbolically by partial integration for given values of $t$
and can be expressed using special functions.
A general closed form seems out of reach, though.
We will list numeric approximations for small sample sizes below.

\paragraph{Solution of the recurrence}

Although the above expression for $t_v(n)$ contains terms of order $n^2 \lg^2 n$ and 
$n^2 \lg n$, in all examined cases, these higher-order terms canceled
and left $t_v(n) \sim c n^2$ for an explicitly computable constant $c>0$.
We conjecture that this is always the case, but we did not find a simple proof.
We therefore need the technical assumption that indeed $t_v(n) = \Theta(n^2)$.
Under that assumption,
we obtain an asymptotic approximation for $v(n)$ from \weqref{eq:variance-recurrence} using the CMT (\wref{thm:CMT})
with $\cmtA=2$ and $\cmtB=0$.
Note that the shape function $w(z)$ of the recurrence is exactly the same as for the expected costs
(see \wref{sec:shape-function}).
We thus compute 
\begin{align*}
		H
	&\wwrel=
		1-\int_0^1 z^2\,w(z) \:dz 
\\	&\wwrel= 
		1 - \int_0^1 2\,\left[\tfrac{\alpha}{1+\alpha} < z < \tfrac12 \bin\vee z > \tfrac1{1+\alpha}\right]\, 
					\frac{z^{t+2}(1-z)^{t}}{\BetaFun(t+1,t+1)} \: dz
\\	&\wwrel= 
		1 - 2\frac{(t+1)^{\overline 2}}{(k+1)^{\overline 2}}
				\int_0^1 \left[\tfrac{\alpha}{1+\alpha} < z < \tfrac12 \bin\vee z > \tfrac1{1+\alpha}\right]\,
				\frac{z^{t+2}(1-z)^{t}}{\BetaFun(t+3,t+1)} \: dz
\\	&\wwrel= 
		1 - \frac{t+2}{k+2}\Bigl( I_{\frac{\alpha}{1+\alpha},\frac12}(t+3,t+1)
			+ I_{\frac1{1+\alpha},1}(t+3,t+1)\Bigr)
\numberthis\label{eq:Hv}.
\end{align*}
Since $\frac{t+2}{k+2} \le \frac23$ and the integral over the entire unit interval would be exactly $1$,
we have $H > 0$ for all $\alpha$ and $t$.
So by Case~1 of the CMT, the variance of QuickXsort is
\begin{align*}
		v(n)
	&\wwrel\sim
		\frac{t_v(n)}{H}
\end{align*}
and in particular it is quadratic in $n$, and the leading coefficient can 
be computed symbolically.
\end{proof}

\subsection{Variance for methods with optimal leading term}

Below, we give the leading-term coefficient for the variance
(\ie, $c$ in the terminology of \wref{thm:transfer-variance})
for several values of $\alpha$ and $k$. 
We fix $a=1$, \ie, we consider methods X with optimal leading term;
the constant $b$ of the linear term in $x(n)$ does not influence the 
leading term of the variance.
In the results, we keep $a_v$ as a variable, although for the methods X
of most interest, namely \Mergesort and \algorithmname{ExternalHeapsort},
we actually have $a_v=0$.

\begin{table}[htb]
	\plaincenter{
	\begin{tabular}{r  >{\(} c <{\)} >{\(} c <{\)} >{\(} c <{\)}}
	\toprule
		                    & k=1                 & k=3                 & k=9                   \\
	\midrule
		       $\alpha = 1$ & 0.4344 + 0.2000 a_v & 0.1119 + 0.2195 a_v & 0.01763 + 0.2477 a_v  \\
		$\alpha = \sfrac12$ & 0.4281 + 0.2941 a_v & 0.1068 + 0.3234 a_v & 0.01572 + 0.3632 a_v  \\
		$\alpha = \sfrac14$ & 0.3134 + 0.4413 a_v & 0.0728 + 0.4550 a_v & 0.00988 + 0.4483 a_v \\
	\bottomrule
	\end{tabular}
	}
	\caption{%
		Leading term coefficients of the variance of \QuickXsort.
	}
	\label{tab:variance}
\end{table}

\subsection{Variance in Mergesort}
\label{sec:variance-mergesort}

First note that since \Mergesort's costs differ by $\Oh(n)$ for the best%%
\footnote{%
	We assume here an unmodified standard \Mergesort variant that executes 
	all merges in any case.
	In particular we assume the following folklore trick is \emph{not} used:
	One can check (with one comparison) whether the two runs are already sorted prior to
	calling the merge routine and skip merging entirely if they are. 
	This optimization leads to a linear best case and will increase the variance.
}
and worst case, the variance is obviously in $\Oh(n^2)$.
A closer look reveals that Mergesort's costs are indeed much more concentrated
and the variance is of order $\Theta(n)$:
For a given size $n$, the overall costs are the sum of independent contributions
from the individual merges, each of which has constant variance.
Indeed, the only source of variability in the merge costs is that we do not need 
further comparisons once one of the two runs is exhausted.

More precisely,
for standard top-down mergesort,
$X_n$ can be characterized by (see~\cite{FlajoletGolin1994})
\begin{align*}	
		X_n
	&\wwrel\eqdist
		X_{\lceil n/2\rceil} + X_{\lfloor n/2\rfloor} + n - L_{\lceil n/2\rceil,\lfloor n/2\rfloor}
\\[.3em]
		\Prob{L_{m,n} \le \ell}
	&\wwrel=
		\frac{\binom{n+m-\ell}m + \binom{n+m-\ell}n}{\binom {n+m}m}.
\end{align*}
Following Mahmoud~\cite[eq.~(10.3), eq.~(10.1)]{Mahmoud2000}, we find that 
the variance of the costs for a single merge is constant:
\begin{align*}
		\E{L_{m,n}}
	&\wwrel=
		\frac m{n+1} + \frac n{m+1}
	\wwrel=
		m^{\underline 1} n^{\underline{-1}}
		+n^{\underline 1} m^{\underline{-1}}
\\
		\E{L_{m,n}^{\underline 2}}
	&\wwrel=
		2m^{\underline 2} n^{\underline{-2}}
		+2n^{\underline 2} m^{\underline{-2}}
\\	
		\Var{L_{m,n}}
	&\wwrel=
		\E{L_{m,n}^{\underline 2}} + \E{L_{m,n}} - \E{L_{m,n}}^2
\\	&\wwrel=
		2m^{\underline 2} n^{\underline{-2}}
		+2n^{\underline 2} m^{\underline{-2}}
		+m^{\underline 1} n^{\underline{-1}}
		+n^{\underline 1} m^{\underline{-1}}
		-\Bigl(m^{\underline 1} n^{\underline{-1}}
			+n^{\underline 1} m^{\underline{-1}}
		\Bigr)^2
\\	&\wwrel\le
		2, \qquad\text{for $|m-n|\le 1$}
\end{align*}
which gives an upper bound of $2n$ for the variance.
Precise asymptotic expansions have been
computed by Hwang~\cite{Hwang1998}:
\begin{align*}
		\Var{X_n}
	&\wwrel=
		n \phi(\lg(n)) - 2 + o(1)
\end{align*}
for a periodic function $\phi(x) \in [0.30,0.37]$.

\subsection{Variance in QuickMergesort}

Since the variance of Mergesort is subquadratic,
\wref{thm:transfer-variance} would be applied with $a_v=0$,
and we obtain, \eg, a variance of $0.4281 n^2$ for $k=1$ 
and $0.1068 n^2$ for $k=3$.
Interestingly, these results do not depend on our choice for 
the constant $b$ of the linear term of $x(n)$.

\begin{figure}[tbhp]\smaller
	\plaincenter{
	\includegraphics{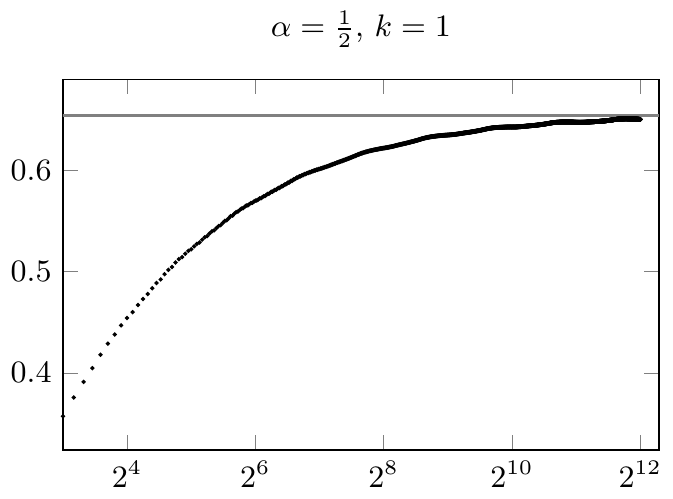}
	\includegraphics{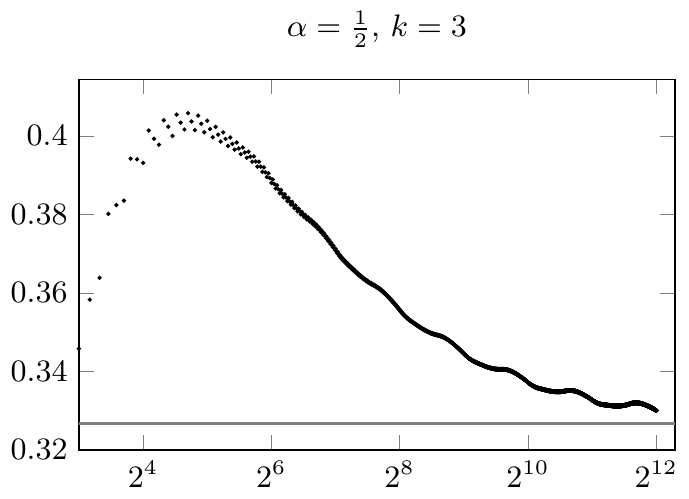}
	}
	\caption{%
		Exact values for the normalized standard deviation in \QuickMergesort
		(computed from the exact recurrence for the second moment)
		and the asymptotic approximation from \wref{tab:variance} (gray line).
		The $x$-axis shows the inputs size $n$ (logarithmic) and the $y$-axis
		is the standard deviation of the number of comparisons divided by $n$.
		The plots show different sample sizes.
	}
	\label{fig:stdevs-exact-values}
\end{figure}

They match empirical numbers quite well.
There is still a noticeable difference in \wref{fig:stdevs-exact-values},
which compares the above approximations with exact values for small $n$
computed from the recurrence.
For large $n$, though, the accuracy is stunningly good,
see \wref{fig:stddevcomps} in the experiments section.

\paragraph{Fine print}
Although our transfer theorem is perfectly valid 
and fits Monte Carlo simulations very well, 
it is formally \emph{not} applicable to \QuickMergesort.
The reason for this are the tiny periodic fluctuations (\wrt $n$) in the cost of \Mergesort
in both expected costs and their variance.

For the expected values, we could use upper and lower bounds for $x(n)$ to derive 
upper and lower bounds for the costs of \QuickXsort.
Determining the precise influence of fluctuations in \QuickXsort's expected cost 
is an interesting topic for future research,
but since the bounds are so close, our approach taken in this paper is certainly
sufficient on practical grounds.
For the variance, this is different.
The variance of \QuickMergesort is influenced by the periodic terms
of the \emph{expected} costs of Mergesort, and simple arguments do not yield
rigorous bounds.

Intuitively \QuickMergesort acts as a smoothing on the costs of Mergesort
since subproblem sizes are random.
It is therefore quite expected to find very smooth periodic influences of small amplitude.
The fact that our estimate does not depend on $b$ or the precise variance of Mergesort 
at all, gives hope that is a very good approximation. But it remains heuristic approximation.

% -*- mode:LaTeX; TeX-master: "quickXsort_journal" -*-
% !TeX spellcheck = en-US

\section{QuickMergesort with base cases}\label{sec:basecase}

In \QuickMergesort, we can improve the number of comparisons even further by sorting 
small subarrays with yet another algorithm Z.
The idea is to use Z only for tiny subproblems, so that
it is viable methods that require extra space and have otherwise prohibitive cost
for other operations like moves.
Obvious candidates for Z are \algorithmname{Insertionsort} and 
\algorithmname{MergeInsertion}.

If we use $\Oh(\log n)$
elements for the base case of \algorithmname{Mergesort}, we have to call Z at most
$\Oh(n/ \log n)$ times. In this case we can allow an overall $\Oh(n^2)$ running time for Z
and still obtain only $\Oh((n/
\log n)\cdot \log^2 n ) = \Oh(n \log n)$ overhead in \QuickMergesort.
We note that for
the following result, we only need that the size of the base cases grows with $n$,
but not faster than logarithmic.

We start by bounding the costs of \Mergesort base case Z.
Reinhardt~\cite{Reinhardt1992} proposes this idea 
using \algorithmname{MergeInsertion} for base cases of constant size
and essentially states the following result,
but does not provide a proof for~it.

\begin{theorem}[\algorithmname{Mergesort} with Base Case]\label{thm:MSbase}
Let Z be some  sorting algorithm with
	$z(n) = n \lg n + (b \pm \epsilon) n + o(n)$ comparisons on average and other operations
taking at most $\Oh(n^2)$ time. If base cases of size $\Oh(\log n)$ are sorted with Z, \ algorithmname{Mergesort} uses at most
 $n \lg n +(b \pm \epsilon)n + o(n)$ comparisons and $\Oh(n\log n)$ other instructions on average.
\end{theorem}

\begin{proof}%[Proof of \wref{thm:MSbase}]
 Since Z uses $ z(n) = n \lg n +(b \pm \epsilon)n + o(n)$ comparisons on average, for every $\delta >0 $ we have $\abs{z(n) -( n\lg n + b n)}\leq (\epsilon + \delta)\cdot n$ for $n$ large enough. Let $k \geq 6$ be large enough such that this bound is satisfied for all $k/2 \leq n \leq k$
 and let $x_{k}(m)$
 denote the average case number of comparisons of \algorithmname{Mergesort} with
 base cases of size $k$ sorted with Z, \ie, $x_{k}(n) = z(n)$ for $n \leq k$.  
 \newcommand{\err}{e}
 
 By induction we will show that 
 \begin{align*}
 		\abs{x_{k}(n) -( n\lg n + b n)} 
	\wwrel\leq 
 		\left(\epsilon + \delta + \frac{ 8}{k} \right)\cdot n -4 
 	\wwrel\equalscolon 
 		\err_k(n)
 \end{align*}
for $n\geq k/2$.

 For $k/2 \leq n \leq k$ this holds by hypothesis, so assume that $n >k$. 
 We have 
 \[x_{k}(n) =   x_{k}(\ceil{n/2}) +x_{k}(\floor{n/2}) + n - \eta(n)\] for some $\eta$ with $1 \leq \eta(n) \leq 2$ for all $n$ (see \eg\ \cite[p.\ 676]{FlajoletGolin1994}). It follows that
 \begin{align*}
 		\abs{x_{k}(n) -( n\lg n + b n)} 	
 	&\wwrel= 
 		\abs{\vphantom{k^k} x_{k}(\ceil{n/2}) +x_{k}(\floor{n/2}) + n - \eta(n) - ( n\lg n + b n)}
\\	&\wwrel{\relwithtext[r]{[inductive hypothesis]}\leq} 
		\err_k(\ceil{n/2}) + \err_k(\floor{n/2}) + \Bigl| 
			\vphantom{k^k}\ceil{n/2}( \lg \ceil{n/2} + b ) 
 \\*[-1.5ex]	&\wwrel\ppe \quad{}
 			+ \floor{n/2}( \lg \floor{n/2} + b )  + n - \eta(n) - ( n\lg n + b n) \vphantom{k^k}
 		\Bigr| 
 \\	&\wwrel\leq 
 		\err_k(n) - 4 + \Bigl| 
 			\vphantom{k^k}\ceil{n/2}( \lg (n/2) + b )
 \\	&\wwrel\ppe \quad{}
 			+ \floor{n/2}( \lg (n/2) + b )  + n - \eta(n) - ( n\lg n + b n) \vphantom{k^k}\Bigr| + 2
 \\	&\wwrel\leq 
 		\err_k(n) -2 + \eta(n) \wwrel\leq 	\err_k(n)						
 \end{align*}
 Notice here that $\lg\ceil{n/2} - \lg (n/2) \leq \frac{1}{\ln(2)\cdot (n+1)} \leq \frac{2}{n} $. This can be easily seen by the series expansion of the logarithm.
 By choosing $k=\lg n$, the lemma follows.
\end{proof}

\Mergesort with base cases can thus be very comparison efficient, but is an external algorithm.
By combining it with \QuickMergesort, we obtain an internal method with essentially the same
comparison cost.
Using the same route as in the proof of \wref{cor:quickmergesort-average-case},
we obtain the formal result.
\begin{corollary}[\algorithmname{QuickMergesort} with Base Case]
\label{cor:QMSbase}
	Let Z be some  sorting algorithm with
	$z(n) = n \lg n + (b \pm \epsilon) n + o(n)$ comparisons on average and other operations
	taking at most $\Oh(n^2)$ time. If base cases of size $\Theta(\log n)$ are sorted with Z, 
	\algorithmname{QuickMergesort} uses at most
	$n \lg n +(b \pm \epsilon)n + o(n)$ comparisons and $\Oh(n\log n)$ other instructions on average.
\end{corollary}

Base cases of growing size
always lead to a constant factor overhead in running time if an algorithm
with a quadratic number of total operations is used. Therefore, in the
experiments we also consider constant size base cases which
offer a slightly worse bound for the number of comparisons, but are
faster in practice. 
A modification of our proof above allows to bound the impact on the number of comparisons,
but we are facing a trade-off between comparisons
and other operations, so the
best threshold for Z depends on the type of data to be sorted and 
the system on which the algorithms run.

\subsection{Insertionsort}\label{sec:is}
We know study the average cost of the natural candidates for Z.
We start with \algorithmname{Insertionsort}, since
it is an elementary method and its analysis is used as part of our
average-case analysis of \algorithmname{MergeInsertion} later.
Recall that \algorithmname{Insertionsort} inserts the elements one by one into the already
sorted sequence by binary search.
For the average number of comparisons we obtain the following result. 

\begin{proposition}[Average Case of \algorithmname{Insertionsort}]\label{pro:avgIns}
The sorting algorithm \algorithmname{Insertionsort} needs $n\lg n - 2 \ln
2 \cdot n + c(n)\cdot n + \Oh(\log n)$ comparisons on average where
$c(n) \in [-0.005,0.005]$.
\end{proposition}
Sorting base cases of logarithmic size in \QuickMergesort with \algorithmname{Insertionsort}, we obtain the next result by \wref{cor:QMSbase}:
\begin{corollary}[\algorithmname{QuickMergesort} with Base Case \algorithmname{Insertionsort}]\label{cor:isbase}
\oneline{Median-of-$\sqrt{n}$ \QuickMergesort} with \algorithmname{Insertionsort} base cases uses at most $n \lg n - 1.38 n + o(n)$ comparisons
and $\Oh(n \log n)$ other instructions on average.
\end{corollary}

\begin{figure}[tbh]
	\plaincenter{\small
	\externalizedpicture{insertionsort-periodic-term}
	\begin{tikzpicture}
		\begin{axis}[
				width=.5\textwidth,height=.3\textwidth,
				yticklabel style={/pgf/number format/.cd,fixed,precision=4}
			]
			\addplot[mark=none] table[x=x,y=f(x)] {pics/insertionsort-periodic-term.tab};
		\end{axis}
	\end{tikzpicture}
	}
	\caption{
		The periodic function in \algorithmname{Insertionsort}
		\(x\mapsto x- 2^x + 1 - \ln 2\cdot(2 -x)\cdot 2^x\) for $x= \lg n - \floor{\lg n} \in [0,1)$.
	}
	\label{fig:periodic-insertionsort}
\end{figure}
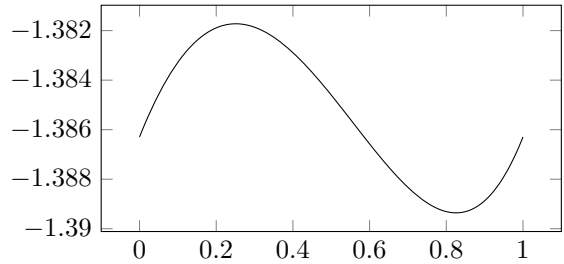

\begin{proof}[\prettyref{pro:avgIns}]
	First, we take a look at the average number of comparisons  $x_{\mathrm{Ins}}(k)$ to insert
	one element into a sorted array of $k-1$ elements by binary insertion.
	To insert a new element into $k-1$ elements either needs $\ceil{\lg
		k}-1$ or $\ceil{\lg k}$ comparisons. There are $k$ positions where
	the element to be inserted can end up, each of which is equally likely. For $2^{\ceil{\lg k}} - k$ of
	these positions $\ceil{\lg k}-1$ comparisons are needed. For the
	other $k - (2^{\ceil{\lg k} } - k ) = 2k -2^{\ceil{\lg k}}$
	positions $\ceil{\lg k}$ comparisons are needed. This means
	\begin{align*}
			x_{\mathrm{Ins}}(k) 
		&\wwrel=
			\frac{(2^{\ceil{\lg k}} - k )\cdot (\ceil{\lg{k}}-1) + (2k -2^{\ceil{\lg k}}) \cdot \ceil{\lg k} }{k}
	\\	&\wwrel= 
			\ceil{\lg k } + 1 - \frac{2^{\ceil{\lg k} }}{k}
	\end{align*}
	comparisons are needed on average. We obtain for the average case for sorting $n$ elements:
	\begin{align*}
			x_{\mathrm{InsSort}}(n) 
		&\wwrel= 
			\sum_{k=1}^{n}  x_{\mathrm{Ins}}(k)= \sum_{k=1}^{n}  \left(\ceil{\lg k } + 1- \frac{2^{\ceil{\lg k}}}{k}\right) 
	\\	&\wwrel{\relwithtext[r]{\cite[5.3.1--(3)]{Knuth1998}}=} 
			n\cdot\ceil{\lg n} -  2^{\ceil{\lg n}} + 1 + n - \sum_{k=1}^{n}  \frac{2^{\ceil{\lg k}}}{k}
			.
	\end{align*}
	We examine the last sum separately. As before we write $H_n =
	\sum_{k = 1}^n\frac{1}{k} = \ln n + \gamma \pm \Oh(\frac{1}{n})$ for 
	the harmonic numbers where $\gamma\in \R$ is Euler's
	constant.
	\begin{align*}
			\sum_{k=1}^{n}  \frac{2^{\ceil{\lg k}}}{k} 
		&\wwrel= 
			1 + \sum_{i=0}^{\ceil{\lg n}-2} \sum_{\ell = 1}^{2^i}\  \frac{2^{i+1}}{2^i+\ell} \;\; 
			+ \sum_{\ell = 2^{\ceil{\lg n}-1} + 1}^{n}  \frac{2^{\ceil{\lg n}}}{\ell}
	\\	&\wwrel= 
			1+\left(\sum_{i=0}^{\ceil{\lg n}-2} 2^{i+1}\cdot\Bigl(H_{2^{i+1}} - H_{2^i}\Bigr) \right)  
			+  2^{\ceil{\lg n}}\cdot\left(H_n - H_{2^{\ceil{\lg n}-1}}\right)
	\\	&\wwrel= 
			\sum_{i=0}^{\ceil{\lg n}-2} 2^{i+1}\cdot\left(\ln \left(2^{i+1}\right) 
			+ \gamma- \ln\left(2^i\right)-\gamma \right)
	\\*	&\wwrel\ppe\quad{}
			+ \left(\ln\left(n\right) + \gamma -\ln\bigl(2^{\ceil{\lg n}-1}\bigr) 
			- \gamma \right)\cdot 2^{\ceil{\lg n}} 
			\wbin\pm \Oh(\log n)
	\\	&\wwrel= 
			\ln 2 \cdot \sum_{i=0}^{\ceil{\lg n}-2} 2^{i+1} \;
			+ \Bigl(\lg(n)\cdot \ln 2  -({\ceil{\lg n}-1})\cdot \ln 2\Bigr)\cdot 2^{\ceil{\lg n}} 
			\wbin\pm \Oh(\log n)
	\\	&\wwrel= 
			\ln 2 \cdot \left( 2\cdot \bigl(2^{\ceil{\lg n}-1} -1\bigr) 
				+ (\lg n -\ceil{\lg n} + 1)\cdot 2^{\ceil{\lg n}}\right) 
			\wbin\pm \Oh(\log n)
	\\	&\wwrel= 
			\ln 2 \cdot \bigl( 2  + \lg n -\ceil{\lg n}\bigr)\cdot 2^{\ceil{\lg n}} 
			\wbin\pm \Oh(\log n).
	\end{align*}
The error term of $\Oh(\log n)$ is due to the fact that for any $C$ we have $\sum_{i=0}^{\ceil{\lg n} - 2}2^{i+1}\cdot \frac{C}{2^i}= 2C(\ceil{\lg n} - 2)$. Hence, we have 
	\begin{align*}
			x_{\mathrm{InsSort}}(n)  
		&\wwrel=
			n\cdot\ceil{\lg n} -  2^{\ceil{\lg n}}  
			+ n- \ln 2 \cdot \left( 2  + \lg n -\ceil{\lg n}\right)\cdot 2^{\ceil{\lg n}}
			\wbin+ \Oh(\log n).
	\end{align*}
	In order to obtain a numeric bound for $x_{\mathrm{InsSort}}(n)$,
	we compute $(x_{\mathrm{InsSort}}(n) - n\lg n)/n$ and then
	replace $\ceil{\lg n} - \lg n$ by $x$. This yields a function
	\[x\wwrel\mapsto x- 2^x + 1 - \ln 2\cdot(2 -x)\cdot 2^x,\]
	which oscillates between $-1.381$ and $-1.389$ for $0\leq x < 1$; see also \wref{fig:periodic-insertionsort}. 
	For
	$x=0$, its value is $2 \ln 2 \approx 1.386$.
\end{proof}

\subsection{MergeInsertion}
\label{sec:MI}\label{sec:mi}

\algorithmname{MergeInsertion} by Ford and Johnson \cite{FordJ59} 
is one of the best sorting algorithms in terms of number of comparisons. Applying it for sorting base cases of \algorithmname{QuickMergesort} yields even better results than \algorithmname{Insertionsort}. W give a brief description of the algorithm and analyze its average case for a simplified version.
Algorithmically, \algorithmname{MergeInsertion}$(s_0,\ldots,s_{n-1})$  
can be
described as follows (an intuitive example for $n=21$ can be found in~\cite{Knuth1998}):

\begin{enumerate}
\item Arrange the input such that $s_i \geq
  s_{i+\floor{n/2}}$ for $0 \leq i < \floor{n/2}$ with one comparison
  per pair. Let $a_i = s_i$ and $b_i = s_{i+\floor{n/2}} $ for
$0 \leq i < \floor{n/2}$, and $b_{\floor{n/2}} = s_{n-1}$ if $n$ is odd.
\item Sort the values $a_0{,}...{,}a_{\lfloor n/2 \rfloor-1}$
  recursively with \algorithmname{MergeInsertion}.
\item 
Rename the solution as follows:  $b_0 \le a_0 \le a_1 \le \dots \le a_{\lfloor n/2
   \rfloor-1}$ and insert the elements  $b_1,\ldots,b_{\lceil n/2
  \rceil-1}$ via binary insertion, following the ordering $b_2$,
$b_1$; $b_4$, $b_3$; $b_{10}$, $b_9, \dots,b_5,\dots$; $b_{t_{k-1}-1}, \dots b_{t_{k-2}}$; $b_{t_{k}-1}, \dots$ 
into the main chain, where  $t_k = (2^{k+1}+(-1)^k)/3$ using (at most) $k$ comparisons for the elements $b_{t_{k}-1}, \dots, b_{t_{k-1}}$.
\end{enumerate}

While the description is simple, \algorithmname{MergeInsertion} is not easy to implement efficiently because of the different renamings, the recursion, and the
insertion in the sorted list. 
Our proposed implementation of \algorithmname{MergeInsertion} is based on a
tournament tree representation with weak heaps as in \cite{Dut93,EW00}. It uses quadratic time and requires $n \lg n + n$ extra bits.

When inserting some of the $b_i$ with $t_{k-1}\leq i\leq t_k-1$ in the already sorted chain, we know that at most $k$ comparisons are needed. During an actual execution of the algorithm, it might happen, that only $k-1$ comparisons are needed  (if the insertion tree is balanced at least $k-1$ comparisons are needed). This decreases the average number of comparisons. Since the analysis is involved, we analyze a simplified variant, where all elements of one insertion block (i.\,e.\ elements $b_{t_{k}-1}, b_{t_{k-1}-1}, \dots b_{t_{k-1}}$) are always inserted into the same number of elements. Thus, for the elements of the $k$-th block always $k$ comparisons are used ~-- except for the last block $b_{\ceil{n/2}-1}, \dots b_{t_{k}}$. In our experiments we evaluate the simplified and the original variant.

\begin{theorem}[Average Case of \algorithmname{MergeInsertion}]\label{thm:avgMergeIns}\label{thm:mi}
Simplified \algorithmname{MergeInsertion} needs $n\lg n - c(n)\cdot
n + \Oh(\log n)$ comparisons on average, where $c(n) \geq
1.3999$.
\end{theorem}

When applying \algorithmname{MergeInsertion} to sort base cases of size $\Oh(\log n)$ in \algorithmname{QuickMergesort}, we obtain the next corollary from \wref{cor:QMSbase} and \wref{thm:avgMergeIns}.

\begin{corollary}[\algorithmname{QuickMergesort} with Base Case \algorithmname{MergeInsertion}]
\label{cor:mibase}
\oneline{Median-of-$\sqrt{n}$ \algorithmname{QuickMergesort}} with \algorithmname{MergeInsertion}
for base cases needs at most $n \lg n - 1.3999 n + o(n)$ comparisons
and $\Oh(n \log n)$ other instructions on average.
\end{corollary}

Instead of growing-size base cases, we also can sort constant-size base cases with \algorithmname{MergeInsertion}. When the size of the base cases is reasonably small, we can hard-code the MergeInsertion algorithm to get a good practical performance combined with a lower number of comparisons than just \algorithmname{QuickMergesort}. In our experiments we also test one variant where subarrays up to nine elements are sorted with \algorithmname{MergeInsertion}.

\begin{proof}[\prettyref{thm:avgMergeIns}]
	According to Knuth \cite{Knuth1998}, \algorithmname{MergeInsertion} requires at most $W(n) =
	n \lg n - (3- \lg 3) n + n (y +1 - 2^y) +\Oh(\log n)$ comparisons
	in the worst case, where $y =y(n)= \ceil{\lg(3n/4)} - \lg(3n/4) \in [0,1)$. 
	In the following we want to analyze the average savings relative to the worst case. We use the simplified version meaning that the average differs from the worst case only for the insertion of the elements of the last block (in every level of recursion).
	Therefore, let $F(n)$ denote the average number of comparisons of the insertion steps of \algorithmname{MergeInsertion}, i.\,e., all comparisons minus the number of comparisons $P(n)$ for forming pairs (during all recursion steps). It is easy to see that $P(n) = n - \Oh(\log n)$ (indeed, $P(n) = n-1$ if $n$ is a power of two); moreover, it is independent of the actual input permutation. 
	We obtain the recurrence relation
	\begin{align*}
			F(n) 
		&\wwrel= 
			F(\floor{n/2}) + G(\ceil{n/2}),\qquad\text{with}
	\\
			G(m) 
		&\wwrel= 
			(k_m - \alpha_m)\cdot (m - t_{k_m-1}) 
			\bin+ \sum_{j=1}^{k_m-1} j \cdot (t_j - t_{j-1}),
	\end{align*}
	with $k_m$ such that $t_{k_m-1} \le m < t_{k_m}$ and some $\alpha_m \in [0,1]$ (recall that $t_k = (2^{k+1}+(-1)^k)/3$). As we do not analyze the improved version of the algorithm, the insertion of elements with index less or equal $t_{k_m-1}$ requires always the same number of comparisons. Thus, the term $\sum_{j=1}^{k_m-1} j \cdot (t_j - t_{j-1})$ is independent of
	the data. However, inserting an element after $t_{k_m-1}$ may either need $k_m$ or $k_m-1$ comparisons. This is where $\alpha_m$ comes from. Note that $\alpha_m$ only depends on $m$.
	We split $F(n)$ into $F'(n) + F''(n)$ with 
	\begin{align*}
			F'(n) 
		&\wwrel= 
			F'(\floor{n/2}) + G'(\ceil{n/2}) 
		&&\text{and} 
	\\
			G'(m) 
		&\wwrel= 
			(k_m - \alpha_m)\cdot (m - t_{k_m-1})
		&&\text{with $k_m$ such that $t_{k_m-1} \le m < t_{k_m}$,}
	\intertext{and}
			F''(n) 
		&\wwrel= 
			F''(\floor{n/2}) + G''(\ceil{n/2})
		&&\text{and} 
	\\
			G''(m) 
		&\wwrel= 
			\sum_{j=1}^{k_m-1} j \cdot (t_j - t_{j-1}) 
		&&\text{with $k_m$ such that $t_{k_m-1} \le m <  t_{k_m}$.}
	\end{align*}
	
	For the average case analysis, we have that $F''(n)$ is independent of
	the data. For $n \approx (4/3)\cdot 2^k$ we have $G'(n)\approx 0$, and hence, $F'(n)\approx 0$. Since otherwise $G'(n)$ is positive, this shows that approximately for $n \approx (4/3)\cdot 2^k$ the average case matches the worst case and otherwise it is better.

	Now, we have to estimate $F'(n)$ for arbitrary $n$.
	We have to consider the
	calls to binary insertion more closely.
	To insert a new element into an array of $m-1$ elements either needs $\ceil{\lg m}-1$ or $\ceil{\lg m}$ comparisons. For a moment assume that the element is inserted at every position with the same probability. Under this assumption the analysis in the proof of \prettyref{pro:avgIns} is valid, which states that
	\begin{align*}
		x_{\mathrm{Ins}}(m) 
	&\wwrel= 
		\ceil{\lg m } + 1 - \frac{2^{\ceil{\lg m} }}{m}
	\end{align*}
	comparisons are needed on average.

	The problem is that in our case the probability at which position an element is inserted is not uniformly distributed. However, it is monotonically decreasing with the index in the array (indices as in the description in \prettyref{sec:MI}).
	Informally speaking, this is because if an element is inserted further to the left, then for the following elements there are more possibilities to be inserted than if the element is inserted on the right.
	
	Now, \Binaryinsert{} can be implemented such that for an odd number of positions the next comparison is made such that the larger half of the array is the one containing the positions with lower probabilities. (In our case, this is the part with the higher
	indices.)
	That means the less probable positions lie on rather longer paths in the search tree, and hence, the average path length is better than in the uniform case.
	Therefore, we may assume a uniform distribution as an upper bound in the following. 
	
	In each	of the recursion steps we have $\ceil{n/2}-t_{k_{\ceil{n/2}}-1}$ calls to binary
	insertion into sets of size $\ceil{n/2}+t_{k_{\ceil{n/2}}-1}-1$ elements each where as before $t_{k_{\ceil{n/2}}-1} \le \ceil{n/2} < t_{k_{\ceil{n/2}}}$. We write $u_{\ceil{n/2}} = t_{k_{\ceil{n/2}}-1}$.
	Hence, for inserting one element, the difference between the average and the worst case is 
	\[\frac{2^{\ceil{\lg( \ceil{n/2} + u_{\ceil{n/2}})}}}{\ceil{n/2}+u_{\ceil{n/2}}} - 1.\] 
	Summing up, we obtain for the average savings $S(n) = W(n) - (F(n)+ P(n)))$ (recall that $P(n)$ is the number of comparisons for forming pairs) w.\,r.\,t.\ the worst case number $W(n)$ the recurrence
	\[
			S(n) 
		\wwrel\geq 
			S(\floor{n/2}) + \bigl(\ceil{n/2} -u_{\ceil{n/2}}\bigr) \cdot \left(
				\frac{2^{\ceil{\lg (\ceil{n/2}+u_{\ceil{n/2}})}}}{\ceil{n/2}+u_{\ceil{n/2}}} - 1
			\right)
			.
	\]
	For $m\in \R_{>0}$ we write $m = 2^{\ell_m -  \lg 3 + x}$ with $\ell_m \in \Z$ and $x \in [0,1)$ and we set
	\[
			f(m) 
		\wwrel= 
			( m - 2^{\ell_m -  \lg 3})\cdot\left(\frac{2^{\ell_m}}{m + 2^{\ell_m -  \lg 3}} -1\right).
	\]
	Recall that we have $t_k = (2^{k+1}+(-1)^k)/3$ meaning that $k_m - 1$ is the largest exponent such that $2^{k_m - \log 3} + (-1)^k/3 \leq m$. Therefore, $u_m =  2^{\ell_m -  \lg 3}$ and $k_m-1 = \ell_m$ except for the case $m= t_k$ for some odd $k\in\Z$. 
	Assume $m \neq t_k$ for any odd $k\in\Z$; then we have
	\[
			\ceil{\lg (m+u_{m})}
		\wwrel= 
			\ceil{\lg (2^{\ell_m -  \lg 3 + x}+2^{\ell_m -  \lg 3})} 
		\wwrel= 
			\ell_m + \ceil{\lg ((2^{x} + 1)/3))} 
		\wwrel= 
			\ell_m
	\]
	and, hence, $f(m) = ( m - u_m)\cdot\left(\frac{2^{\ceil{\lg( m+u_m)}}}{m+u_m} - 1\right)$.
	On the other hand, if $m= t_k$ for some odd $k\in\Z$, we have $k_m = \ell_m$ and
	\[
			f(t_k) 
		\wwrel\leq 
			t_k \cdot\left(  \frac{2^{k}}{t_k + 2^k/3}-1 \right) %= t_k \cdot\left(  \frac{3\cdot 2^{k}}{2^{k+1}-1 + 2^k}-1 \right) 
		\wwrel= 
			t_k \cdot\left(  \frac{3\cdot 2^{k}}{2^{k+1}-1 + 2^k}-1 \right) 
		\wwrel=
			\frac{t_k}{3\cdot 2^{k}-1 } 
		\wwrel\leq 
			1.
	\]

	Altogether this implies that $f(m)$ and $ ( m - u_m)\cdot\left(\frac{2^{\ceil{\lg( m+u_m)}}}{m+u_m} - 1\right)$ differ by at most some constant (as before $u_m = t_{k_m-1}$). Furthermore, $f(m)$ and $f(m+1/2)$ differ by at most a constant. Hence, we have:
	\[S(n) \wwrel\geq S(n/2) + f(n/2) \bin\pm  \Oh(1).\]
	Since we have $f(n/2) = f(n) /2$, this resolves to 
	\[
			S(n) 
		\wwrel\geq 
			\sum_{i>0} f(n/2^i) \bin\pm \Oh(\log n)
		\wwrel=  
			\sum_{i>0}  f(n) / 2^i \bin\pm \Oh(\log n)
		\wwrel= 
			f(n) \bin\pm \Oh(\log n).
	\]
	With $n = 2^{k -  \lg 3 + x}$ this means 
	\begin{align*}
			\frac{S(n)}{n} 
		&\wwrel=  
			\frac{ 2^{k -  \lg 3 + x} - 2^{k -  \lg 3}}{2^{k -  \lg 3 + x}}\cdot\left(\frac{2^{k}}{2^{k -  \lg 3 + x} 
			+ 2^{k -  \lg 3}} -1\right) \wbin\pm\Oh(\log n / n)
		\\&\wwrel=
			(1-2^{-x})\cdot \left(\frac{3 }{2^{  x} +1} -1\right) \wbin\pm\Oh(\log n / n).
	\end{align*}
Recall that we wish to compute $F(n)+ P(n) \leq W(n)- S(n)$.	
	Writing $F(n)+ P(n)= n\lg n - c(n)\cdot
	n $ with $c (n) \in \Oh(1)$, we obtain with \cite[5.3.1 Ex.\ 15]{Knuth1998}
	\[ 
			c(n) 
		\wwrel\geq 
			- (F(n) - n \lg n )/ n 
		\wwrel= 
			(3- \lg 3)- (y +1 - 2^y)    + S(n)/n,
	\]
	where $y = \ceil{\lg(3n/4)} - \lg(3n/4) \in [0,1)$, i.\,e., $n = 2^{\ell - \lg 3-y}$ for some $\ell \in \Z$. With $y = 1-x$ it follows
	\begin{align}
			c(n) 
		&\wwrel\geq 
			(3- \lg 3) - (1-x +1 - 2^{1-x}) + (1-2^{-x})\cdot\left(\frac{3 }{2^{  x} +1} -1\right) 
		\wwrel> 
			1.3999.
	\label{eq:mibound}
	\end{align}
	This function reaches its minimum in $[0,1)$ for 
	\[x \wwrel= \lg\left(\ln 8 -1+\sqrt{(1-\ln 8)^2-1}\right) \wwrel\approx 0.5713.\] 
\end{proof}

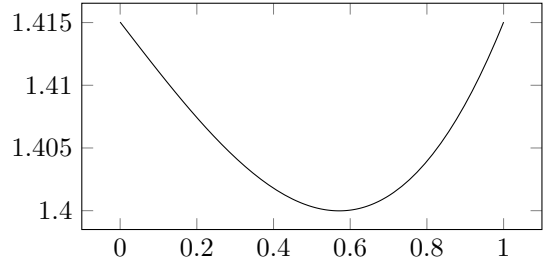
\begin{figure}
	\plaincenter{\small
	\externalizedpicture{mergeinsertion-periodic-term}
	\begin{tikzpicture}
		\begin{axis}[
				width=.5\textwidth,height=.3\textwidth,
				yticklabel style={/pgf/number format/.cd,fixed,precision=4}
			]
			\addplot[mark=none] table[x=x,y=f(x)] {pics/mergeinsertion-periodic-term.tab};
		\end{axis}
	\end{tikzpicture}
	}
	\caption{
		The periodic function in \algorithmname{MergeInsertion}
		\(x\mapsto (3- \lg 3) - (2-x - 2^{1-x}) + (1-2^{-x})\cdot\bigl(\frac{3 }{2^{  x} +1} -1\bigr)\)  for $x= \lg 3n - \floor{\lg 3n} \in [0,1)$.
	}
	\label{fig:periodic-mergeinsertion}
\end{figure}

\begin{remark}[Worst $n$ for MergeInsertion]
	We know that for \Mergesort the optimal input sizes are powers of two. 
	Is the same true for \algorithmname{MergeInsertion}?
	We know that for the worst case, the best $n$ are (close to) $\frac13\cdot 2^k$ for an integer $k$.
	For the average case, we only have the upper bound of \weqref{eq:mibound}. Nevertheless, this should give a reasonable approximation.
	It is not difficult to observe that $c(2^k) = 1.4$:
	For the linear coefficient $e(n)$
	in the worst case costs, $W(n) = n \lg n - e(n) \cdot n + \Oh(\log n)$, we have
	$e(2^k) = (3 - \lg 3) - (y +1 -2^y)$, where
	$y = \bigl\lceil \lg ((3/4) \cdot 2^k) \bigr\rceil - \lg ((3/4) \cdot 2^k)$.
	We know that $y$ can be rewritten as 
	$y = \bigl\lceil\lg (3) + \lg(2^k/4)\bigr\rceil - (\lg 3 + \lg(2^k/4)
	= \ceil{\lg 3} - \lg 3 = 2 - \lg 3$.
	Hence, we have $e(n) = 4/3$.  Finally, we are interested in the value $W(n)-S(n)
	= W(2^k)-S(2^k) = -4/3n - 1/15n = -1.4n$.
	
	Thus, for powers of two the proof of \wref{thm:mi} gives almost the worst bounds, so presumably these are among the worst input sizes for \algorithmname{MergeInsertion} (which also can be seen from the plot in \wref{fig:periodic-mergeinsertion}).
\end{remark}

\begin{remark}[Better bounds?]
	Can one push the coefficient $-1.3999$ even 
	further? 
	Clearly, the non-simplified version of \algorithmname{MergeInsertion} will have a coefficient below $-1.4$ as we can see in our experiments in \wref{fig:base_cases}. A formal proof is lacking, but it should not be very difficult. 
	
	For the simplified version studied here, the empirical numbers from
	\wref{sec:experiments} seem to suggest that our bound is tight. However, there is one step
	in the proof of \wref{thm:mi}, which is not tight (otherwise, we loose only $\Oh(\log
	n)$): in order to estimate the costs of the binary search, we approximated the probability
	distribution where the elements are inserted by a uniform distribution. We conjecture that
	difference between the approximation and the real values is a very small linear term
	meaning that the actual coefficient of the linear term can be still just above or below
	$-1.4$.
	
	Also notice that the exact number of comparisons of the algorithm depends on a 
	small implementation detail: in the binary search it is not completely specified 
	which is the first elements to compare with. 
\end{remark}

\subsection{Combination of (1,2)-Insertion and MergeInsertion}

Iwama and Teruyama~\cite{IwamaTeruyama2017} propose
an improvement of \algorithmname{Insertionsort}, which
inserts a (sorted) \emph{pair} of elements in one step.
The main observation is that the binary searches are good only
if $n$ is close to a power of two, but become more wasteful
for other $n$.
Inserting two elements together helps in such cases.

On the other hand, \algorithmname{MergeInsertion}
is much better than the upper bound in \weqref{eq:mibound} when
$n$ is close to $\frac43\cdot 2^k$ for an integer $k$
(see \wref{fig:periodic-mergeinsertion}).
Using their new (1,2)-\algorithmname{Insertionsort} unless
$n$ is close to $\frac43$ times a power of two,
Iwama and Teruyama obtain a portfolio algorithm ``\algorithmname{Combination}'', 
which needs $n\lg n - c(n)\cdot n + \Oh(\log n)$ comparisons on average, 
where $c(n) \geq 1.4106$. 
(This bound is based on \weqref{eq:mibound} in the proof of \wref{thm:mi}).
The running time of the portfolio algorithm
is at most $\Oh(n^2)$ (in a naive implementation), so that we 
can also use this algorithm as a base case sorter Z.

\begin{corollary}[\algorithmname{QuickMergesort} with Base Case \algorithmname{Combination}]\label{cor:QMSbaseTI}
\oneline{Median-of-$\sqrt{n}$ \algorithmname{QuickMergesort}} with Iwama and Teruyama's
\algorithmname{MergeInsertion/(1,2)-Insertionsort} method for base cases needs at most $n \lg
n - 1.4106 n + o(n)$ comparisons
and $\Oh(n \log n)$ other instructions on average. 
\end{corollary}
In contrast to the original method of Iwama and Teruyama, 
\QuickMergesort with their method for base cases is an internal sorting method
with $\Oh(n\log n)$ running time.

With this present champion in terms of the 
average-case number of comparisons,
we close our investigation of asymptotically optimal sorting methods.
In the following, we will take a look at their actually running times
on realistic input sizes.

\section{Experiments}
\label{sec:experiments}

In this section, we report on studies with efficient implementations
of our sorting methods. 
We conducted two sets of experiments:
	First, 
		we compare our asymptotic approximations 
		with experimental averages for finite $n$
		to assess the influence of lower order terms for 
		realistic input sizes.
	Second,
		we conduct an extensive running-time study to 
		compare \QuickMergesort with other sorting methods from 
		the literature.

\paragraph{Experimental setup} 
We ran thorough experiments with implementations in C++ with different kinds of input
permutations. The experiments are run on an Intel Core i5-2500K CPU (3.30GHz, 4 cores,
32KB L1 instruction and data cache, 256KB L2 cache per core and 6MB L3
shared cache) with 16GB RAM and operating system Ubuntu Linux 64bit
version 14.04.4.  We used GNU's \texttt{g++} (4.8.4); optimized with
flags \texttt{-O3 -march=native}.
For time measurements, we used \texttt{std\dd chrono\dd high\_resolution\_clock}, for generating random inputs, the Mersenne
Twister pseudo-random generator \texttt{std:$\!$:mt19937}. 
All experiments, except those in \wref{fig:runtimeSlow}, were conducted with random permutations of 32-bit integers.

\paragraph{Implementation details}
The code of our implementation of \algorithmname{QuickMergesort} as well as the other
algorithms and our running time experiments is available at
\url{https://github.com/weissan/QuickXsort}. 
In our implementation of \algorithmname{QuickMergesort}, we use the merging procedure from
\cite{ElmasryKS12}, which avoids branch mispredictions. We use the partitioner from the
GCC implementation of \stdsort. For all running time experiments in
\algorithmname{QuickMergesort} we sort base cases up to 42 elements with
\algorithmname{StraightInsertionsort}. When counting the number of comparisons
\algorithmname{StraightInsertionsort} is deactivated and \algorithmname{Mergesort} is used
down to arrays of size two. We also test one variant where base cases up to nine elements
are sorted by a hard-coded \algorithmname{MergeInsertion} variant. 
The median-of-$\sqrt{n}$ variants are always implemented with $\alpha = 1/2$ (notice that
different values for $\alpha$ make very little difference as the pivot is almost always
very close to the median). Moreover, they switch to pseudomedian-of-25 (resp.\ 
pseudomedian-of-9, resp.\ median-of-3) pivot selection for $n$ below 20\,000 (resp.\ 800,
resp.\ 100).

\begin{figure}[htbp]
	\plaincenter{
		\externalizedpicture{qX.base.comp}
		\scalebox{0.79}{\hspace{-1.5mm}\input{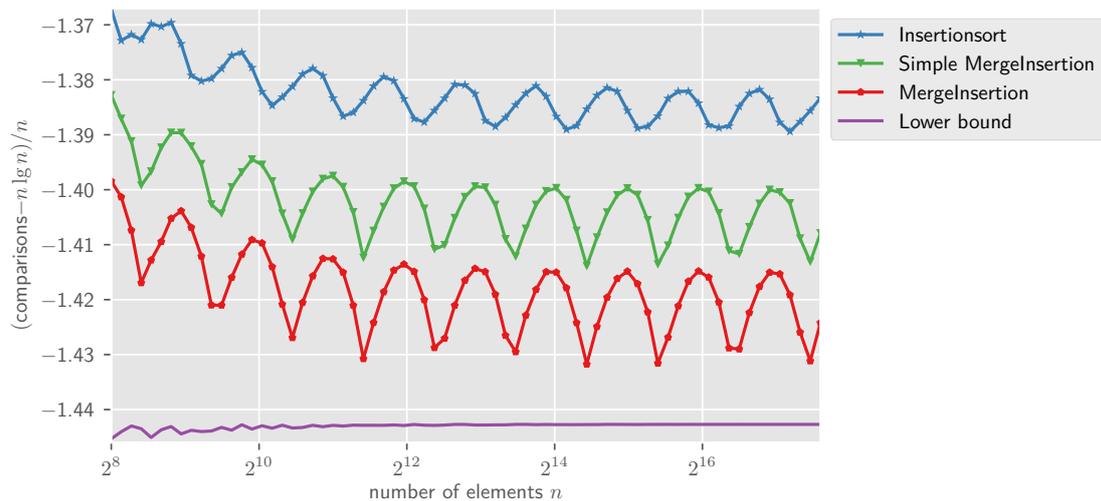}\hspace{35mm}}
	}
	\caption{%
		Coefficient of the linear term of the number of comparisons of 
		\algorithmname{MergeInsertion}, its simplified variant and \algorithmname{Insertionsort}
		(for the number of comparisons $n\lg n + b n$ the value of $b$ is
		displayed).%
	}
	\label{fig:base_cases}
\end{figure}

\subsection{Comparison counts}
\label{sec:comparison-experiments}

The first set of experiments uses our efficient implementations
to obtain empirical estimates for the number of comparisons used.

\paragraph{Base case sorters} 
First, we compare the different algorithms we use as base cases: 
\algorithmname{MergeInsertion}, its simplified variant, and 
\algorithmname{Insertionsort}%
\footnote{%
	For these experiments we use a different experimental setup: 
	depending on the size of the arrays the displayed numbers are 
	averages over 10\,--\,10\,000 runs.% 
}. 
The results can be seen in \wref{fig:base_cases}. 
It shows that both \algorithmname{Insertionsort} and  \algorithmname{MergeInsertion} match
the theoretical estimates very well. Moreover,
\algorithmname{MergeInsertion} achieves results for the
coefficient of the linear term in the range
of $[-1.43,-1.41]$ (for some values of $n$ are even smaller than
$-1.43$). We can see very well the oscillating linear term of
\algorithmname{Insertionsort} (as predicted in \wref{pro:avgIns}) and
\algorithmname{MergeInsertion} (\wref{eq:mibound} for the simple variant).

\paragraph{Number of comparisons of QuickXsort variants}

\begin{figure}[htbp]
	\plaincenter{
		\externalizedpicture{qX.comp}
		\scalebox{0.79}{\hspace{-1.5mm}\input{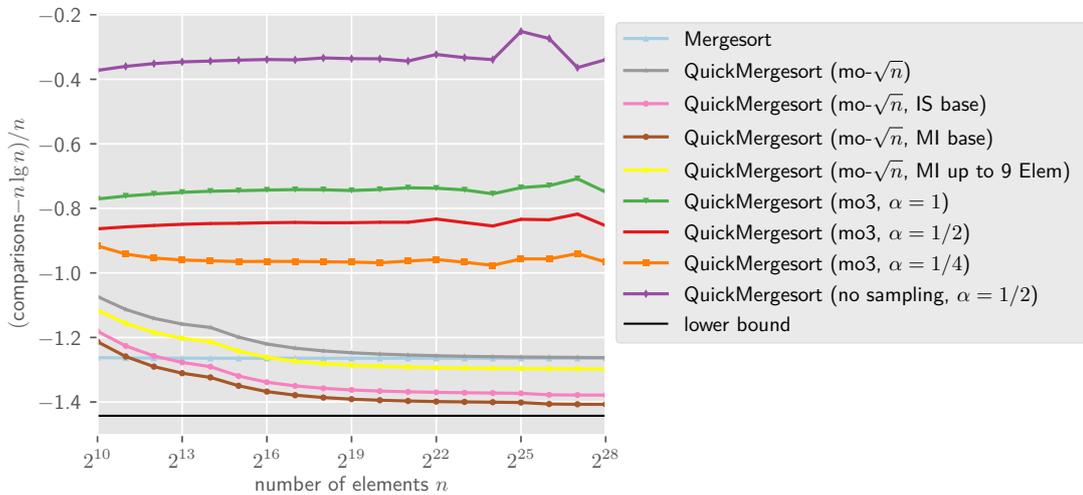}\hspace{58mm}}
	}
%\vspace{16mm}
	\caption{Coefficient of the linear term of the number of comparisons ($(\text{comparisons} - n\lg n)/n$). Median-of$\sqrt{n}$ \algorithmname{QuickMergesort} is always with $\alpha = 1/2$.}\label{fig:comps}
\end{figure}

\begin{figure}[htbp]
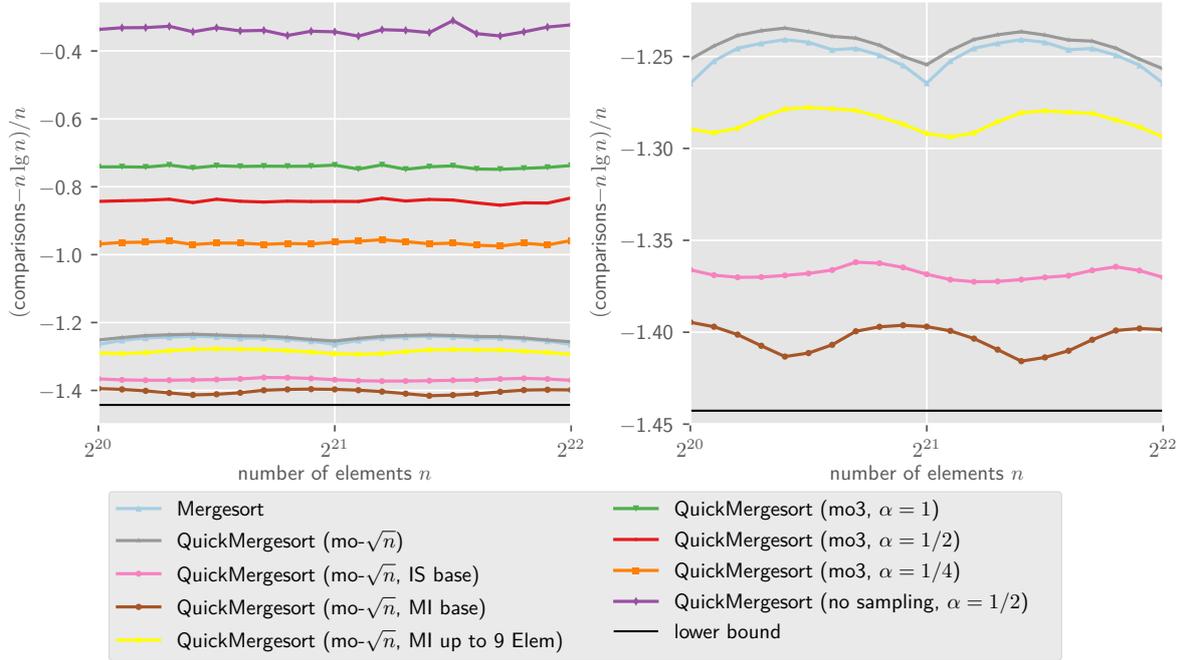

	\plaincenter{
		\externalizedpicture{qX.comp}%
		\scalebox{0.79}{\hspace{-.5mm}\input{pics/qX.comp.dense.pgf}}%
		\externalizedpicture{qX.comp.dense}%
		\scalebox{0.79}{\hspace{-3mm}\input{pics/qX.comp.dense.detail.pgf}}
	}
	\vspace{18mm}
	\caption{%
		Detailed view of the coefficient of the linear term of the number of 
		comparisons ($(\text{comparisons} - n\lg n)/n$) for $n \in [2^{20}..2^{22}]$. 
		Enlarged view of bottom part of the plot.
	}\label{fig:comps_detail}
\end{figure}

We counted the number of comparisons of different \QuickMergesort variants. 
We also include an implementation of top-down \Mergesort which agrees in all relevant
details with the \Mergesort part of our  \QuickMergesort implementation.
The results can be seen in \wref{fig:comps}, \wref{fig:comps_detail}, and
\wref{tab:comps}. Here each data point is the average of 400 measurements (with
deterministically chosen seeds for the random generator) and for each measurement at least
128MB of data were sorted~-- so the values for $n\leq 2^{24}$ are actually averages of
more than 400 runs. From the actual number of comparisons we subtract $n \lg n$ and then
divide by $n$. Thus, we get an approximation of the linear term $b$ in the number of
comparisons $n \lg n + b n + o(n)$.

\begin{table}[htbp]%
	\plaincenter{
		\small	\begin{tabular}{l|cc|cl|c}
		\toprule
			\multirow{2}{*}{Algorithm} &    \multicolumn{2}{c|}{absolute}     &   \multicolumn{2}{c|}{empirical $b$}    &       theoretical $b$        \\
			                           &   $n=2^{22}$    &     $n=2^{28}$     & $n=2^{22}$ & \multicolumn1{c|}{$n=2^{28}$} &        ($n\to\infty$)        \\
		\midrule
			$k=1$, $\alpha = 1/2$      & $90\,919\,646	$ & $7\,425\,155\,999$ &  $-0.323$  & $ -0.339 \pm 0.037$           &     $-0.3407 \pm 0.0119$     \\
			mo3, $\alpha = 1$          & $89\,181\,407	$ & $7\,314\,997\,953$ &  $-0.737$  & $ -0.750 \pm 0.017$           &     $-0.7456 \pm 0.0119$     \\
			mo3, $\alpha = 1/2$        & $88\,780\,825	$ & $7\,287\,011\,306$ &  $-0.833$  & $-0.854 \pm 0.016$            &     $-0.8476 \pm 0.0119$     \\
			mo3, $\alpha = 1/4$        & $88\,254\,970	$ & $7\,256\,806\,284$ &  $-0.958$  & $-0.966 \pm 0.013$            &     $-0.9560 \pm 0.0119$     \\
			mo-$\sqrt{n}$              & $87\,003\,696	$ & $7\,177\,302\,635$ &  $-1.257$  & $-1.262 \pm 4.1\cdot 10^{-5}$ &     $-1.2526 \pm 0.0119$     \\
			mo-$\sqrt{n}$, IS          & $86\,527\,879	$ & $7\,146\,103\,511$ &  $-1.370$  & $-1.379 \pm 5.3\cdot 10^{-6}$ & $-1.3863 \pm 0.005\phantom0$ \\
			mo-$\sqrt{n}$, MI          & $86\,408\,550	$ & $7\,138\,442\,729$ &  $-1.399$  & $-1.407\pm 4.6\cdot 10^{-6} $ &        $\leq -1.3999$        \\
		\bottomrule
		\end{tabular}
	}
	\caption{%
		Absolute numbers of comparisons and linear term 
		($b = (\text{comparisons} - n\lg n)/n$) of \QuickMergesort variants for 
		$n=2^{22}$ and  $n=2^{28}$. 
		We also show the asymptotic regime for $b$ due to 
		\wref{tab:q}, \wref{cor:quickmergesort-average-case}, \wref{cor:isbase} and 
		\wref{cor:mibase}. The $\pm$-terms for the theoretical $b$ represent our lower and upper bound. For the experimental $b$, the $\pm$-terms are the standard error of the mean (standard deviation of the measurements divided by the square-root of the number of measurements).%
	}
	\label{tab:comps}
\end{table}%

In \wref{tab:comps}, we also show the theoretical
values for $b$. We can see that the actual number of comparisons matches the theoretical
estimate very well. In particular, we experimentally confirm that the sublinear terms in
our estimates are negligible for the total number of comparisons (at least for larger
values of $n$). The experimental number of comparisons of \QuickMergesort with
\algorithmname{MergeInsertion} base cases is better than the theoretical estimate because
we analyzed only the simplified variant of \algorithmname{MergeInsertion}. 

For constant-size samples we see that even with $400$ measurements the plots still look a bit bumpy,
particularly for the largest inputs.
Also the difference to the theoretical values is larger for $n=2^{28}$ 
than for $n = 2^{22}$ in \wref{tab:comps}~-- presumably because the average is taken over more
measurements (see setup above).
We note however that the deviations are still within the range we
could expect from the values of the standard deviation (both established theoretically and
experimentally~-- \wref{tab:stddev}): 
for 400 runs, we obtain a standard deviation of approximately $0.65n/\sqrt{400} = 0.0325$. 
Even the largest ``bump'' is thus only slightly over two standard deviations.

In \wref{fig:comps}, we see that median-of-$\sqrt{n}$ \QuickMergesort uses almost 
the same number of comparisons as \Mergesort for larger values of $n$. This shows that
the error terms in \wref{thm:quickXsort} are indeed negligible for practical issues.
The difference between experimental and theoretical values 
for median-of-$\sqrt n$ \QuickMergesort
is due to the fact that the bound holds for arbitrary $n$, but 
the average costs of \Mergesort are actually minimal for powers of two. 

In \wref{fig:comps_detail} we see experimental results for problem sizes which are not
powers of two. 
The periodic coefficients of the linear terms of \algorithmname{Mergesort},
\algorithmname{Insertionsort} and \algorithmname{MergeInsertion} can be observed~-- even
though these algorithms are only applied in \QuickXsort (and for the latter two even only
as base cases in \QuickMergesort). The version with constant size 9 base cases seems to
combine periodic terms of \algorithmname{Mergesort} and \algorithmname{MergeInsertion}.
For the median-of-three version, no significant periodic patterns are visible.
We conjecture that the higher variability of subproblem sizes makes 
the periodic behavior disappear in the noise.

\paragraph{Standard deviation}

Since not only the average running time (or number of comparisons) is of interest, but
also how far an algorithm deviates from the mean running time, we also measure the
standard deviation of the running time and number of comparisons of
\algorithmname{QuickMergesort}. For comparison we also measured two variants of
\algorithmname{Quicksort} (which has a standard deviation similar to
\algorithmname{QuickMergesort}): the GCC implementation of the C++ 
standard sorting function \stdsort (GCC version 4.8.4) and a modified version where the
pivot is excluded from recursive calls and otherwise agreeing with \stdsort. We call the
latter variant simply \algorithmname{Quicksort} as it is the more natural way to implement
\algorithmname{Quicksort}. Moreover, from both variants we remove the final
\algorithmname{StraightInsertionsort} and instead use Quicksort down to size three arrays.

\begin{figure}[htbp]
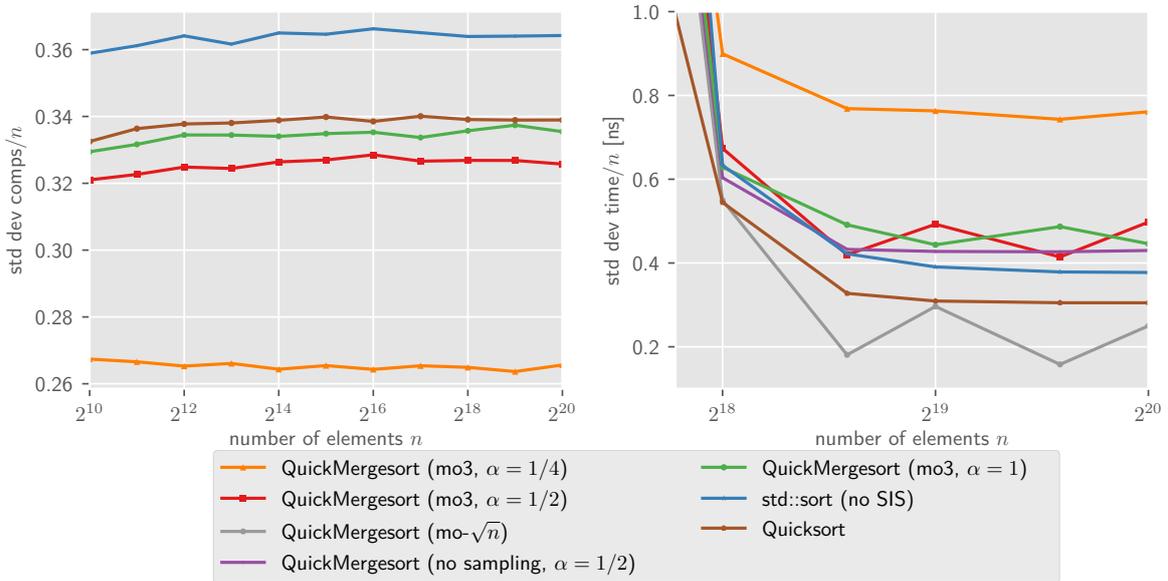

	\rlap{\plaincenter{
		\externalizedpicture{qX.std.comp}%
		\scalebox{0.79}{\hspace{1mm}\input{pics/qX.std.comp.pgf}}%
		\externalizedpicture{qX.std.time}%
		\scalebox{0.79}{\hspace{-4mm}\input{pics/qX.std.time.pgf}}
	}}
	
	\vspace{18mm}
	\caption{Standard deviation of the number of comparisons (left) and the running times (right). For the number of comparisons, median-of-$\sqrt{n}$ \QuickMergesort and \QuickMergesort without pivot sampling are out of range.}\label{fig:stddevcomps}
\end{figure}

In order to get a meaningful estimate of the standard deviation we need many more
measurements than for the mean values. Therefore, we ran each algorithm $40\,000$ times (for
every input size) and compute the standard deviation of these. Moreover, for every
measurement only one array of the respective size is sorted. For each measurement we use a
pseudo-random seed (generated with \texttt{std\dd random\_device}). The results can be
seen in \wref{tab:stddev} and \wref{fig:stddevcomps}.

In \wref{tab:stddev} we also compare the experiments to the theoretical values from
\wref{tab:variance}. Although these theoretical values are only approximate values
(because \wref{thm:transfer-variance} is not applicable to \QuickMergesort), they match
the experimental values very well. This shows that increase in variance due to the
periodic functions in the linear term of the average number of comparisons is negligible.

 Furthermore, we see that choosing the pivot as median-of-3 halves the standard deviation
compared to no sampling. This gives another good reason to always use at least the
median-of-3 version. While the difference between $\alpha = 1$ and $\alpha= 1/2$ is rather
small, $\alpha = 1/4$ gives a considerably smaller standard deviation. Moreover, selecting
the pivot as median-of-$\sqrt{n}$ is far better than median-of-3 (for $N=2^{20}$ the
standard deviation is only around one hundredth).

  All algorithms have a rather large standard deviation of running times for small inputs
(which is no surprise because measurement imprecisions etc.\ play a bigger role here).
Therefore, we only show the results for $n\geq 2^{18}$. Also, while \QuickMergesort with
$\alpha=1/4$ has the smallest standard deviation for the number of comparisons (except
median-of-$\sqrt{n}$) it has the \emph{largest} standard deviation for the running 
time for large~$n$. 
This is probably due to the fact that (our implementation of) Reinhardt's merging
method is not as efficient as the standard merging method. 
Although median-of-$\sqrt{n}$ \QuickMergesort has the smallest standard deviation of
running times, the difference is by far not as large as for the number of comparisons.
This indicates that other factors than the number of comparisons are more relevant for
standard deviation of running times.

We also see that including the pivot into recursive calls in \algorithmname{Quicksort} 
should be avoided.
It increases the standard deviation of both the number of comparisons and the running
time, and also for the average number of comparisons (which we do not show here).

\begin{table}[htbp]%
	\plaincenter{
		\small	\begin{tabular}{l|rr|r}
			\toprule
			\multirow{2}{*}{Algorithm}  									& \multicolumn{2}{c|}{empirical} 	&\multirow{2}{*}{theoretical}   \\
																			&$n=2^{16}$		&$n=2^{20}$ &	\\			
			\midrule
			\algorithmname{Quicksort}  	(mo3)								&0.3385	&0.3389			& 0.3390  	\\
			\algorithmname{Quicksort}  (\stdsort, no SIS)					&0.3662	&0.3642			& --  	\\
			\algorithmname{QuickMergesort}	(no sampling, $\alpha = 1/2$)	&0.6543	&0.6540 		&0.6543	\\
			\algorithmname{QuickMergesort}	(mo3, $\alpha = 1$) 			&0.3353	&0.3355 		&0.3345	\\
			\algorithmname{QuickMergesort}	(mo3, $\alpha = 1/2$) 			&0.3285 &0.3257 		&0.3268	\\
			\algorithmname{QuickMergesort}	(mo3, $\alpha = 1/4$) 			&0.2643	&0.2656 		&0.2698	\\
			\algorithmname{QuickMergesort}	(mo-$\sqrt{n}$) 				&0.0172	&0.00365		& -- 	\\
			\bottomrule
		\end{tabular}
	}
	\caption{%
		Experimental and theoretical values for the standard deviation divided 
		by $n$ of \algorithmname{QuickMergesort} and \algorithmname{Quicksort} 
		(theoretical value for \algorithmname{Quicksort} by \cite[p.~331]{Hennequin89} and 
		for \QuickMergesort by \wref{tab:variance}).
		Recall that for \QuickMergesort, the theoretical value is only a heuristic approximation 
		as \wref{thm:transfer-variance} is not formally applicable with periodic linear terms.
		In light of this, the high precision of all these predictions is remarkable.%
	}
	\label{tab:stddev}
\end{table}%

\subsection{Running time experiments}\label{sec:running-time}

We compare \algorithmname{QuickMergesort} and \algorithmname{QuickHeapsort} with \algorithmname{Mergesort} (our own implementation which is identical with our implementation of \algorithmname{QuickMergesort}, but with using an external buffer of length $n/2$), \algorithmname{Wikisort} \cite{wikisort} (in-place stable Mergesort based on \cite{KimK08}), \stdstablesort (a bottom-up Mergesort, from GCC version 4.8.4), \algorithmname{InSituMergesort} \cite{ElmasryKS12} (which is essentially \QuickMergesort where always the median is used as pivot), and \stdsort (median-of-three Introsort, from GCC version 4.8.4).

\begin{figure}[htbp]
	\plaincenter{
		\externalizedpicture{qX.time.onlyQM.int}
		\scalebox{0.79}{\hspace{-3.5mm}\input{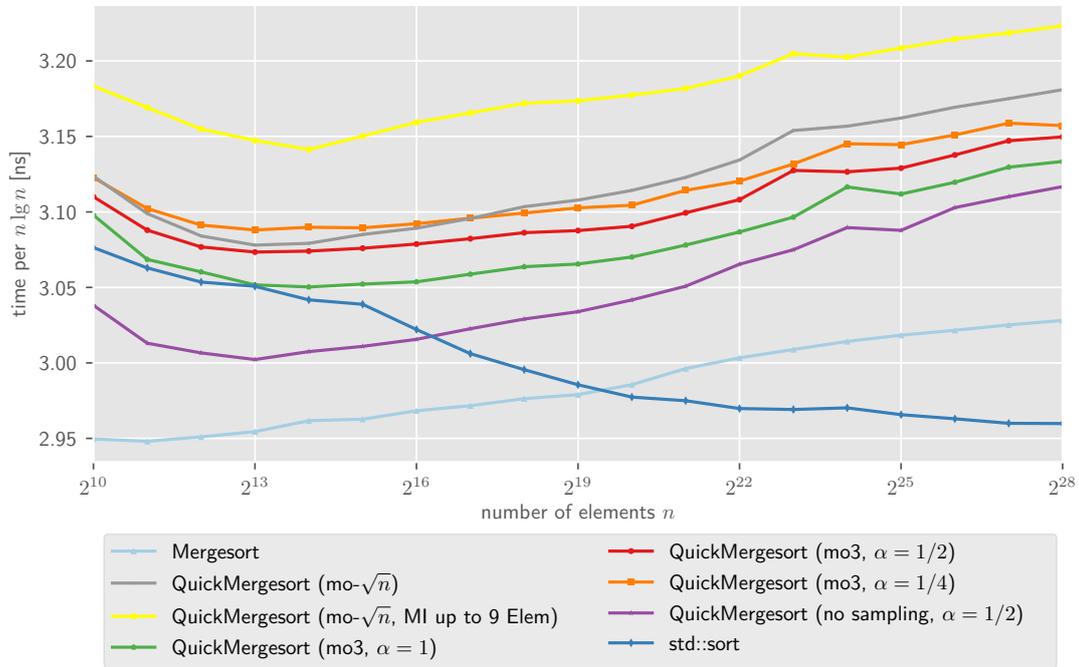}}
	}
	\vspace{15mm}
	\caption{Running times of \algorithmname{QuickMergesort} variants, \Mergesort, and \stdsort when sorting random permutations of integers. }\label{fig:runtimeQM}
\end{figure}

\begin{figure}[htbp]
	\plaincenter{
		\externalizedpicture{qX.time.slow.int}
		\scalebox{0.79}{\hspace{-1.5mm}\input{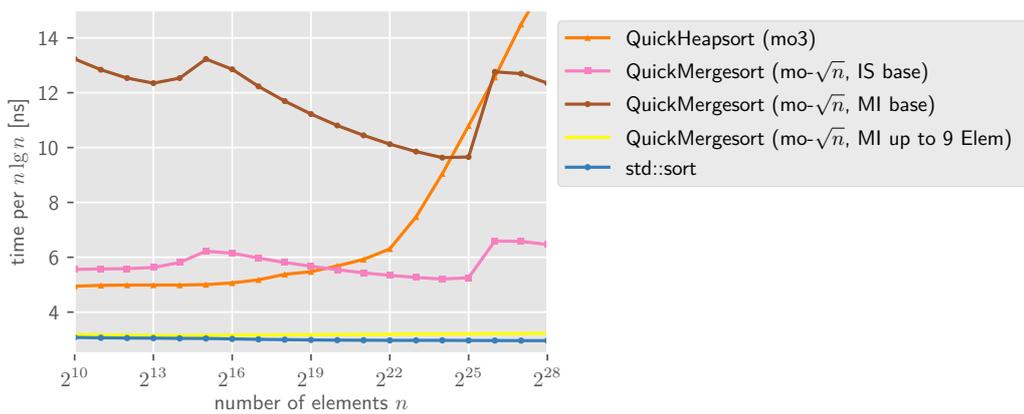}\hspace{58mm}}
	}
	\caption{%
		Running times of \algorithmname{QuickMergesort} variants with base cases and
		\algorithmname{QuickHeapsort} when sorting random permutations of integers.%
	}
	\label{fig:runtimeQMbase}
\end{figure}

\begin{figure}[htbp]
	\plaincenter{
		\externalizedpicture{qX.time.int}
		\scalebox{0.79}{\hspace{-1.5mm}\input{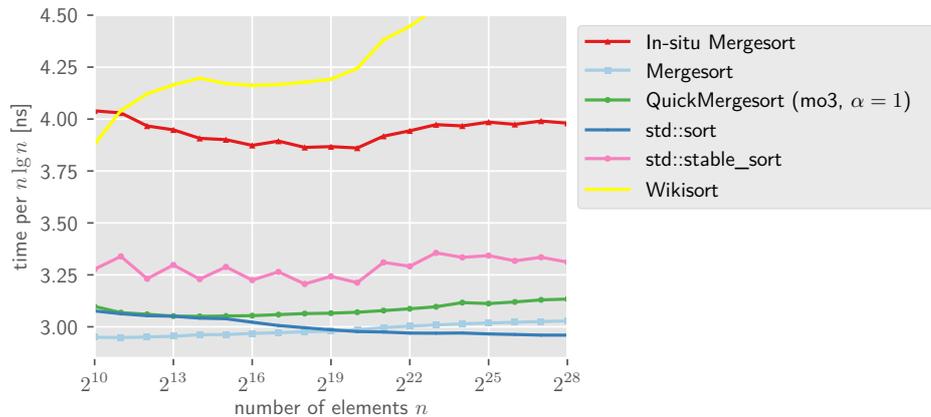}\hspace{58mm}}
	}
	\caption{Running times when sorting random permutations of integers.}
	\label{fig:runtimeOther}
\end{figure}

All time measurements were repeated with the same 100 deterministically chosen
seeds~-- the displayed numbers are the averages of these 100 runs. 
Moreover, for each time measurement, at least 128MB of data were sorted~-- if the array size is smaller, then for this time
measurement several arrays have been sorted and the total
elapsed time measured. The results for sorting 32-bit integers are displayed in  \wref{fig:runtimeQMbase}, \wref{fig:runtimeQM}, and \wref{fig:runtimeOther}, which all contain the results of the same set of experiments~-- we use three different figures because of the large number of algorithms and different scales on the y-axes.

\wref{fig:runtimeQM} compares different \QuickMergesort variants to \Mergesort and
\stdsort. In particular, we compare median-of-3 \QuickMergesort with different values of
$\alpha$. While for the number of comparisons a smaller $\alpha$ was beneficial, it turns
out that for the running time the opposite is the case: the variant with $\alpha=1$ is the
fastest. Notice, however, that the difference is smaller than 1\%. The reason is
presumably that partitioning is faster than merging: for large $\alpha$ the problem sizes
sorted by \Mergesort are reduced and more ``sorting work'' is done by the partitioning.
As we could expect our \Mergesort implementation is faster than all \QuickMergesort
variants~-- because it can do simply moves instead of swaps. 
Except for small $n$, \stdsort beats \QuickMergesort. However, notice that for $n=2^{28}$
the difference between \stdsort and \QuickMergesort without sampling is only approximately
5\%, thus, can most likely be bridged with additional tuning efforts (\eg\ block partitioning \cite{EdelkampW16}).

In \wref{fig:runtimeQMbase} we compare the \QuickMergesort variants with base cases with
\algorithmname{QuickHeapsort} and \stdsort.
 While \algorithmname{QuickHeapsort} has still an acceptable speed for small $n$, it
becomes very slow when $n$ grows. This is presumably due to the poor locality of memory accesses 
in \algorithmname{Heapsort}.  
The variants of \QuickMergesort with growing size base cases
are always quite slow. This could be improved by sorting smaller base cases with the
respective algorithm~-- but this opposes our other aim to minimize the number of
comparisons. Only the version with constant size \algorithmname{MergeInsertion} base cases
reaches a speed comparable to \stdsort (as it can be seen also in \wref{fig:runtimeQM}).

\wref{fig:runtimeOther} shows median-of-3 \QuickMergesort together with the other
algorithms listed above. As we see, \algorithmname{QuickMergesort} beats the other
in-place \algorithmname{Mergesort} variants \algorithmname{InSituMergesort} and
\algorithmname{Wikisort} by a fair margin. However, be aware that
\algorithmname{QuickMergesort} (as well as \algorithmname{InSituMergesort}) neither
provides a guarantee for the worst case nor is it a stable algorithm.

\begin{figure}[htbp]
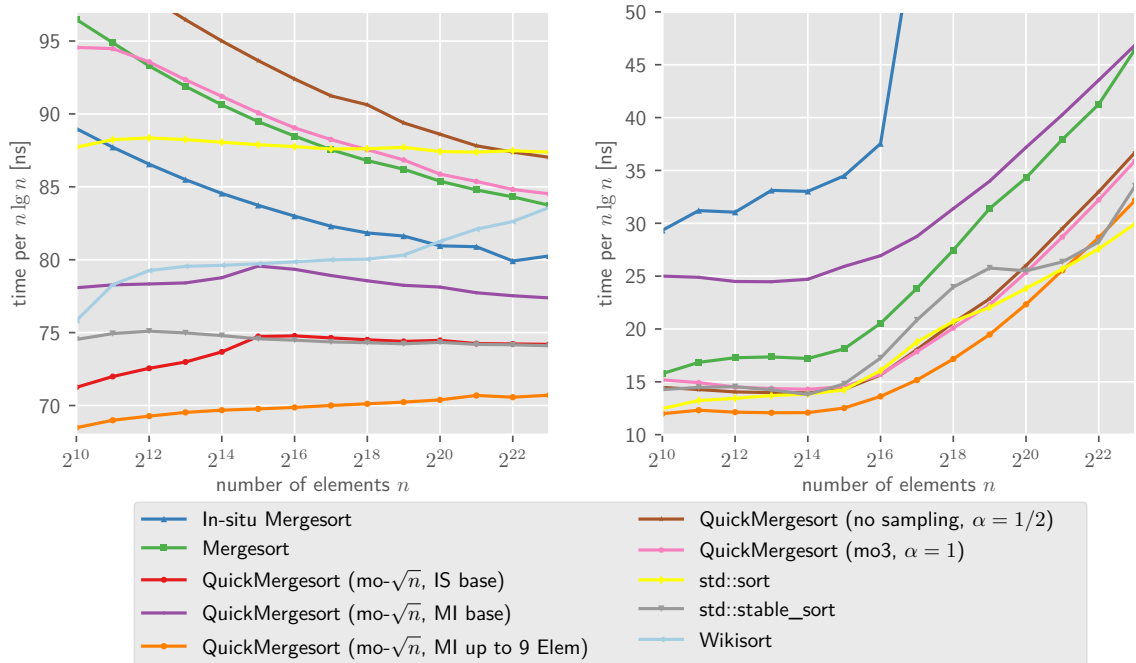

	\plaincenter{
		\externalizedpicture{qX.time.Logint}%
		\scalebox{0.79}{\hspace{-.5mm}\input{pics/qX.time.Logint.pgf}}%
		\externalizedpicture{qX.time.PointerRecord}%
		\scalebox{0.79}{\hspace{-4mm}\input{pics/qX.time.PointerRecord.pgf}}
	}
	\vspace{18mm}
	\caption{%
		Running times when sorting random permutations of ints with special comparison function 
		(computing the log in every comparison~-- left) and pointers to Records (right). 
		\algorithmname{Wikisort} did not run for sorting pointers and 
		\algorithmname{QuickMergesort} with \algorithmname{Insertionsort} base cases is out of range.%
	}
	\label{fig:runtimeSlow}
\end{figure}

\paragraph{Other data types}
While all the previous running time measurements were for sorting 32-bit integers, in
\wref{fig:runtimeSlow}  we also tested two other data types: 
(1) 32-bit integers with a special comparison function which before every comparison computes the logarithm of the
operands, and 
(2) pointers to records of 40 bytes which are compared by the first 4 bytes.
Thus in both cases, comparisons are considerably more expensive than for standard
integers. 
Each record is allocated on the heap with \texttt{new}~-- since we do this in
increasing order and only shuffle the pointers, 
we expect them to reside memory in close-to-sorted order.  

For both data types,
\algorithmname{QuickMergesort} with constant size \algorithmname{MergeInsertion} base
cases is the fastest (except when sorting pointers for very large $n$). 
This is plausible since it combines the best of two worlds: 
on one hand, it has an almost minimal number of comparisons, on the other
hand, it does not induce the additional overhead for growing size base cases. Moreover,
the bad behavior of the other \algorithmname{QuickMergesort} variants (``without'' base
cases) is probably because we sort base cases up to 42 elements with
\algorithmname{StraightInsertionsort}~-- incurring many more comparisons (which we did not
count in \wref{sec:comparison-experiments}).

\section{Conclusion}\label{sec:conclusion}

Sorting $n$ elements remains a fascinating topic for computer
scientists both from a theoretical and from a practical point of view. 
With \QuickXsort we have described a procedure to convert an 
external sorting algorithm into an internal one introducing 
only a lower order term of additional comparisons on average. 

We examined \algorithmname{QuickHeapsort} and \algorithmname{QuickMergesort} 
as two examples for this construction.
\algorithmname{QuickMergesort}  is close to the lower bound for the 
average number of comparisons and at the same time is efficient 
in terms of running time, even when the comparisons are fast.

Using \algorithmname{MergeInsertion} to sort base cases of growing size 
for \algorithmname{QuickMergesort}, we derive an 
an upper bound of
$n \lg n - 1.3999n + o(n)$ comparisons for the average case. 
Using the recent algorithm by Iwama and Teruyama~\cite{IwamaTeruyama2017} 
this can be improved even further to $n \lg n - 1.4106n + o(n)$,
without causing the overall operations to become more than $\Oh(n\log n)$. Thus, the average of our best
implementation has a proven gap of at most
$0.0321n + o(n)$ comparisons to the lower bound.
Of course, there is still room in closing the gap to the lower bound of
$n \lg n - 1.44n + \Oh(\log n)$ comparisons.

This illustrates one underlying strength of the framework architecture of
\algorithmname{QuickXsort}: by applying the transfer results as shown in this paper 
\algorithmname{QuickXsort} directly participates in advances to the performance of
algorithm X. Moreover, our experimental results suggest that the bound of 
$n \lg n - 1.43n + \Oh(\log n)$ element comparisons may be beaten at least 
for some values of $n$. This very close gap between the lower and upper bound manifested
in the
second order (linear) term makes the sorting problem a fascinating topic and mainstay for
the analysis of algorithms in general. 

We were also interested in the practical performance of
\algorithmname{QuickXsort} and study variants with smaller sampling sizes for the pivot in
great detail.
Besides average-cases analyses, variances were analyzed. 
The established close mapping of the theoretical results with the empirical findings
should be taken as a convincing arguments for the preciseness of the mathematical
derivations.

\paragraph{Open questions}

Below, we list some possibilities for extensions of this work.

\begin{itemize}
\item 
	By \wref{thm:quickXsort} for the average number of comparisons sample sizes of
	$\Theta(\sqrt{n})$ are optimal among all polynomial size samples. However, it remains open
	whether $\Theta(\sqrt{n})$ sample sizes are also optimal among \emph{all} (also
	non-polynomial) sample sizes.
\item 
	In all theorems, we only use $\Theta$ (or $\Oh$) notation for sublinear terms and only give upper and lower bounds for the periodic linear terms. Exact
	formulas for the average number of comparisons of \QuickXsort are still open and also
	would be a tool to find the exact optimal sample sizes.
\item 
	In this work the focus was on expected behavior. Nevertheless, in practice often also
	guarantees for the worst case are desired. In \wref{thm:QuickXYsort}, we did a first step
	towards such guarantees. Moreover, in \cite{EdelkampWeiss2018MQMS}, we examined the same
	approach in more detail. Still there are many possibilities for good worst-case guarantees
	to investigate.
\item 	
	In \wref{thm:transfer-variance} we needed the technical conjecture that the variance is
	in $\Oh(n^2)$ since we only could show it for special values of $k$ and $\alpha$. Hence,
	it remains to find a general proof (or disproof) that the variance is always in $\Oh(n^2)$
	for constant size samples.
	This issue becomes even more interesting when fluctuations in the expected costs
	of X are taken into account.
\item 
	What is the order of growth of the variance of \QuickXsort for growing size samples for
	pivot selection.
\item 
	We only analyzed the simplified variant of \algorithmname{MergeInsertion}. The average
	number of comparisons of the original variant still is an open problem and seems rather
	difficult to attack. Nevertheless, better bounds than just the simplified version should
	be within reach.
\item 
	Further future research avenues are to improve the empirical behavior for large-scale inputs and to study options for parallelization.
\end{itemize}

\bibliographystyle{plainurl}
\bibliography{heaps,shortstrings,sort,quick-mergesort}

\clearpage
\appendix
\section*{Appendix}

\section{Notation}
\label{app:notation}

\subsection{Generic mathematics}
\begin{notations}
\notation{$\N$, $\N_0$, $\Z$, $\R$}
	natural numbers $\N = \{1,2,3,\ldots\}$, 
	$\N_0 = \N \cup \{0\}$,
	integers $\Z = \{\ldots,-2,-1,0,1,2,\ldots\}$,
	real numbers $\R$.
\notation{$\R_{>1}$, $\N_{\ge3}$ etc.}
	restricted sets $X_\mathrm{pred} = \{x\in X : x \text{ fulfills } \mathrm{pred} \}$.
%\notation{$0.\overline 3$}
%	repeating decimal; $0.\overline3 = 0.333\ldots = \frac13$; \\
%	numerals under the line form the repeated part of 
%	the decimal number.
\notation{$\ln(n)$, $\lg(n)$, $\log n$}
	natural and binary logarithm; $\ln(n) = \log_e(n)$, $\lg(n) = \log_2(n)$.
	We use $\log$ for an unspecified (constant) base in $\Oh$-terms
\notation{$X$}
	to emphasize that $X$ is a random variable it is Capitalized.
\notation{$[a,b)$}
	real intervals, the end points with round parentheses are excluded, 
	those with square brackets are included.
\notation{$[m..n]$, $[n]$}
	integer intervals, $[m..n] = \{m,m+1,\ldots,n\}$;
	$[n] = [1..n]$.
\notation{$[\text{stmt}]$, $[x=y]$}
	Iverson bracket, $[\text{stmt}] = 1$ if stmt is true, $[\text{stmt}] = 0$ otherwise.
\notation{$\harm n$}
	$n$th harmonic number; $\harm n = \sum_{i=1}^n 1/i$.
\notation{$x \pm y$}
	$x$ with absolute error $|y|$; formally the interval $x \pm y = [x-|y|,x+|y|]$;
	as with $\Oh$-terms, we use one-way equalities $z=x\pm y$ instead of $z \in x \pm y$.
\notation{$\binom nk$}
	binomial coefficients; $\binom nk = n^{\underline k} / k!$.
\notation{$\BetaFun(\betL,\betR)$}
	for $\betL,\betR\in\R_+$;
	the beta function, $\BetaFun(\betL,\betR) = \int_0^1 z^{\betL-1}(1-z)^{\betR-1}\, dz$;
	see also \wpeqref{eq:betaFun}
\notation{$I_{x,y}(\betL,\betR)$}
	the regularized incomplete beta function;
	$I_{x,y}(\betL,\betR)=
			\int_x^y \frac{z^{\betL-1}(1-z)^{\betR-1}}{\BetaFun(\betL,\betR)} \, dz$
			for $\betL,\betR\in\R_+$, $0\le x\le y\le 1$.
\notation{$a^{\underline b}$, $a^{\overline b}$}
	factorial powers;
	``$a$ to the $b$ falling resp.\ rising'';
	\eg, $x^{\underline 3} = x(x-1)(x-2)$, $x^{\underline{-3}} = 1/((x+1)(x+2)(x+3))$.
\end{notations}

\subsection{Stochastics-related notation}
\begin{notations}
\notation{$\Prob{E}$, $\Prob{X=x}$}
	probability of an event $E$ resp.\ probability for random variable $X$ to
	attain value $x$.
\notation{{$\E{X}$}}
	expected value of $X$; we write $\E{X\given Y}$ for the conditional expectation
	of $X$ given $Y$, and $\Eover X{f(X)}$ to emphasize that expectation is taken 
	\wrt random variable $X$.
\notation{$X\eqdist Y$}
	equality in distribution; $X$ and $Y$ have the same distribution.
\notation{$\uniform(a,b)$}
	uniformly in $(a,b)\subset\R$ distributed random variable. 
\notation{$\betadist(\betL,\betR)$}
	Beta distributed random variable with shape parameters $\betL\in\R_{>0}$ and $\betR\in\R_{>0}$.
\notation{$\binomial(n,p)$}
	binomial distributed random variable with $n\in\N_0$ trials and success probability $p\in[0,1]$.\\
\notation{$\betaBinomial(n,\betL,\betR)$}
	beta-binomial distributed random variable;
	$n\in\N_0$, $\betL,\betR\in\R_{>0}$;
\end{notations}

\subsection{Specific notation for algorithms and analysis}
\begin{notations}
\notation{$n$}
	length of the input array, \ie, the input size.
\notation{$k$, $t$}
	sample size $k\in \N_{\ge1}$, odd; $k=2t+1$, $t\in\N_0$;
	we write $k(n)$ to emphasize that $k$ might depend on $n$.
\notation{$w$}
	threshold for recursion, for $n\le w$, we sort inputs by X;
	we require $w\ge k-1$.
\notation{$\alpha$}
	$\alpha\in[0,1]$;
	method X may use buffer space for $\lfloor \alpha n\rfloor$ elements.
\notation{$c(n)$}
	expected costs of \QuickXsort; see \wref{sec:recurrence}.
\notation{$x(n)$, $a$, $b$}
	expected costs of X,
	$x(n) = a n \lg n + b n \pm o(n)$; see \wref{sec:recurrence}.
\notation{$J_1$, $J_2$}
	(random) subproblem sizes; $J_1+J_2 = n-1$;
	$J_1 = t + I_1$;
\notation{$I_1$, $I_2$}
	(random) segment sizes in partitioning;
	$I_1\eqdist \betaBinomial(n-k, t+1,t+1)$;
	$I_2 = n-k-I_1$;
	$J_1 = t + I_1$
\notation{$R$}
	(one-based) rank of the pivot; $R=J_1+1$.
\notation{$s(k)$}
	(expected) cost for pivot sampling, \ie, cost for choosing median of $k$ elements.
\notation{$A_1$, $A_2$, $A$}
	indicator random variables; $A_1 = [\text{left subproblem sorted recursively}]$;
	see \wref{sec:recurrence}.
\end{notations}

\end{document}